\documentclass[sigconf]{acmart}

\usepackage{amssymb}
\usepackage{amsfonts}
\usepackage{graphicx}
\usepackage{bbm}
\usepackage{diagbox}
\usepackage{algpseudocode}
\usepackage{amsmath}
\usepackage{float}
\usepackage{tabularx}
\usepackage{slashed}
\usepackage{algorithm}
\usepackage{algorithmicx} 
\usepackage{subfigure}
\usepackage{libertine}
\usepackage{tikz}
\usepackage{verbatim}
\usepackage{tablefootnote}
\usepackage{enumitem}
\usepackage{color}
\usepackage{xspace}
\usepackage{hyperref}
\usepackage[noabbrev,capitalize,nameinlink]{cleveref}
\usepackage{multirow}
\usepackage{threeparttable}

\newcommand{\ourname}{QuerySplit\xspace}

\newcommand{\postgres}{PostgreSQL\xspace}
\newcommand{\RelationshipCenter}{Foreign-Key-Center\xspace}
\newcommand{\RCenter}{FK-Center\xspace}

\newcommand{\EntityCenter}{Primary-Key-Center\xspace}
\newcommand{\ECenter}{PK-Center\xspace}

\AtBeginDocument{
  \providecommand\BibTeX{
        {
            \normalfont B
            \kern-0.5em
            {
                \scshape i
                \kern-0.25em b
            }
            \kern-0.8em
            \TeX
        }
    }
}
\newtheorem{Definition}{Definition}
\newtheorem*{Theorem}{Theorem}
\newtheorem*{Proof}{Proof}

\setcopyright{acmcopyright}
\copyrightyear{2022}
\acmYear{2022}
\acmDOI{XXXXXXX.XXXXXXX}

\acmConference[SIGMOD '23]{June 18--23, 2023}{Seattle, WA, USA}

\begin{document}

\acmPrice{15.00}
\acmISBN{978-1-4503-XXXX-X/18/06}

\title{Efficient Query Re-optimization with Judicious Subquery Selections}

\author{Junyi Zhao}
\email{zhaojy20@mails.tsinghua.edu.cn}
\affiliation{
  \institution{Institute for Interdisciplinary Information Science, Tsinghua University}
  \city{Beijing}
  \country{China}
}

\author{Huanchen Zhang}
\email{huanchen@tsinghua.edu.cn}
\affiliation{
    \institution{Institute for Interdisciplinary Information Science, Tsinghua University}
    \city{Beijing}
    \country{China}
}
\affiliation{
    \institution{Shanghai Qi Zhi Institute}
    \city{Shanghai}
    \country{China}
}

\author{Yihan Gao}
\email{gaoyihan@mail.tsinghua.edu.cn}
\affiliation{
    \institution{Institute for Interdisciplinary Information Science, Tsinghua University}
    \city{Beijing}
    \country{China}
}

\begin{abstract}
Query re-optimization is an adaptive query processing technique
that re-invokes the optimizer at certain points in query execution.
The goal is to dynamically correct the cardinality estimation errors
using the statistics collected at runtime to adjust the query plan
to improve the overall performance.
We identify a key weakness in existing re-optimization algorithms:
their subquery division and re-optimization trigger strategies
rely heavily on the optimizer's initial plan, which can be far away
from optimal.
We, therefore, propose \ourname, a novel re-optimization algorithm
that skips the potentially misleading global plan and instead generates
subqueries directly from the logical plan as the basic re-optimization
units.
By developing a cost function that prioritizes the execution of less
``damaging'' subqueries, \ourname successfully postpones (sometimes
avoids) the execution of complex large joins to maximize their
probability of having smaller input sizes.
We implemented \ourname in \postgres and compared our solution against
four state-of-the-art re-optimization algorithms using the Join Order
Benchmark.
Our experiments show that \ourname reduces the benchmark execution time
by $35\%$ compared to the second-best alternative.
The performance gap between \ourname and an optimal optimizer is within
$4\%$.

% Query re-optimization has been proposed for decades to solve the cardinality estimation problem. Re-optimization executes the query partially, obtains some true cardinalities and runtime statistics, and then invokes the optimizer again to refine the remaining physical plan. Currently, re-optimization each time chooses a subtree from the global physical plan as the partial query to be executed. However, if the global physical plan is far away from optimal, the subtree chosen from a physical plan may itself be a bad plan, and damage the subsequent re-optimization steps. To handle this weakness, we propose a novel re-optimization framework, \ourname. It skips the potentially misleading physical plan, extracts subqueries directly from the logical plan and each time chooses the simplest subquery to execute. By integrating \ourname into \postgres, we get faster execution speed on Join Order Benchmark than other baselines and achieve near-optimal execution time.
\end{abstract}

\maketitle

\section{Introduction}
\label{S:intro}
Given a query, a cost-based optimizer in a relational database management system (DBMS) enumerates
a subset of valid plans through dynamic programming and computes the cost for each plan by feeding
the estimated cardinalities of the intermediate results to the cost model.
If such estimations are way off, no matter how precise the cost model is, the optimizer is likely to choose
a sub-optimal plan, thus slowing down the query~\cite{leis2015good}.
Unfortunately, it is difficult to get accurate cardinality estimations (CE) consistently, especially for joins
because columns are often correlated in real-world data sets~\cite{wu2016sampling,perron2019learned}. Researchers have proposed new approaches beyond conventional histograms, including multidimensional histogram~\cite{gunopulos2005selectivity},
sketch~\cite{rusu2008sketches, cai2019pessimistic}, sampling~\cite{leis2017cardinality, wu2016sampling} and machine learning~\cite{wang2020we} to improve on the CE accuracy. None of them, however, is robust enough to be able to declare victory in
solving the problem~\cite{leis2015good, leis2018query, perron2019learned}.

The intrinsic difficulty of cardinality estimation calls for alternative approaches to query optimizations.
One of such is re-optimization~\cite{kabra1998efficient, markl2004robust, kaftan2018cuttlefish, perron2019learned, neumann2013taking}.
The idea is straight-forward: if the optimizer cannot make accurate predictions of the cardinalities upfront,
we will have to correct its mistakes dynamically at runtime.
Therefore, the process of re-optimization is an interleaving of query execution and query optimization:
it executes the query partially, obtains some true cardinalities and runtime statistics, and then invokes the optimizer again,
hoping to improve the efficiency of the remaining plan.
The recent investigation by Perron et al. shows that even a basic re-optimization strategy could remedy
a significant portion of the performance losses caused by cardinality mis-estimations~\cite{perron2019learned}.

The key problem of designing a re-optimization strategy is to decide
(1) which subquery to execute next and
(2) when to materialize the intermediate results and re-invoke the optimizer.
Existing solutions rely heavily on the optimizer's initial plan~\cite{kabra1998efficient, markl2004robust, perron2019learned, neumann2013taking}. They repeatedly extract subtrees from the complete plan to execute and then
use the results to refine the remaining parts.
The initial plan, however, can be far away from optimal because of inaccurate
cardinality estimations. In this case, the DBMS is likely to choose the ``wrong''
subplan (e.g., costly itself or generate large results) to execute first,
and such a mistake is often unrecoverable by subsequent re-optimization steps.

Meanwhile, these solutions are ``reactive'' in terms of when to trigger re-optimization,
i.e., the re-optimization frequency depends heavily on the initial physical plan.
For example, mid-query re-optimization by Kabra et al. only materializes results at
pipeline breakers (e.g., a sort operation)~\cite{kabra1998efficient}.
Consequently, for a left-deep join tree where each join is a nested-loop join,
re-optimization is never triggered.
On the contrary, \textit{Pop} aggressively materializes the output at every nested-loop join,
causing a large performance and space overhead because of re-optimization~\cite{markl2004robust}.

In this paper, we propose a novel re-optimization algorithm, called \ourname, to address
the above issues. The key idea is to skip the potentially misleading global plans and instead
extract \textit{subqueries} directly from the logical plan as the basic units for re-optimization.
Join operators in the logical plan are grouped into subqueries according to heuristics developed from
the primary-foreign-key relationships to bound/minimize the output sizes of intermediate results.

Such a ``query split'' algorithm is more robust to balance the gains and costs of re-optimization
than those operating on the physical plan.
\ourname then adopts a greedy algorithm to select a subquery with the smallest cost and output cardinality
to execute first.
The intuition is that the performance of a complex query is often determined by a few large joins
(e.g., fact-fact table join). By executing ``simpler'' (or ``less-damaging'') subqueries first and
re-optimizing the rest, we increase the probability of delaying the execution of those large joins
and thus approaching an optimal plan.
Notice that the re-optimization points are purely determined by the logical plan.
The optimizer is only invoked for the subqueries to get/update their costs and output cardinalities 
at each iteration, and the subqueries are usually simple enough for existing optimizers to generate
reasonably good plans quickly.

We implemented \ourname in \postgres and compared our algorithm against
the state-of-the-art re-optimization solutions~\cite{kabra1998efficient, markl2004robust, neumann2013taking, perron2019learned},
robust query processing techniques~\cite{hertzschuch2021simplicity, cai2019pessimistic, wolf2018robustness, wolf2018calculation},
and learned cardinality estimation algorithms~\cite{yang2020neurocard, hilprecht2019deepdb, kipf2018learned}
on the Join Order Benchmark (JOB)~\cite{leis2015good} 
(as well as TPC-H~\cite{TPCH} and Decision Support Benchmark (DSB)~\cite{ding2021dsb}).
Our experiments show that \ourname reduces the JOB execution
time by $35\%$ as opposed to the second-best alternative algorithm.
Moreover, compared to an optimal optimizer
(i.e., an optimizer fed by the true cardinality of each operator),
\ourname slows down the benchmark execution by less than $4\%$.

The contributions of this paper are as follows.
First, we identified that relying on the sub-optimal initial plan is a key weakness of existing re-optimization strategies.
Second, we proposed the \ourname algorithm that extracted subqueries directly from the logical plan based on the
primary-foreign-key relationships to achieve a robust re-optimization efficiency.
Finally, we integrated \ourname into PostgreSQL and demonstrated the superiority of our algorithm by comparing it to
state-of-the-art solutions.
    
\section{Background \& Motivation}
\label{S:mov}
\subsection{Cardinality Estimation}
\label{S21}

Cardinality estimation refers to the process of estimating the number of rows generated by
each operator at query optimization. It is used as an input parameter to the optimizer's cost model.
Improved cardinality estimation enables more accurate cost estimation, thus helping the optimizer
select an efficient plan.
Most DBMSs maintain table/column-level statistics such as histograms and the number of distinct values,
from which they derive the selectivity of basic single-column predicates.
The more challenging tasks are to estimate the selectivity of conjunctive predicates involving
multiple columns and to estimate the join cardinality.
Because it is too costly to maintain a relatively complete set of multi-column statistics,
the optimizer has to make assumptions about the correlation between columns in these cases.

Most widely-used DBMSs such as PostgreSQL and MySQL assume independent data distributions
between columns~\cite{leis2015good, MySQLDoc}.
It is probably one of the best strategies an optimizer could apply given the lack of statistics.
In reality, however, highly correlated columns are common, and this approach is likely to
deliver underestimated cardinalities~\cite{leis2015good}.
Although the accuracy of cardinality estimation for complex queries can be improved through
sampling~\cite{leis2017cardinality, wu2016sampling} and machine learning techniques~\cite{wang2020we},
none of the approaches is robust enough, and their intrinsic overhead is hardly justified
in real-world database applications~\cite{perron2019learned}.
What makes it worse is error propagation.
For an N-way natural join, for example, the cardinality estimation error at each join step
could grow exponentially with N~\cite{ioannidis1991propagation}.
This theoretical result matches what we have observed in practice.

\subsection{Re-optimization}
\label{S22}

\begin{figure}[t!]
    \centering
    \includegraphics[width=\linewidth]{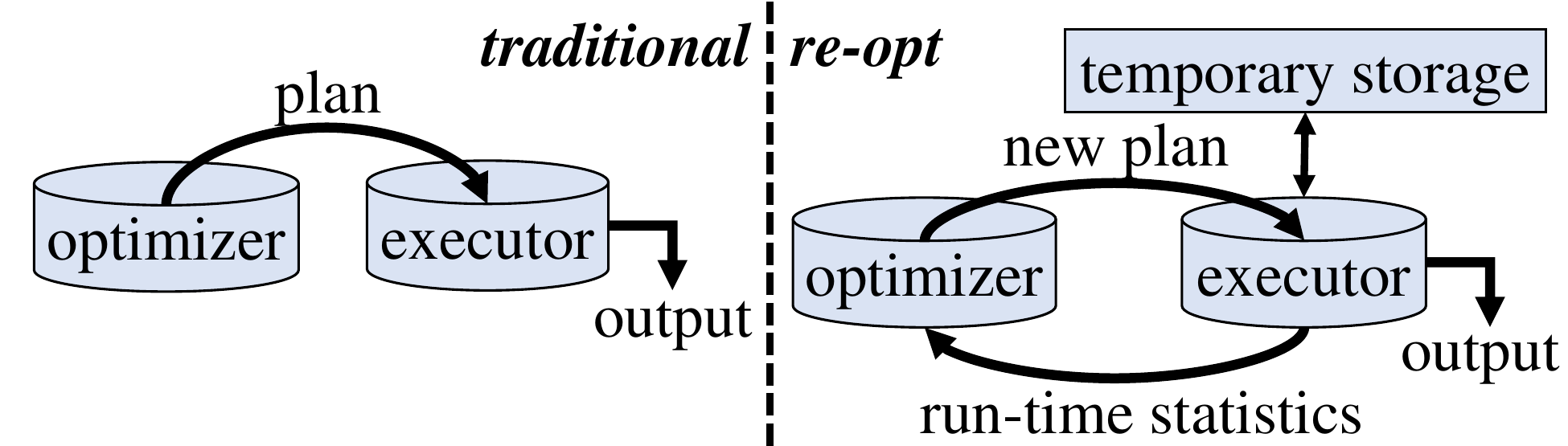}
    \caption{Structure difference between traditional query processing and re-optimization}
    \label{F1}
    \Description{}
\end{figure}

Query re-optimization is a technique of adaptive query processing~\cite{babu2005adaptive, deshpande2007adaptive}
where the optimizer is (re)invoked at execution time to correct potential bad plans.
Specifically, the optimizer selects a few operations in the physical plan to materialize the intermediate results.
It then compares the true cardinality (i.e., statistics from the actual intermediate results) against the previously
estimated value. If those values differ too much, the optimizer would re-plan the remaining part of the query
using the true cardinality.
Although re-optimization itself brings overheads, the revised plan is almost guaranteed to be at least as good as
the original one.
Re-optimization, therefore, is a process of interleaving the execution of the query engine and the optimizer,
as shown in \cref{F1}.

Existing re-optimization algorithms~\cite{kabra1998efficient, markl2004robust, perron2019learned, neumann2013taking} operate directly on global physical plans.
They choose a subtree from the plan obtained from the previous re-optimization cycle and execute that
subplan to decide whether further re-optimization is needed.
Such a strategy of ``selecting a partial query to execute'' can be problematic if the referencing
global plan deviates largely from an optimal one.
And once an undesirable subplan (typically involving large tables but with an underestimated cardinality)
is chosen, the damage often propagates through later re-optimization iterations.
    
\begin{figure}[t!]
    \subfigure[Optimal plan]
    {
        \begin{minipage}[t]{0.3\linewidth}
            \includegraphics[width=\linewidth]{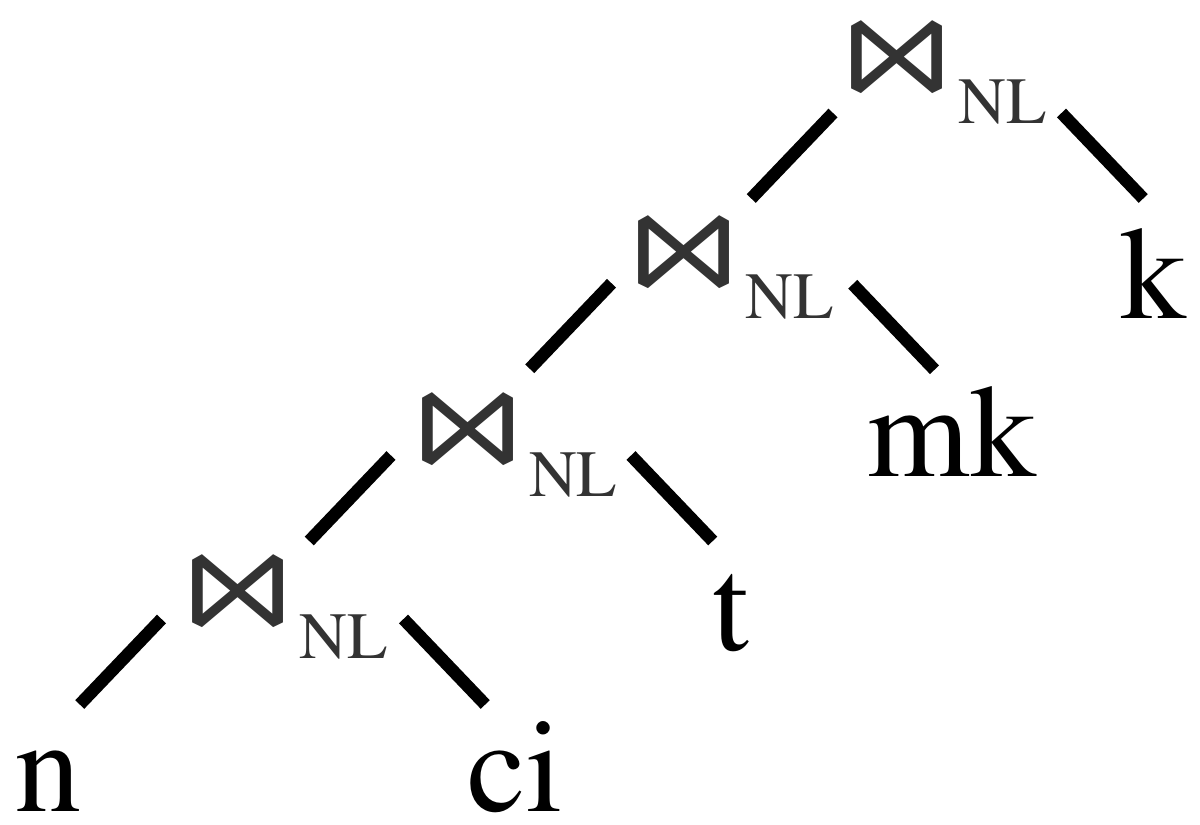}
            \label{F21a}
        \end{minipage}
    }
    \subfigure[Initial plan]
    {
        \begin{minipage}[t]{0.3\linewidth}
            \includegraphics[width=\linewidth]{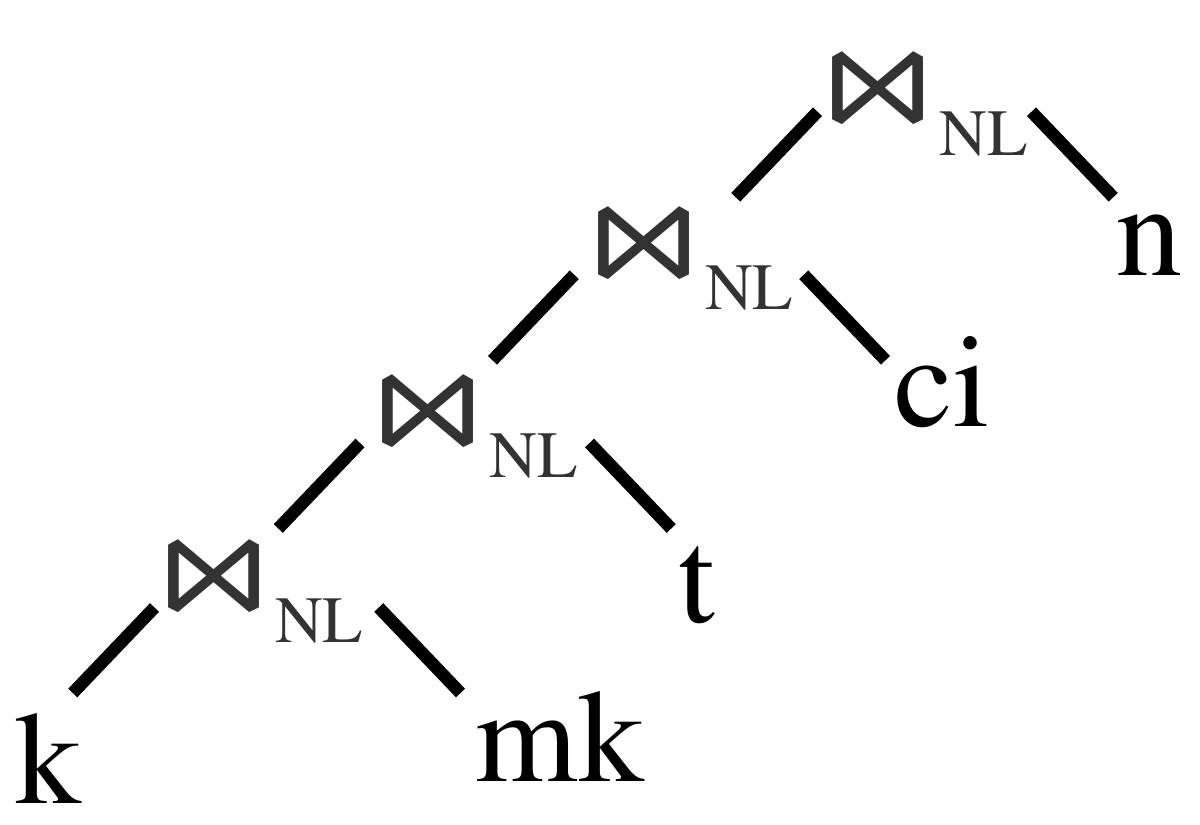}
            \label{F21b}
        \end{minipage}
    }
    \subfigure[Re-Optimized plan after the first join]
    {
        \begin{minipage}[t]{0.3\linewidth}
            \includegraphics[width=\linewidth]{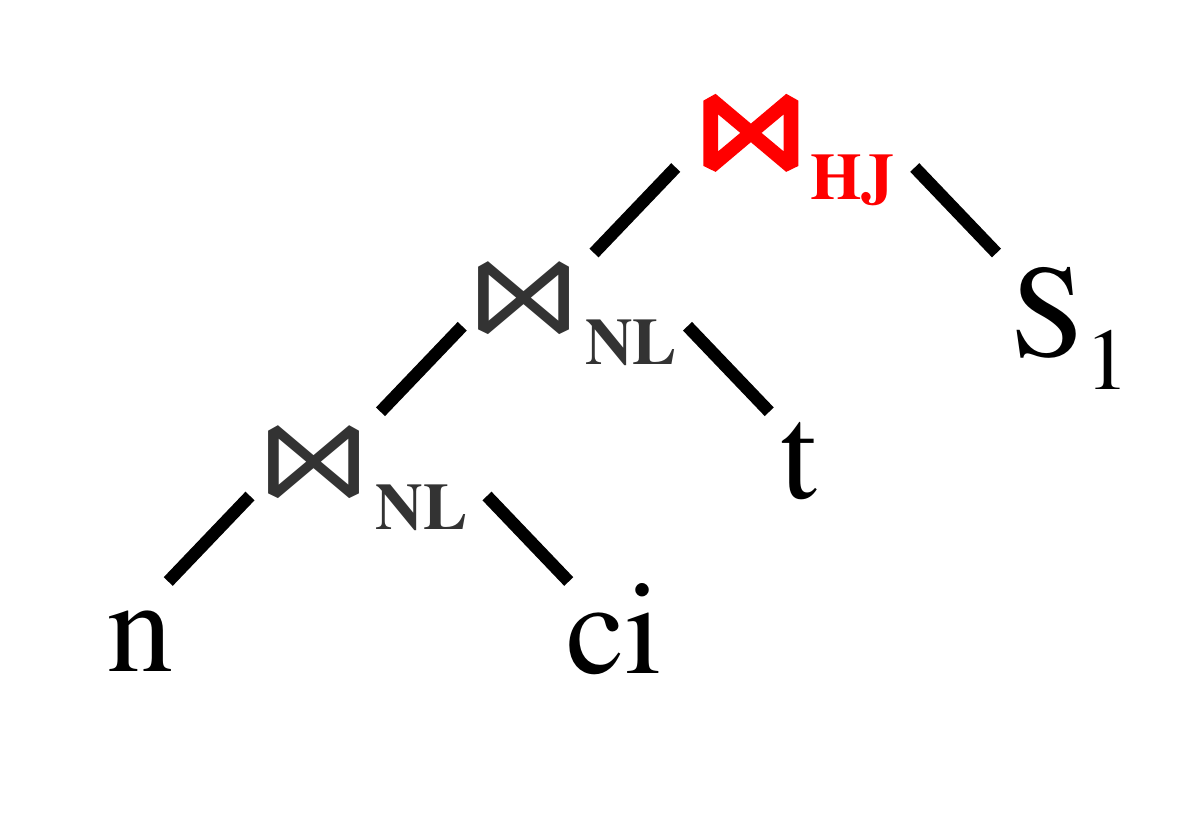}
            \label{F21c}
        \end{minipage}
    }
    \centering
    \caption{The example of how a bad global plan influence re-optimization}
    \label{F21}
    \Description{}
\end{figure}

\cref{F21} shows an example to illustrate such an unrecoverable subplan execution.
The query is a 5-way join extracted from the JOB benchmark.
There is an index built for each join column.
As shown in \cref{F21a}, the optimal plan joins table \texttt{n} and \texttt{ci} first and uses
the results to probe table \texttt{t} in an index nested-loop join (denoted as \texttt{NL} in the figure).
The actual initial plan (\cref{F21b}), however, underestimates the cardinality of \texttt{k} $\Join$ \texttt{mk}
and thus chooses to execute this subplan first.
We use \texttt{S$_1$} to denote the intermediate result of \texttt{k} $\Join$ \texttt{mk}.
Once we discover that \texttt{S$_1$} is much larger than the prior estimation, we trigger the optimizer
to re-plan the rest of the query.
However, the best the optimizer can do at this point is shown in \cref{F21c}.
Because the large temporary table \texttt{S$_1$} does not have an index, the DBMS must perform a hash join
(highlighted in bold red) for the last step, which could be orders of magnitude slower than probing
the existing indexes of the two base tables.

\begin{figure}[t!]
    \subfigure[Similarity = 0]
    {
        \begin{minipage}[t]{0.3\linewidth}
            \includegraphics[width=\linewidth]{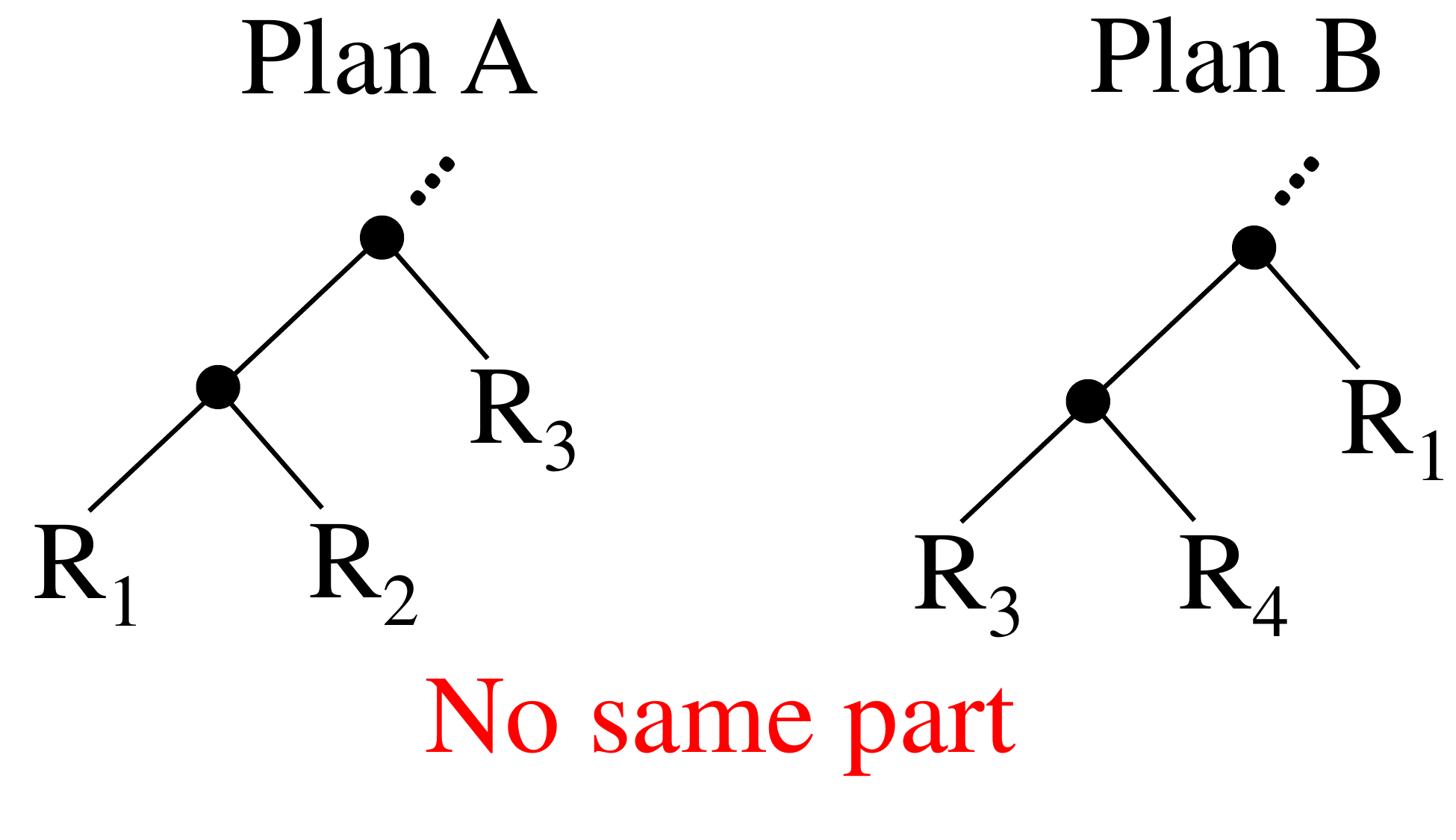}   
            \label{F2a}
        \end{minipage}
    }
    \subfigure[Similarity = 1]
    {
        \begin{minipage}[t]{0.3\linewidth}
            \includegraphics[width=\linewidth]{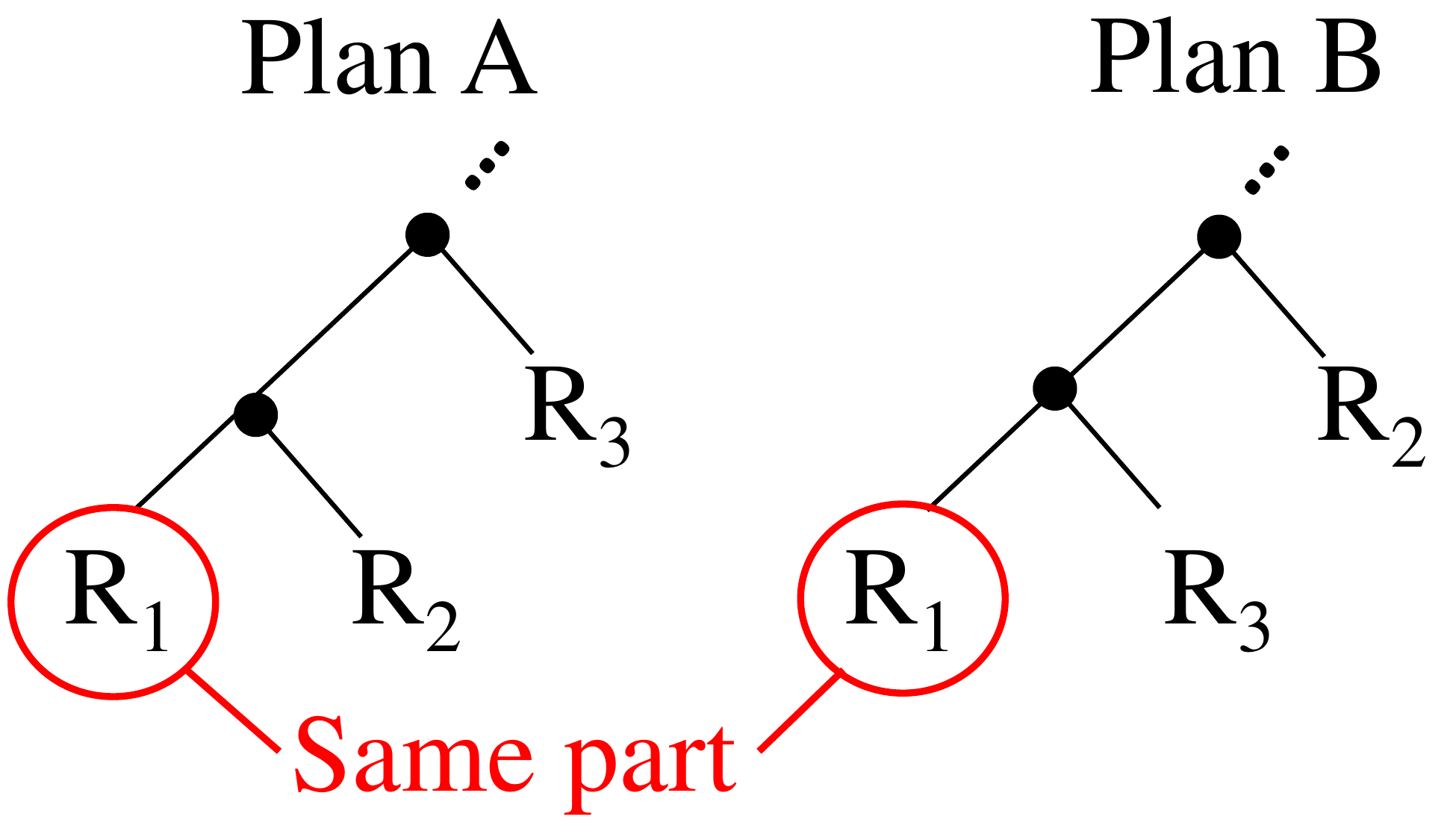}
            \label{F2b}
        \end{minipage}
    }
    \subfigure[Similarity = 2]
    {
        \begin{minipage}[t]{0.3\linewidth}
            \includegraphics[width=\linewidth]{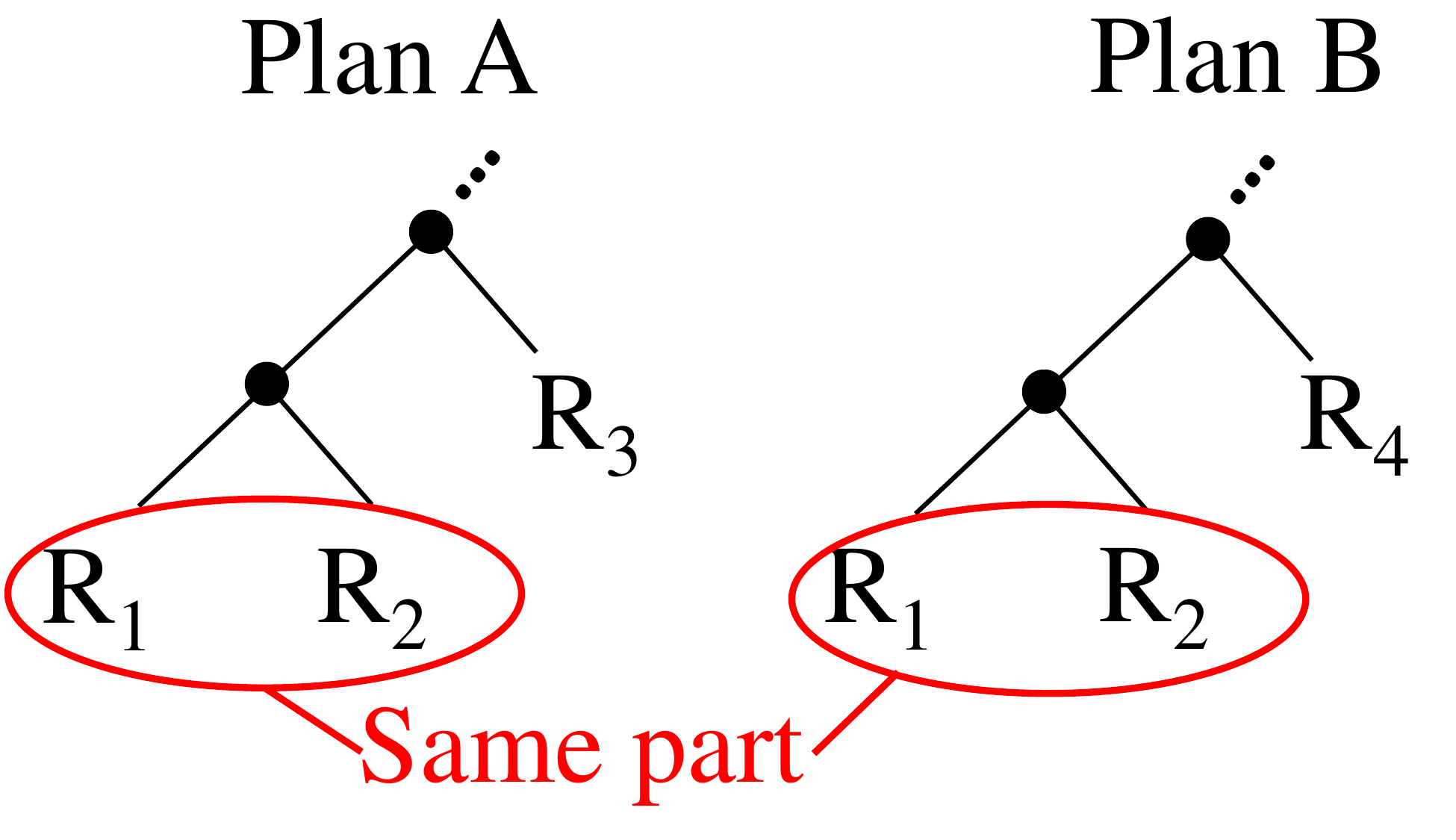}
            \label{F2c}
        \end{minipage}
    }
    \centering
    \caption{The example of different similarity}
    \label{F2}
    \Description{}
\end{figure}

To show how much an initial global plan can deviate from optimal,
we investigate the query plans in the Join Order Benchmark using the optimizer from PostgreSQL.
We define the similarity score of two plans as the number of leaf nodes included in their
largest common subtree. For example, as shown in \cref{F2}, if the first joins of the two plans
differ completely, they have a similarity score of 0 (\cref{F2a}); if the probe side scans the
same table (but joins a different one), the similarity of the plans is 1 (\cref{F2b});
similarity = 2 means that the plans differ after the first join (\cref{F2c}).
In \cref{T1}, we demonstrate how often initial global plans diverge from the optimal ones early in the execution.
We observe that more than half of the JOB queries have initial plans whose optimality does not
``survive'' after one join, among which a quarter of the plans even made mistakes on the first join.
    
\begin{table}[t!]
    \caption{The ratio of queries whose global plans deviate from the optimal plan with different degrees}
    \label{T1}
    \begin{tabular}{c|cccc}
        \toprule
        Similarity & 0 & 1 & 2 & > 2    \\
        \midrule
        Ratio & 13\% & 12\% & 32\% & 43\% \\
        \bottomrule
    \end{tabular}
\end{table}
    
The second problem of existing algorithms is that the re-optimization decision is made
reactively according to the types of physical plan nodes.
Such a heuristic-based approach often leads to extreme re-optimization frequencies.
If the system triggers re-optimization only at pipeline breakers,
it never gets a chance to change the ordering of the nested-loop joins in a left-deep plan.
On the other hand, if the system invokes re-optimization at every join,
the overhead of materializing intermediate results might be intimidating:
it essentially converts the execution from the Volcano model to the fully-materialized one.

\subsection{A Proactive Strategy}
\label{S23}

As shown in \cref{S22}, a suboptimal global plan can cause irrecoverable damages
to the effectiveness of existing re-optimization algorithms.
We, therefore, argue that a better strategy is to examine the query's logical plan
and decide \textit{proactively} when to materialize results (and re-invoke the optimizer)
before execution.
We call this scheme \ourname.
Specifically, we divide the logical plan into subqueries based on the
primary-foreign-key relationships and optimize them separately.
Because each subquery is relatively simple, it is less likely that the optimizer
would make serious mistakes as in a global plan.
We then choose one of the subqueries to run and materialize its output.
Once the execution is finished, we use the updated statistics (e.g., output size)
to re-optimize the remaining relevant subqueries.
This process continues until no subquery is left to be executed.
The execution order is determined by a ``ranking'' function (detailed in \cref{S42})
where subqueries with small costs and output sizes are prioritized.

\begin{figure}[t!]
    \subfigure[Logical Plan]
    {
        \begin{minipage}[t]{0.47\linewidth}
            \includegraphics[width=\linewidth]{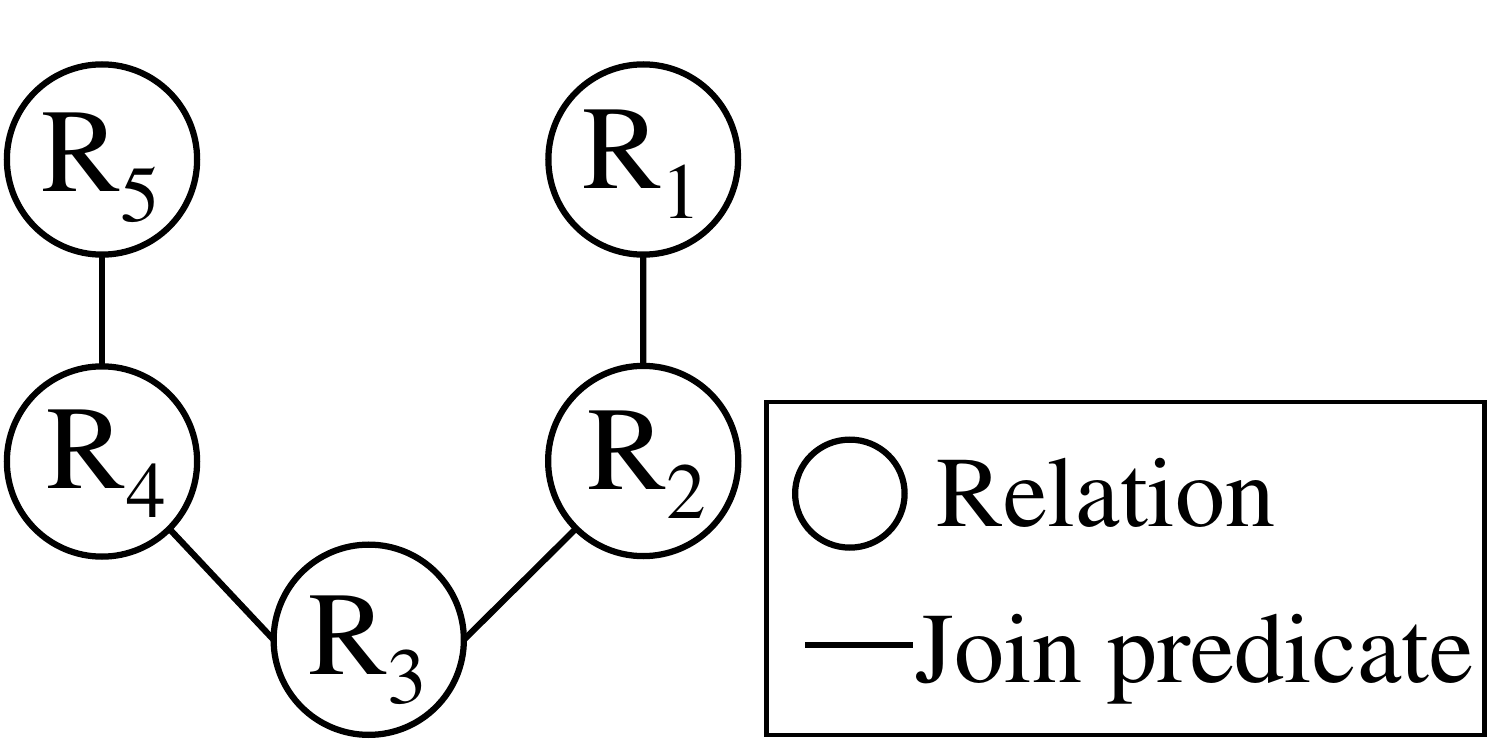}   
        \end{minipage}
    }
    \subfigure[Subqueries \texttt{S$_1$}, \texttt{S$_2$}, and \texttt{S$_3$}]
    {
        \begin{minipage}[t]{0.47\linewidth}
            \includegraphics[width=\linewidth]{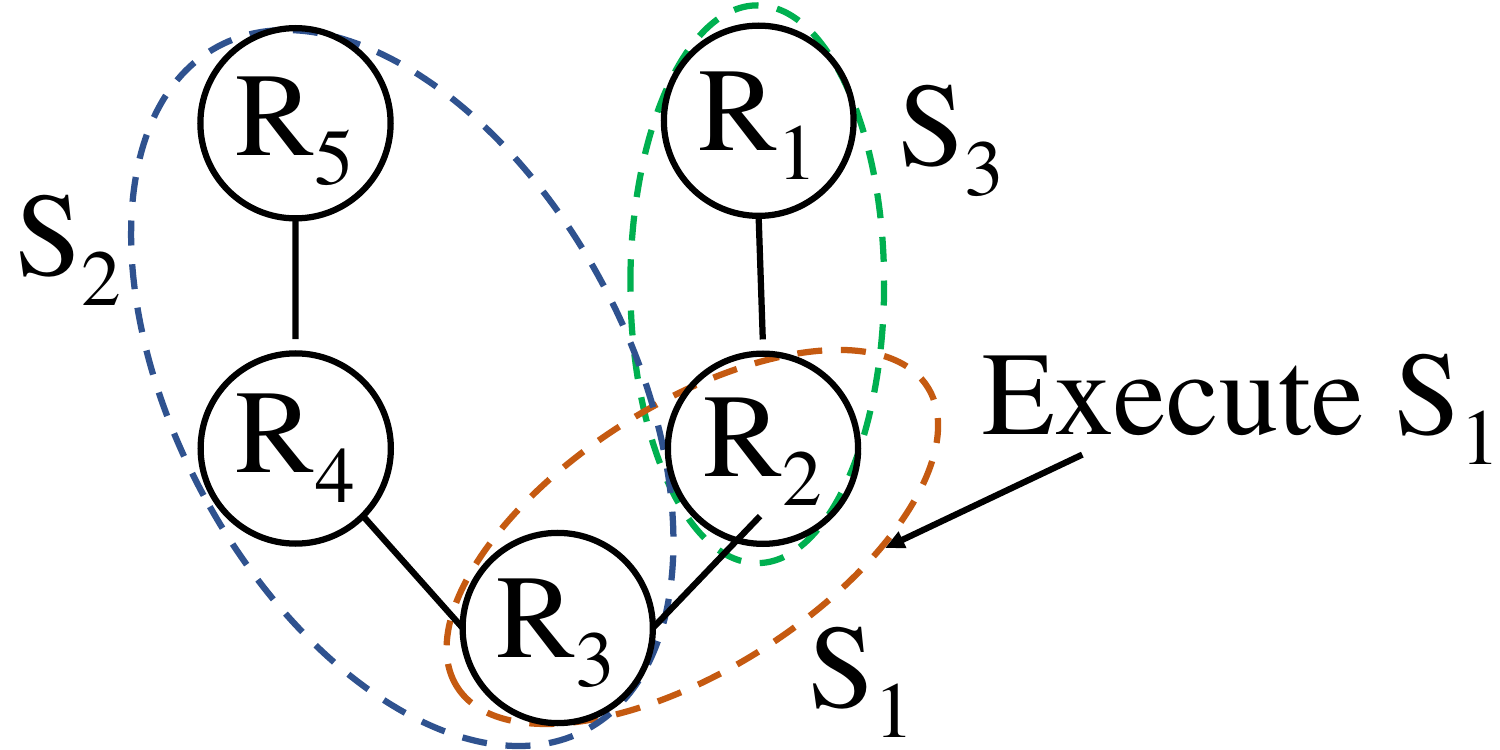}
        \end{minipage}
    }
    \subfigure[Re-optimizing \texttt{S$_2$} and \texttt{S$_3$}]
    {
        \begin{minipage}[t]{0.47\linewidth}
            \includegraphics[width=\linewidth]{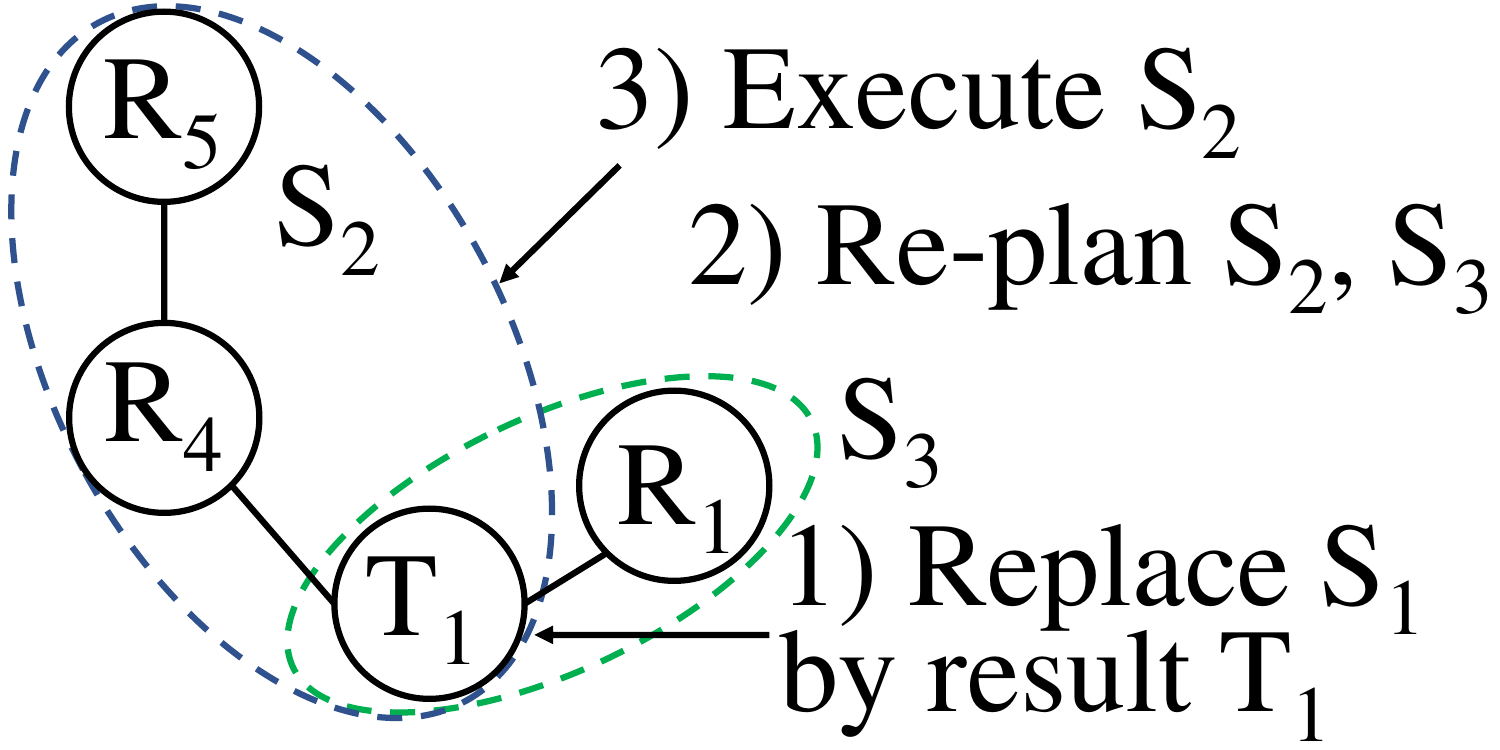}
        \end{minipage}
    }
    \subfigure[Re-optimizing \texttt{S$_3$}]
    {
        \begin{minipage}[t]{0.47\linewidth}
            \includegraphics[width=\linewidth]{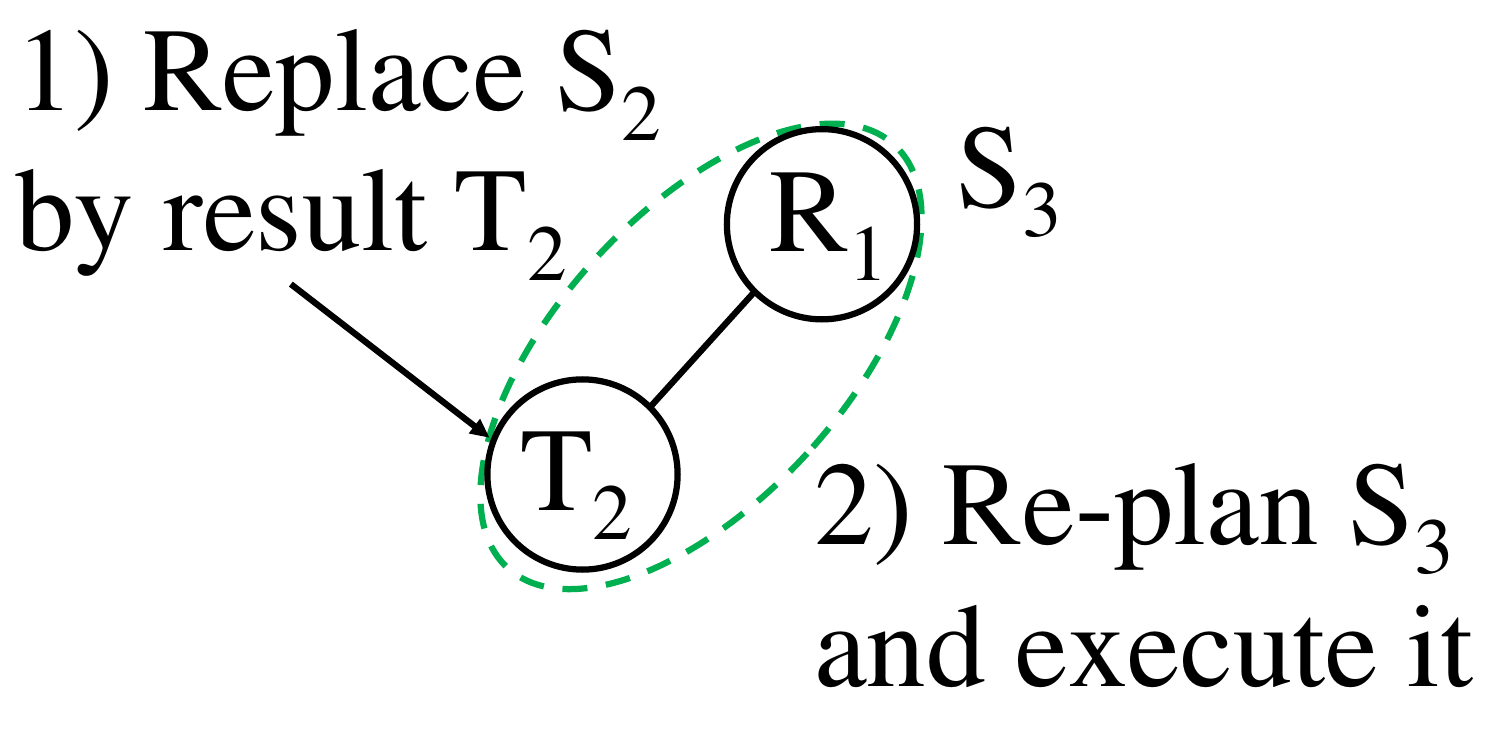}
        \end{minipage}
    }
    \centering
    \caption{\ourname example}
    \label{F3}
    \Description{}
\end{figure}

\cref{F3} shows an example of a 5-way join re-optimized using \ourname.
We first split the query into three subqueries and optimize them separately:
\texttt{S$_1$}=\texttt{R$_2$} $\Join$ \texttt{R$_3$}, 
\texttt{S$_2$}=\texttt{R$_3$} $\Join$ \texttt{R$_4$} $\Join$ \texttt{R$_5$},
and \texttt{S$_3$}=\texttt{R$_1$} $\Join$ \texttt{R$_2$}.
We then choose \texttt{S$_1$} to execute and materialize its output as \texttt{T$_1$}.
Using the statistics of \texttt{T$_1$}, we trigger re-optimization on the remaining subqueries
\texttt{S$_2$}=\texttt{T$_1$} $\Join$ \texttt{R$_4$} $\Join$ \texttt{R$_5$}
and \texttt{S$_3$}=\texttt{R$_1$} $\Join$ \texttt{T$_1$} (\cref{F3}(c)).
\texttt{S$_2$} is selected to run next. The result is materialized in \texttt{T$_2$},
whose statistics is used to re-optimize the final subquery
\texttt{S$_3$}=\texttt{R$_1$} $\Join$ \texttt{T$_2$}.

Compared to traditional re-optimization algorithms, \ourname is more robust at
avoiding suboptimal plans, and has more control over the re-optimization cost.
First, \ourname does not depend on global physical plans.
Because the difficulty of the optimization task grows exponentially as the number of
joins increases,
the optimizer is likely to make mistakes when planning a complex query by entirety.
As discussed in \cref{S22}, ``early mistakes'' are common and they cannot be
recovered through re-optimization.
In \ourname, on the other hand, the optimizer only deals with much simpler subqueries,
and the probability of generating bad plans is reduced dramatically.

Second, \ourname has predictable re-optimization overhead because it determines
where to re-invoke the optimizer before executing the query.
The overhead is also adjustable by modifying the subquery granularity.
In this way, \ourname avoids undesirable re-optimization frequencies
caused by various physical plan shapes, as described in \cref{S22}.
The trade-off, though, is that \ourname might miss certain optimization opportunities
that are only recognizable when examining the query as a whole.
\ourname could be ``myopic'' because the optimizer only operates at the subquery level.
Our detailed evaluation (\cref{S54,S56}), however, demonstrate that
such a trade-off is modest and is outweighed by the aforementioned benefits.

The rest of the paper is organized as follows.
\cref{S:qs} provides an overview of \ourname with correctness proof.
\cref{S:policy} discusses two critical policies that could largely determine the
efficiency of our algorithm.
\cref{S:impl} briefly describes the integration of \ourname into \postgres.
An evaluation of \ourname along with detailed case studies is presented in \cref{S:eva}
followed by the related work in \cref{S:rel}.

\section{\ourname}
\label{S:qs}
    \begin{figure}[t!]
        \centering
        \includegraphics[width=\linewidth]{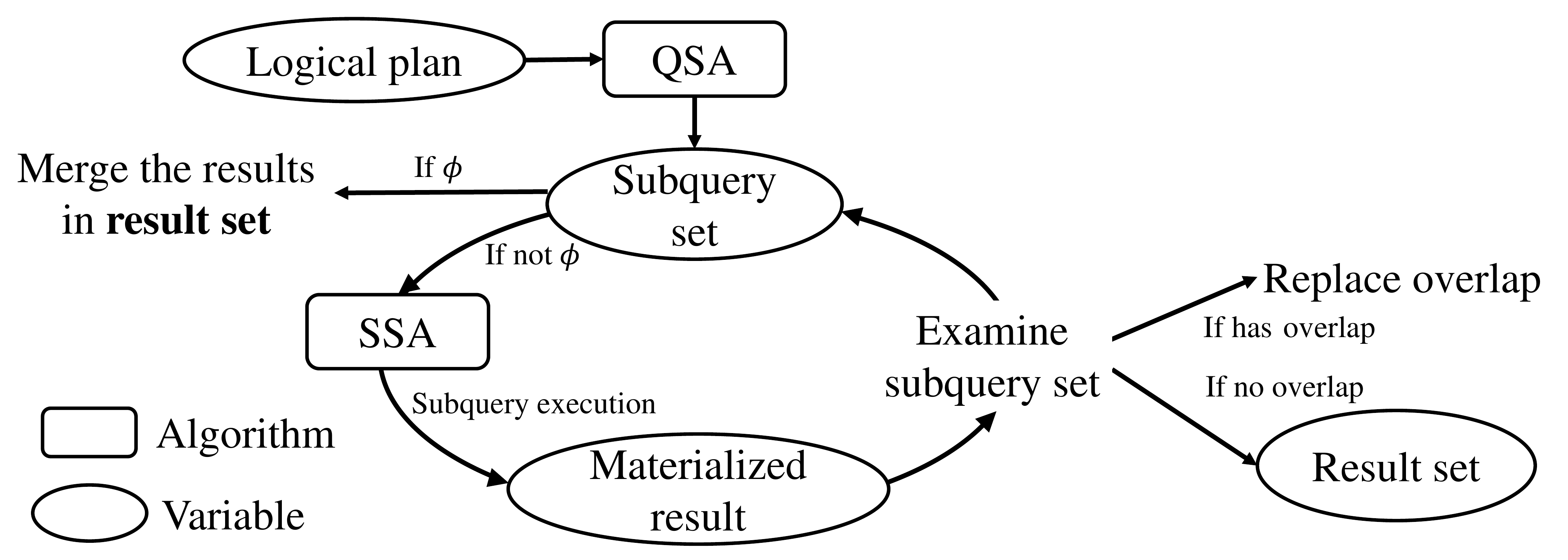}
        \caption{The workflow of \ourname}
        \label{F4}
        \Description{}
    \end{figure}

In this section, we present an overview of the \ourname algorithm
followed by a proof of correctness. 
Algorithm details and implementation are further discussed in \cref{S:policy}.

\subsection{Algorithm Overview}
\label{S31}

As shown in \cref{F4}, \ourname takes in a query's logical plan 
and runs the Query Splitting Algorithm (QSA) against it.
For simplicity, we restrict the input to be select-projection-join (SPJ) queries
only (i.e., queries involving only select, projection, and join operators).
Algorithm extension to support non-SPJ queries is discussed in \cref{S33}.
The goal of QSA is to generate a set of subqueries where a serial execution of
the subqueries would have the same result as executing the original query.
We discuss the requirements of such a valid \texttt{subquery set} in \cref{S32}.

With a valid subquery set, \ourname proceeds to enter a loop.
At each loop iteration, \ourname picks a subquery from the current set to execute
according to the Subquery Selection Algorithm (SSA).
Although SSA does not affect correctness, it has a significant impact on the
efficiency of the entire \ourname algorithm.
We discuss different subquery ranking strategies in detail in \cref{S42}.
The selected subquery is then removed from the set,
and the execution results as well as the associated statistics are materialized.

Next, \ourname examines each of the remaining subqueries in the set:
if it overlaps with the just-executed subquery (i.e., they have shared relations),
the shared relations are replaced with the corresponding materialized results.
If it turns out that the just-executed subquery does not overlap with any of the
subqueries in the set, its execution results are pushed to the \texttt{result set}.
The loop continues until the subquery set becomes empty.
Finally, \ourname merges the results (through Cartesian product) if there are
multiple items in the result set.
    
    \begin{figure*}[t!]
        \subfigure[Original subquery set]
        {
            \begin{minipage}[t]{0.23\linewidth}
                \includegraphics[width=\linewidth]{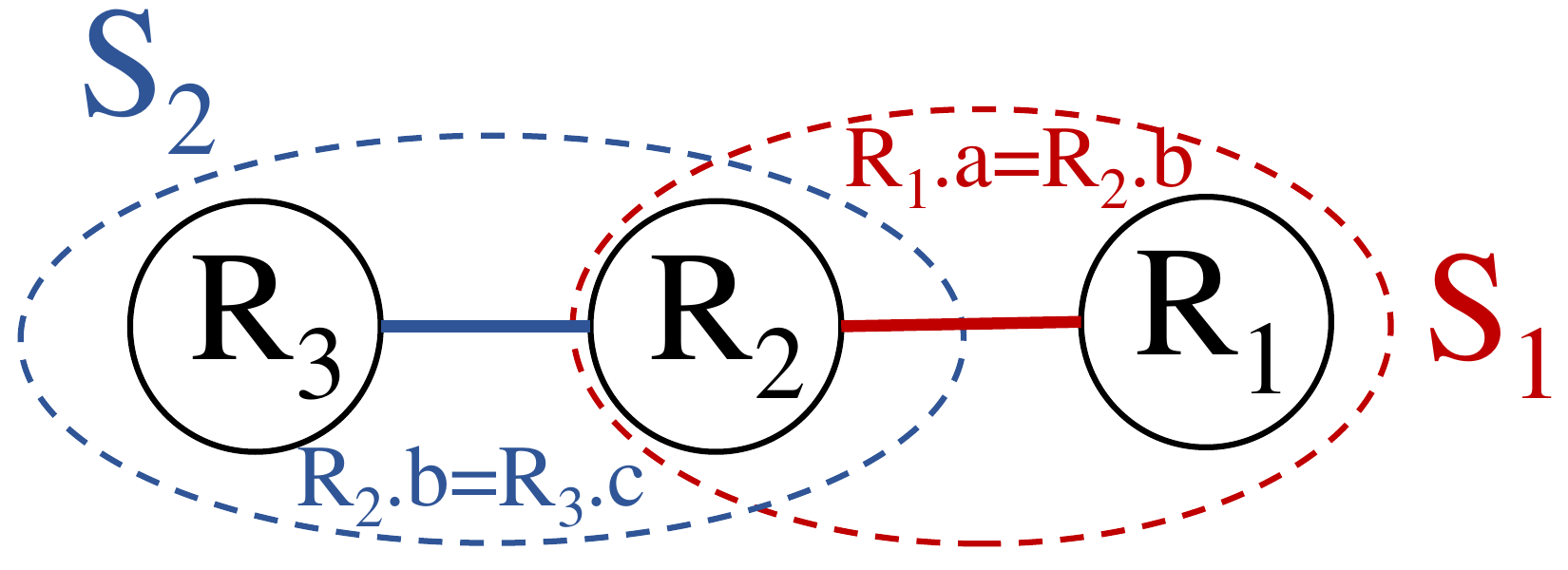}
                \label{F22a}
            \end{minipage}
        }
        \subfigure[Execute \texttt{S$_1$}]
        {
            \begin{minipage}[t]{0.23\linewidth}
                \includegraphics[width=\linewidth]{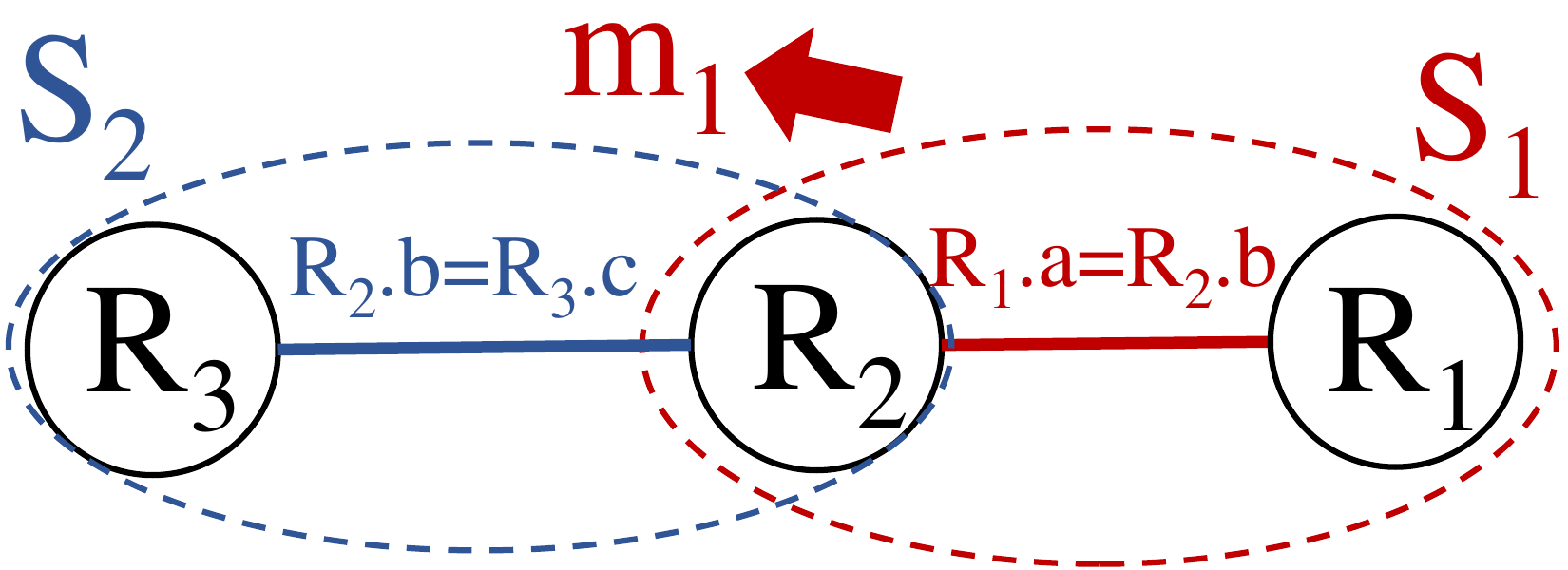}
                \label{F22b}
            \end{minipage}
        }
        \subfigure[Modify \texttt{S$_2$}]
        {
            \begin{minipage}[t]{0.23\linewidth}
                \includegraphics[width=\linewidth]{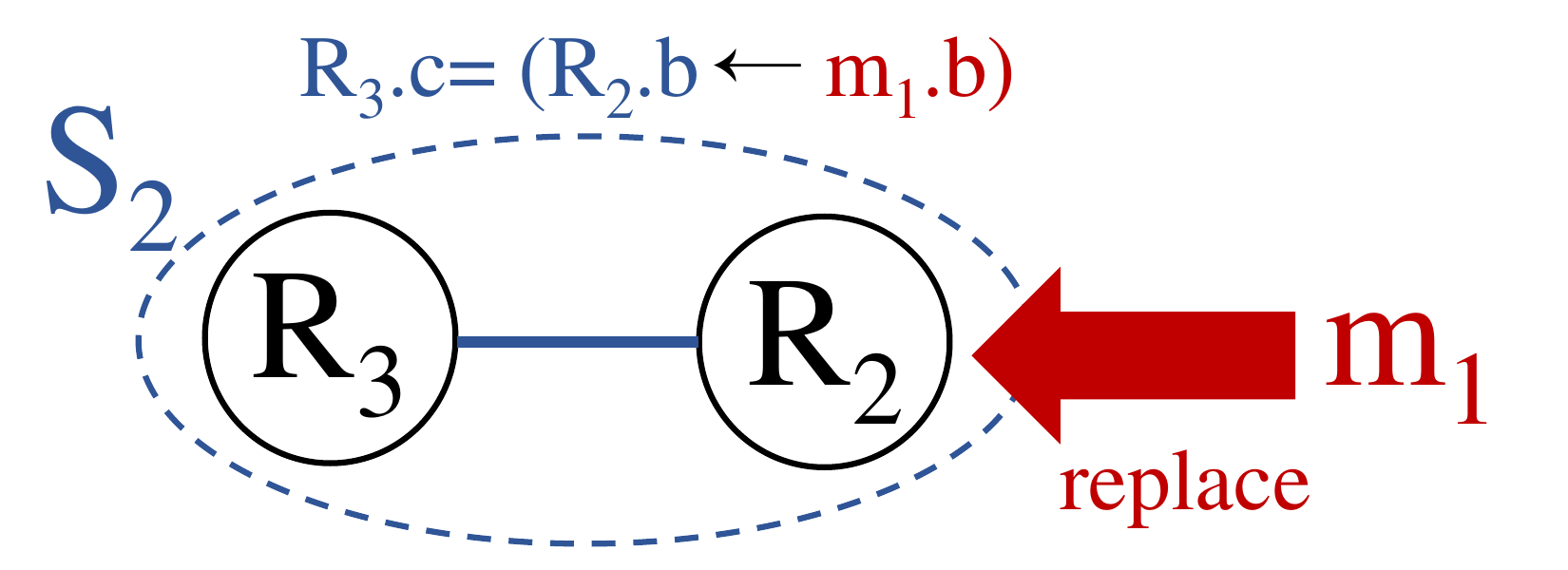}
                \label{F22c}
            \end{minipage}
        }
        \subfigure[Execute \texttt{S$_2$}]
        {
            \begin{minipage}[t]{0.23\linewidth}
                \includegraphics[width=\linewidth]{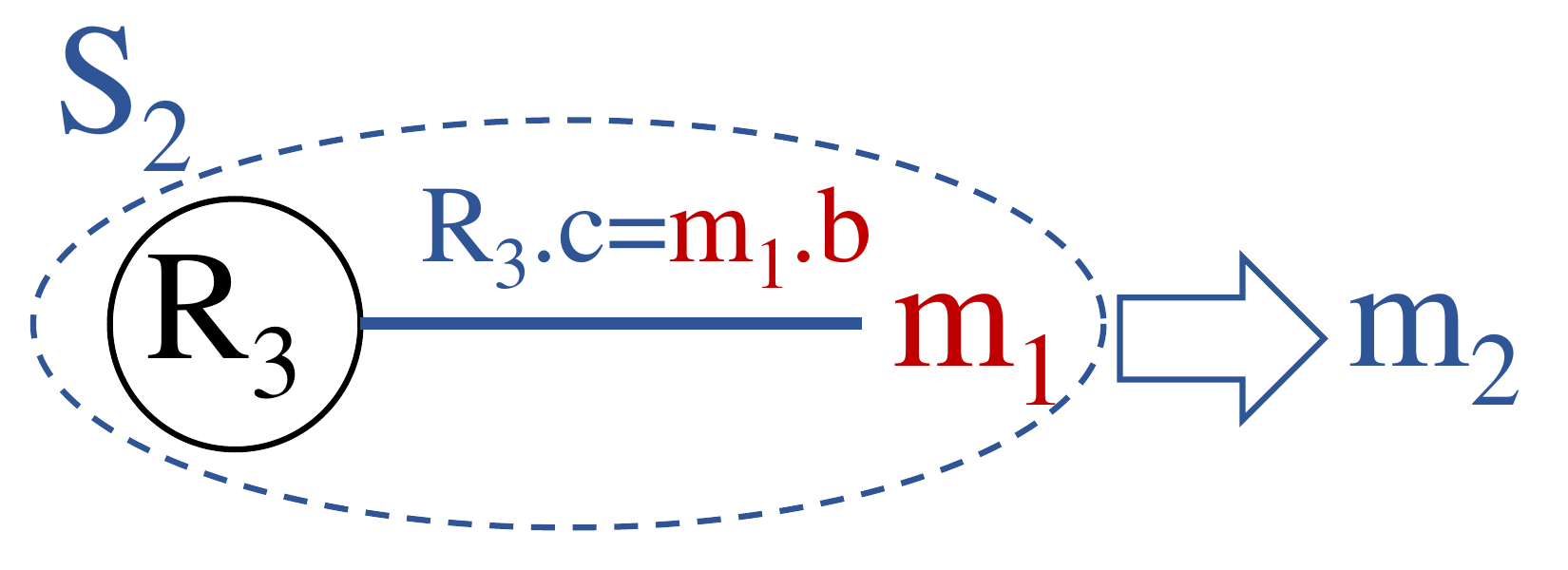}
                \label{F22d}
            \end{minipage}
        }
        \centering
        \caption{An example of how \ourname works}
        \label{F22}
        \Description{}
    \end{figure*}
    
\cref{F22} shows an example.
The original query is $R1 \Join_{R1.a = R2.b} R2 \Join_{R2.b = R3.c} R3$,
where $R1.a$ denotes attribute $a$ from relation $R1$.
After running the QSA, we obtain two subqueries in the set:
$S1 = R1 \Join_{R1.a = R2.b} R2$ and $S2 = R2 \Join_{R2.b = R3.c} R3$.
Suppose the SSA selects $S1$ to execute first, and the materialized result is denoted by $m1$.
We next examine the remaining subquery $S2$.
Because $S2$ and $S1$ share the common relation $R2$, we substitute $m1$ for $R2$
and rewrite $S2$ as $m1 \Join_{m1.b = R3.c} R3$.
We then enter the next iteration and execute $S2$.
Because there is no more subquery in the set, we push the execution result $m2$ to
the result set and complete the algorithm.
    
Notice that the projection operators can be added to the subqueries
following the general projection push-down rules.
We omit projections in the following discussions for presentation clarity.

\subsection{Correctness} \label{S32}
As indicated above, the correctness of \ourname depends on the output of QSA
(i.e., the subquery set).
In this subsection, we define the required properties of the subquery set
and prove that the \ourname algorithm is correct given a valid subquery set.
    
Given a set of relations \textbf{\textit{R}} and a set of predicates \textbf{\textit{P}}
over \textbf{\textit{R}}, we define the normal form of an SPJ query as
$$q(\textbf{\textit{R}},\textbf{\textit{P}})=(\sigma_{\textbf{\textit{P}}}(r_1 \times r_2 \times... \times r_m)),r_i \in \textbf{\textit{R}}$$

A query $q'(\textbf{\textit{R}}', \textbf{\textit{P}}')$ is said to be
a subquery of $q(\textbf{\textit{R}}, \textbf{\textit{P}})$ if
$\textbf{\textit{R}}' \subseteq \textbf{\textit{R}}$ and
$\textbf{\textit{P}}' \subseteq \textbf{\textit{P}}$.
A Query Splitting Algorithm (QSA) takes a query $q$ as input and
produces a set of subqueries of $q$.
\ourname then operates on this subquery set, as described in \cref{S31}.
Intuitively, in order for \ourname to produce the same result as the
original query, the subquery set generated by QSA must ``cover'' all
the relations and predicates in the original query. More formally,

\begin{Definition} \label{D1}
Given an SPJ query $q(\textbf{\textit{R}}, \textbf{\textit{P}})$,
let $\textbf{\textit{Q}} =\{q_1(\textbf{\textit{R}}_1,\textbf{\textit{P}}_1)$, ..., $q_n(\textbf{\textit{R}}_n,\textbf{\textit{P}}_n)\}$
be a set of subqueries of q.
$\textbf{\textit{Q}}$ is said to cover $q$ (denoted as $\textbf{\textit{Q}} \rightharpoonup_c q$) if the following holds: \\
(1) $\cup_{i=1}^n \textbf{\textit{R}}_i = \textbf{\textit{R}}$ \\
(2) $\cup_{i=1}^n \textbf{\textit{P}}_i$ logically implies $\textbf{\textit{P}}$. \footnote{~``A logically implies B" means that each predicate from B can be inferred by A.}
\end{Definition}

The above definition guarantees that each base relation $r_i$ in \textbf{\textit{R}}
and each predicate $p_i$ in \textbf{\textit{P}} must appear at least once in the
subquery set \textbf{\textit{Q}}.
Notice that the definition allows the same $r_i$ or $p_i$ to be included in multiple subqueries.
This does not affect the correctness of the \ourname algorithm because the duplicates
will be removed during the (materialized) result substitution step.
In fact, the ``coverness'' property is sufficient to prove the correctness of
the entire \ourname algorithm.

\begin{Theorem}[1] \label{Th1}
    Let $q(\textbf{\textit{R}}, \textbf{\textit{P}})$ be an SPJ query, $\textbf{\textit{Q}}$ be a set of subqueries of $q$.
    \ourname produces the same output as $q$ if $\textbf{\textit{Q}} \rightharpoonup_c q$.
\end{Theorem}

\textit{Proof sketch:}
We prove the theorem by induction.
For the base case, if there is only one query $q'$ in $\textbf{\textit{Q}}$,
and $\textbf{\textit{Q}} \rightharpoonup_c q$, then $q'$ and $q$ are
equivalent queries.
Suppose the statement is true for $\lvert \textbf{\textit{Q}} \rvert = n - 1$,
we want to prove that it also holds for $\lvert \textbf{\textit{Q}} \rvert = n$.
Let $\textbf{\textit{Q}} = \{q_1, q_2, ..., q_n \}$.
Without loosing generality, suppose the first query executed by \ourname is $q_1$,
and the remaining set is $\textbf{\textit{Q'}} = \{q_2, ..., q_n \}$.
If $q_1$ overlaps (i.e., with at least one shared relation) with a query $q_i$
in $\textbf{\textit{Q'}}$, we can prove that substituting the materialized view
of $q_1$ into $q_i$ does not change the overall query result.
Then, the remaining subquery set (after the substitution)
$\textbf{\textit{Q''}} = \{q_2, ..., q'_i, ...,  q_n \}$
falls back to the induction hypothesis.
On the other hand, if $q_1$ does not overlap with any of the queries in
$\textbf{\textit{Q'}}$, then executing this ``isolated'' subquery first
and pushing its result to the final buffer does not affect the correctness
of the original query.
Again, after executing $q_1$, the remaining subquery set $\textbf{\textit{Q'}}$
falls back to the induction hypothesis.
A detailed proof can be found in \cref{A2}.

\subsection{Extending to Non-SPJ Queries}
\label{S33}

The focus of the \ourname algorithm is on optimizing the join ordering of SPJ queries.
In this subsection, we briefly discuss how to extend \ourname to be compatible with
queries containing Non-SPJ operators.

First of all, we do not distinguish between the join types when running the QSA.
After getting the subquery set, we perform an extra check for each subquery
containing ``special'' joins (i.e., outer join, semi-join, and anti-join)
to determine whether executing it in the current iteration would violate any
ordering constraints.
If a violation is detected, the subquery is temporarily removed from the candidate
set for this iteration.
If the candidate set becomes empty after the check, the execution falls back to
follow a global physical plan.

Besides special joins, a majority of the Non-SPJ optimizations (e.g., subquery flattening/merging
and outer join simplification) are performed at the rule-based transformation stage~\cite{chaudhuri1998overview, MySQLDoc, PostgresDoc, OracleDoc}.
These optimizations happen before the \ourname algorithm, and their results
(i.e., the transformed plan) are consumed by \ourname at the cost-based enumeration stage
in the following way.
Given a Non-SPJ operator (e.g., aggregation, semi join) $\phi$
whose inputs are generated by a set of SPJ subqueries $q_1, q_2, ..., q_n$,
we apply the \ourname algorithm to each of the subqueries
and feed their results to $\phi$.
Unlike the subquery execution within \ourname where the result
must be materialized for re-optimization, the data transfer between
$\phi$ and the $q_i$'s can be pipelined.

For a query plan that contains multiple Non-SPJ operators,
we segment the plan tree according to these operators and
execute them from the bottom up
After completing each Non-SPJ operator, we materialize its
result and treat it as a base relation in the subsequent
\ourname invocations.

\begin{figure}
        \subfigure[Non-SPJ query]
        {
            \begin{minipage}[t]{0.47\linewidth}
                \includegraphics[width=\linewidth]{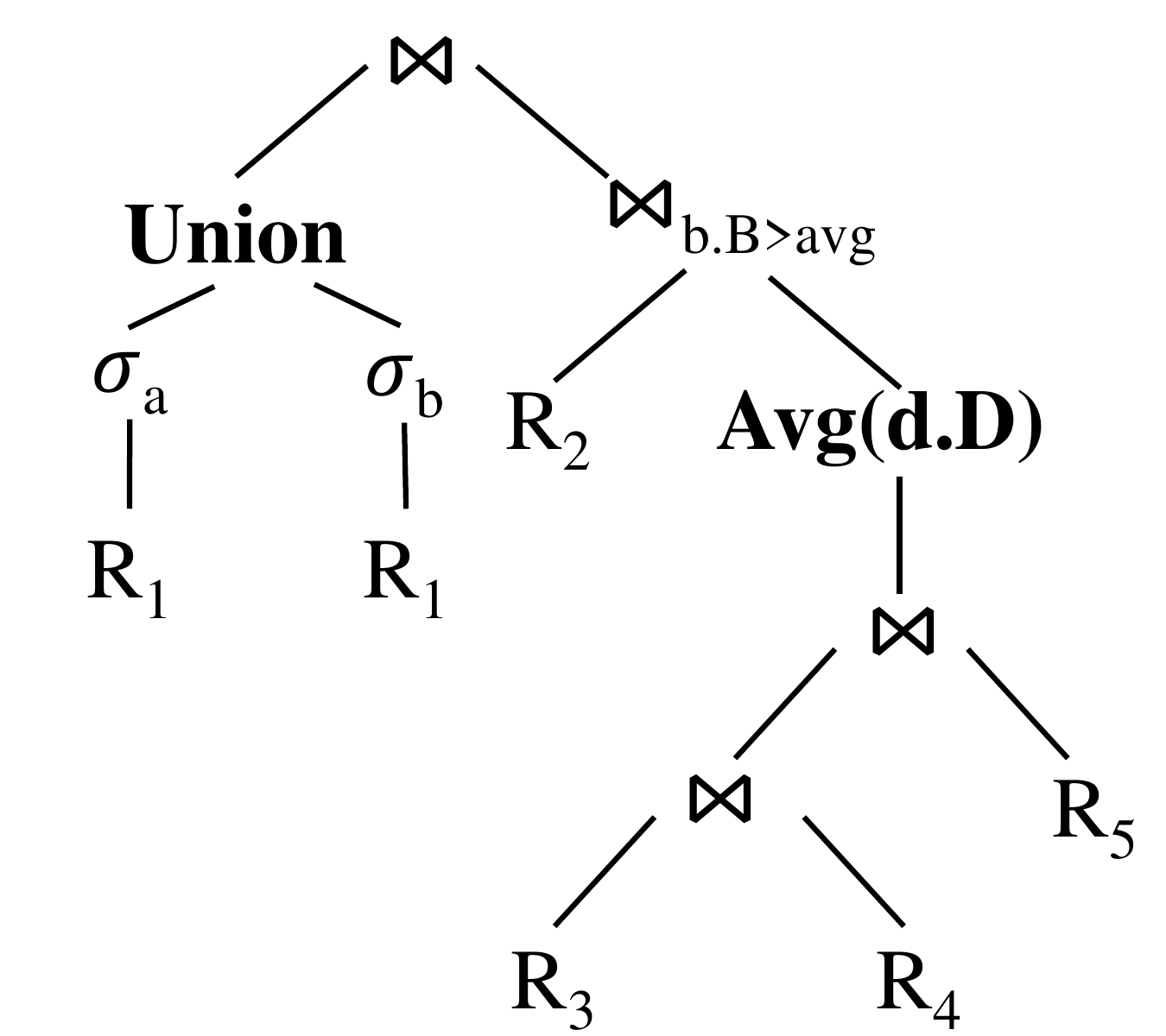}
                \label{F42a}
            \end{minipage}
        }
        \subfigure[New SPJ query]
        {
            \begin{minipage}[t]{0.47\linewidth}
                \includegraphics[width=\linewidth]{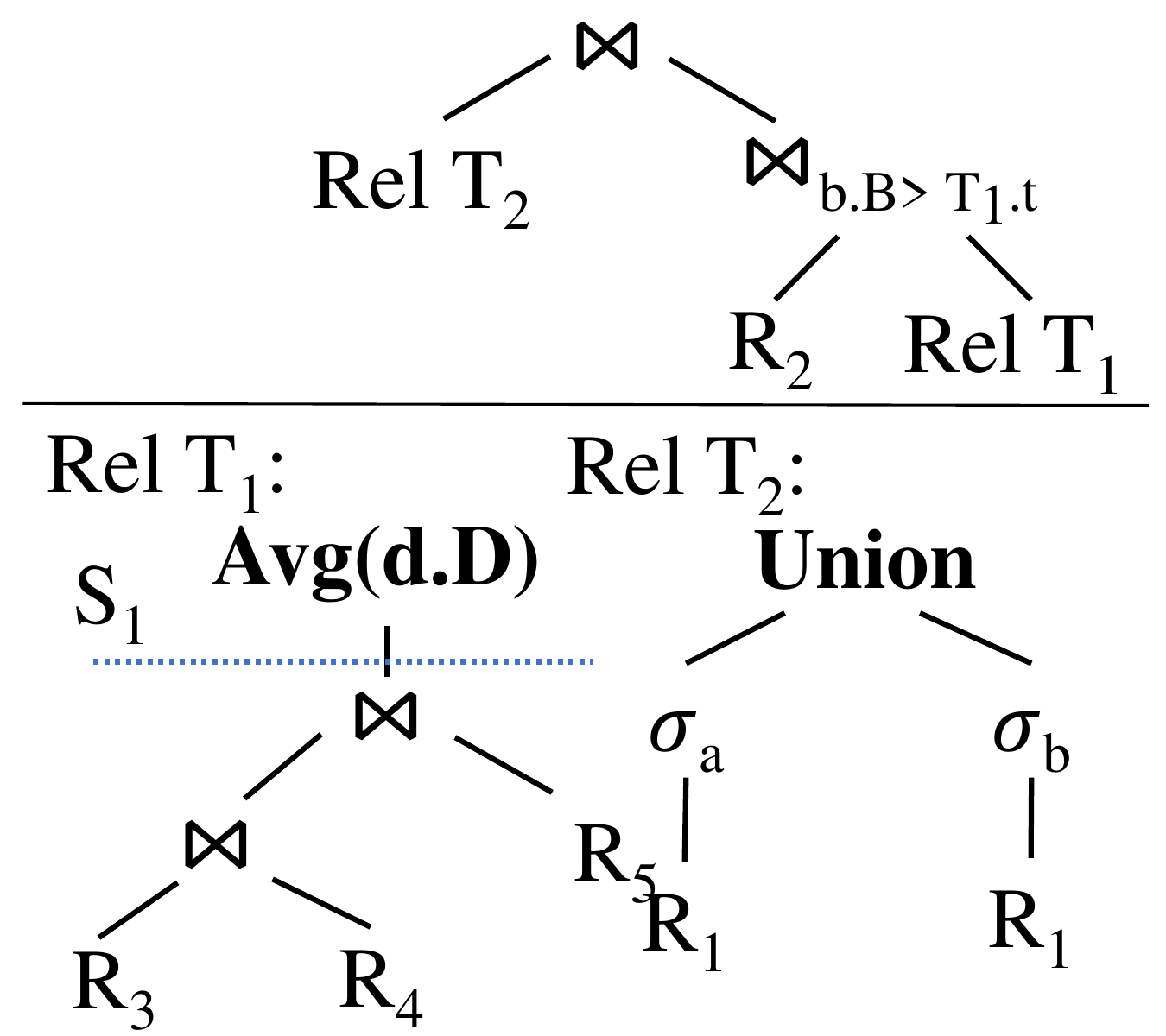}
                \label{F42b}
            \end{minipage}
        }
        \caption{How to deal with non-SPJ query}
        \label{F42}
        \Description{}
    \end{figure}

\cref{F42} shows an example.
On the left, there is a query plan containing two Non-SPJ operators:
a \textbf{Union} and an \textbf{Avg} aggregation.
We first divide the plan tree based on these two operators so that
they become the roots of their own subtrees.
For each of the subtree, we execute the SPJ part of the plan first.
In the \textbf{Avg} subtree, for example, the subquery
$R_3 \Join R_4 \Join R_5$ is first executed through \ourname,
and the result is used as an input to the \textbf{Avg} operator.
We then materialize the output of \textbf{Avg} and \textbf{Union}
as relation $T_1$ and $T_2$, and use those to replace the
corresponding subtrees in the root plan.
Finally, we invoke \ourname again on the root plan to obtain
the final result.

Notice that a potential limitation of adopting the above Non-SPJ extension of \ourname
is that it might miss a few global optimization opportunities such as group-by
pushdown\footnote{As far as we know, neither PostgreSQL 15.1~\cite{PostgresDoc} nor MySQL 8.0~\cite{MySQLDoc} supports group-by pushdown.}
during the cost-based enumeration stage.
Our evaluation on TPC-H~\cite{TPCH} and DSB~\cite{ding2021dsb} (an extension to TPC-DS~\cite{TPCDS})
in \cref{S:eva} shows that such a negative impact is limited and is often outweighed by
the benefit of a more optimized join ordering.
A deeper integration of Non-SPJ operators into \ourname is beyond the scope of this paper,
and we leave it as future work.

\section{Subquery Creation \& Selection Policies}
\label{S:policy}
In \cref{S:qs}, we mainly focused on the correctness of the \ourname algorithm.
To fully exploit \ourname's ability to deliver good query performance,
we discuss two critical policies in this section:
(1) how to pick a subquery set from a (exponentially) large number of valid
choices, and
(2) how to select a subquery from an existing set at each \ourname iteration.

\subsection{Generating a Subquery Set}
\label{S41}

        \begin{figure}[t!]
            \subfigure[Original SQL]
            {
                \begin{minipage}[t]{0.97\linewidth}
                    \includegraphics[width=\linewidth]{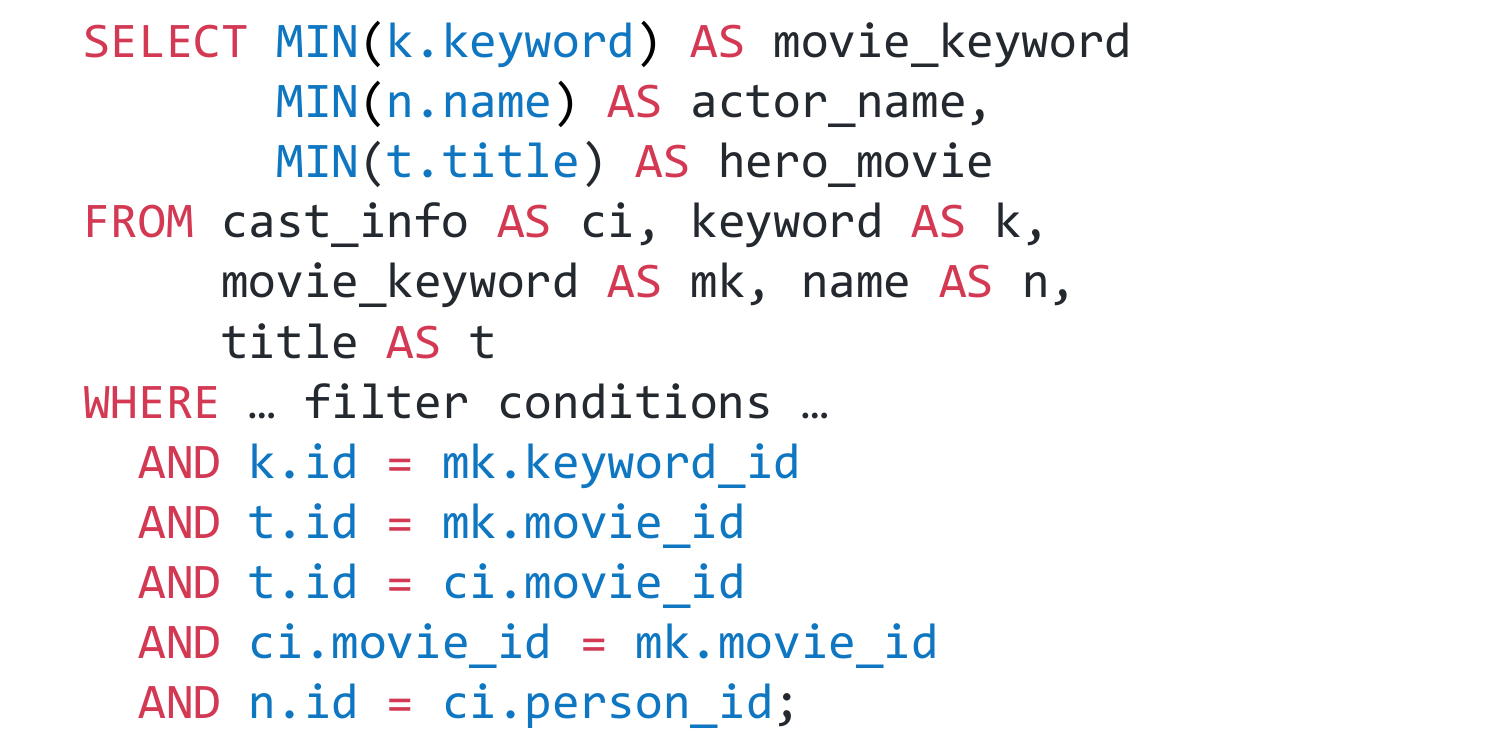}
                    \label{F6a}
                \end{minipage}
            }
            \subfigure[Join schema]
            {
                \begin{minipage}[t]{0.47\linewidth}
                    \includegraphics[width=\linewidth]{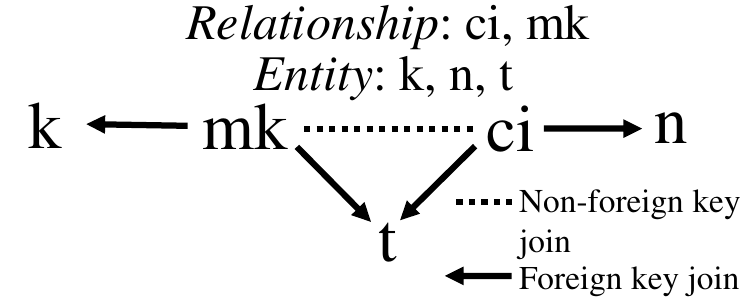}
                    \label{F6b}
                \end{minipage}
            }
            \subfigure[Directed join graph]
            {
                \begin{minipage}[t]{0.47\linewidth}
                    \includegraphics[width=\linewidth]{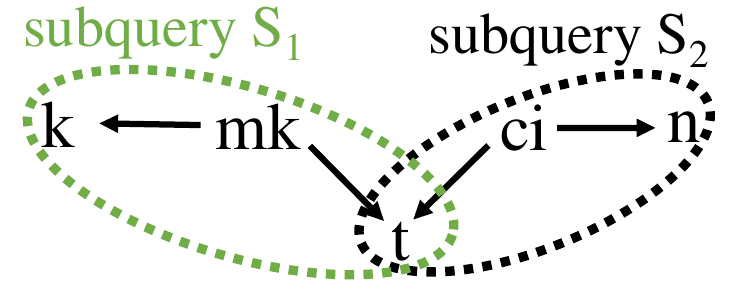}
                    \label{F6c}
                \end{minipage}
            }
            \centering
            \caption{Join graph split by \textit{RelationshipCenter}}
            \label{F6}
            \Description{}
        \end{figure}

As described in \cref{S32}, a subquery set $\textbf{\textit{Q}}$ output by
the Query Splitting Algorithm (QSA) is valid if it ``covers'' the original
query $q$.
The number of the candidate sets, however, grows exponentially with the
complexity of $q$ (e.g., the number relations in $q$).
The goal of the QSA is to perform an efficient search in the candidate space
and produce a subquery set that best serves the overall \ourname algorithm.
    
Intuitively, we prefer subqueries that return small results.
There are two potential advantages.
First, the cost of materializing the output of such a subquery
is small in each \ourname iteration.
Second, because the materialized result will become a new base
relation to participate in subsequent executions,
having a smaller size is beneficial to improving the performance
of succeeding subqueries.

We, therefore, propose a subquery generation strategy, called the
\textit{\RCenter} strategy (i.e., \RelationshipCenter).
\RCenter leverages the concept of non-expanding operators
proposed by Hertzschuch et al.~\cite{hertzschuch2021simplicity}.
A non-expanding operator is defined as an operator that has an
output size smaller than or equal to its input size.
The \RCenter strategy is based on the non-expanding property
of the primary-foreign-key joins.
Given a query, we define \textit{FK-relation} as a relation
that has at least one foreign-key reference to another relation and
\textit{PK-relation} as a relation whose primary key is referenced by
other relations.

The \textit{RCenter} strategy is based on the non-expanding property
of the primary-foreign-key joins.
Many relational database schemas follow the classic entity-relationship
design~\cite{chen1976entity} (i.e., the star schema) where
a ``relationship'' relation (or the ``fact'' table) is at the center
containing foreign keys referring to the surrounding ``entity'' relations
(or ``dimension'' tables).
Therefore, given a query, we define \textit{R-relation} as a relation
that has at least one foreign-key reference to another relation and
\textit{E-relation} as a relation whose primary key is referenced by
other relations.

We then construct a directed join graph for the query where each vertex
represents a relation and each edge represents a join predicate
(single table predicates are associated with the vertices).
For each edge, its direction is from an R-relation to an E-relation.
If the join happens between two relations with the same type,
the edge is bidirectional.
Redundant join predicates (e.g., those form cycles in the join graph)
are deleted with the priority of removing bidirectional edges.

The \textit{RCenter} strategy works by traversing all the vertices in the
directed join graph.
For each vertex that has at least one outgoing edge, we create a
subquery centered at this vertex (i.e., relation) with all the
relations it points to.
We next demonstrate the process with an example.

\cref{F6a} shows the SQL text of Query 6d from the Join Order Benchmark.
Among the five join predicates, four of them
($k \Join mk$, $t \Join mk$, $t \Join ci$, and $n \Join ci$)
are primary-foreign-key joins.
We first build the directed join graph, indicating the primary-foreign-key
relationships, as shown in \cref{F6b}.
Because $mk$, $t$, and $ci$ form a join cycle, we remove the redundant
bidirectional edge between $mk$ and $ci$ from the graph.
We then scan the node list and detect that $mk$ and $ci$ have outgoing edges.
We, therefore, create subquery $S_1 = k \Join mk \Join t$ centered around $mk$
and subquery $S_2 = t \Join ci \Join n$ centered around $ci$,
as illustrated in \cref{F6c}.

We note that for a query that follows a strict star schema, the \RCenter strategy
takes no effect because there is a single center (i.e., the fact table) in the join graph
(and hence, no re-optimization is carried out by the \ourname algorithm).
Star-schema queries, however, are relatively easy to optimize because all the joins
are primary-foreign-key joins (thus non-expanding).
This prevents the optimizer from making ``huge'' mistakes (e.g., exploded intermediate results)
regarding the join ordering.
Our experiments on TPC-H which contains mostly star-schema queries show that
all the state-of-the-art query optimization algorithms, including re-optimization,
learned cardinality estimation, and robust query processing,
produce plans with similar performance (refer to \cref{S54}, \cref{F16}).
Their improvement over the default \postgres optimizer is also limited.
Therefore, it is reasonable in practice to adopt the \RCenter strategy because
it focuses on re-optimizing the inverse star-schema pattern,
which are more likely to have severe cardinality estimation errors~\cite{leis2015good}.

    We further introduce two alternative strategies,
    named \textit{\ECenter} (i.e., \textit{\EntityCenter}) and \textit{MinSubquery},
    for comparison in the evaluation.
    \ECenter is the dual of \RCenter:
    the directed join graph in \ECenter has all edges reversed
    (i.e., from a PK-relation to an FK-relation), making the PK-relation
    at the center of each generated subquery.
    The other alternative MinSubquery refers to the strategy of dividing
    the query into minimum-sized subqueries.
    For each join predicate in the original query, we construct a subquery
    containing only the involved relations with their corresponding filter
    conditions, thus creating the smallest join units.

\subsection{Subquery Execution Order}
\label{S42}

    \begin{table}[t!]
        \caption{Cost Functions for SSA}
        \label{T2}
        \begin{tabular}{c|l}
            \toprule
            Function Name  &  Expression \\
            \midrule
            $\Phi_1$    & $\textbf{C}(q)$ \\
            $\Phi_2$    & $\textbf{C}(q) \cdot \log(\textbf{S}(q))$ \\
            $\Phi_3$    & $\textbf{C}(q) \cdot \sqrt{\textbf{S}(q)}$ \\
            $\Phi_4$    & $\textbf{C}(q) \cdot \textbf{S}(q)$ \\
            $\Phi_5$    & $\textbf{S}(q)$ \\
            \bottomrule
        \end{tabular}
    \end{table}
    
Given a subquery set generated by the \textit{RCenter} strategy,
the second policy decision is how to determine the execution
order of the subqueries
(i.e., the aforementioned Subquery Selection Algorithm, or SSA).
Although this order is irrelevant to the correctness of \ourname,
it can largely affect the overall query performance.
For example, if the largest join in the query is executed early
in the re-optimization process, the overhead of materializing
its output and scanning it in subsequent executions is overwhelming.

As mentioned in the introduction, it is beneficial to run the
``simpler'' subqueries first and delay the execution of large
joins by as much as possible.
In this way, we increase the probability of reducing the input
sizes of those large joins with a modest re-optimization cost.
We, therefore, developed a set of cost functions $\Phi$ to
measure the ``simplicity'' of the subqueries, as shown in \cref{T2}.
At each \ourname iteration, we compute $\Phi$ for each subquery
in the set and select the one with the smallest value to execute.
    
The two metrics used to compute $\Phi$ for a subquery $q$ are
the estimated cost generated from the optimizer
(denoted by $\textbf{C}(q)$)
and the cardinality estimation of $q$'s output
(denoted by $\textbf{S}(q)$).
The intuition is that we want to prioritize a subquery that is fast
to execute and has the most potential to speed up later subqueries.
Correspondingly, the optimizer-generated cost reveals the complexity
of the current subquery, while the output size estimation suggests its
potential ``burden'' on future subqueries.
For both metrics, a smaller value is better.

A combination of $\textbf{C}(q)$ and $\textbf{S}(q)$ indicates the
algorithm's aggressiveness in investing the cost of the present subquery
for potential future benefits.
A larger factor of $\textbf{C}(q)$ suggests a more conservative strategy
that emphasizes picking the easiest subquery to execute at the moment.
On the contrary, putting more emphasis on $\textbf{S}(q)$ means that the
algorithm believes that firing a more complex subquery with a smaller
result set is going to pay off in subsequent executions.
$\Phi_1$ through $\Phi_5$ defined in \cref{T2} indicate five strategies
with an ascending weight of $\textbf{S}(q)$.
Our evaluation in \cref{S52} shows that a balanced SSA strategy
(i.e., $\Phi_4 = \textbf{C}(q) \cdot \textbf{S}(q)$) in \ourname
delivers the best and most robust performance.

\section{Implementation}
\label{S:impl}
\begin{figure}[!t]  
    \includegraphics[width=\linewidth]{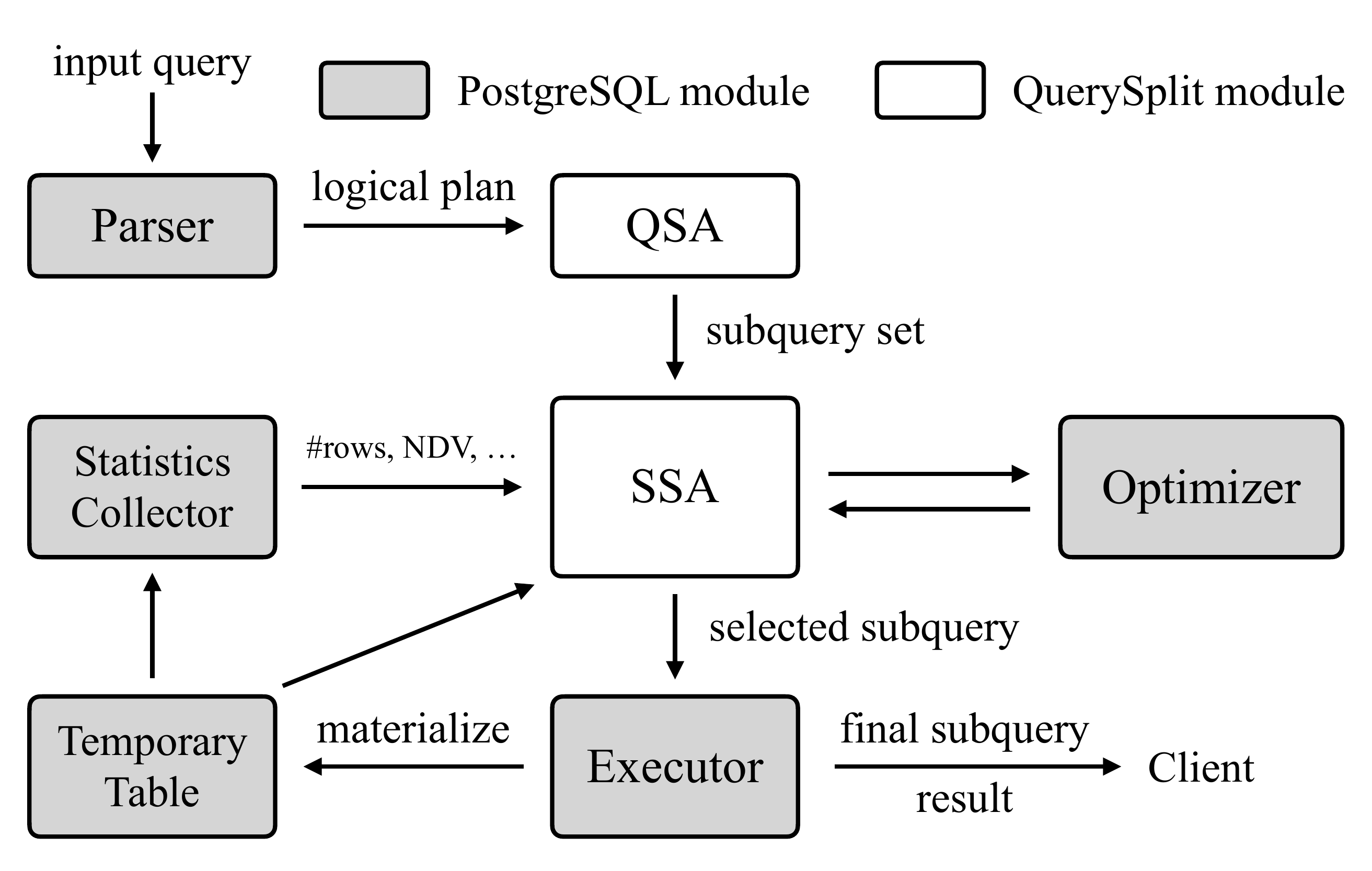}
    \centering
    \caption{Implementing \ourname in \postgres}
    \label{F43}
    \Description{}
\end{figure}

We implemented the \ourname algorithm in \postgres 12.3 in C.
As shown in \cref{F43}, we included two new modules: the
Query Splitting Algorithm (QSA) module and
the Subquery Selection Algorithm (SSA) module.
The data flow is redirected accordingly to perform subquery
execution and re-optimization.

Specifically, the QSA module receives a logical plan from the
built-in parser and runs the RCenter-based algorithm (\cref{S41})
to produce a subquery set that covers the original query.
The subqueries in the set are logical plans of the same type
as the input of QSA. The subquery set is then sent to the
SSA module for execution.

Upon receiving the subquery set, the SSA module starts a loop
to consume one subquery at each iteration.
As described in \cref{S42}, the SSA computes the cost function
$\Phi$ for each subquery and sends the one with the smallest cost
to the execution engine.
During this process, the SSA invokes \postgres's native optimizer
on each subquery to obtain its execution time estimation
(i.e., $\textbf{C}(q)$) and its output size estimation
(i.e., $\textbf{S}(q)$).

The execution result of the selected subquery (except for the last one)
is materialized in a memory buffer by setting the output destination
to a temporary table in \postgres.
The materialized table is then sent to the Statistics Collector,
where a set of \postgres's native routines are performed to compute
the basic statistics about the table such as the number of rows, number
of distinct values, histograms, etc.
After the statistical analysis, both the temporary table and its
associated statistics are sent back to the SSA module to update
the remaining ``overlapping'' subqueries, preparing for the
next iteration.

The source code of \ourname-integrated \postgres is available on Github
through the anonymous link in~\cite{Github}.

\section{Evaluation}
\label{S:eva}
The evaluation of \ourname is organized as follows.
In \cref{S52}, we first examine the policies proposed in \cref{S:policy}
for the QSA and SSA algorithms.
Because the cost function in SSA relies on the optimizer's output,
we then investigate the robustness of our algorithm against varying
cardinality estimation errors.
Next, we show our main results in \cref{S54}, where we compare the
end-to-end performance of \ourname to that of existing baselines
using the Join Order Benchmark (JOB)~\cite{leis2015good}.
A follow-up study on whether to collect run-time statistics on the
materialized intermediate results is presented in \cref{S55} to
show the trade-offs.
In \cref{S56}, We investigate whether the cost functions proposed in~\cref{S42} can boost
the performance of existing re-optimization algorithms.
In \cref{S57}, we conduct detailed case studies to provide further
insights about the reasons why \ourname outperforms the baselines
in a majority of situations.

\subsection{Workload \& Experiment Setup}
\label{S51}

\begin{itemize}[leftmargin = 10pt]
\item \textbf{Join Order Benchmark (JOB)}
JOB is a collection of manually-created queries
over the IMDB data set~\cite{IMDB}.
It has been widely used in prior work to evaluate the optimizer in relational
database management systems (DBMSs)
~\cite{cai2019pessimistic, hertzschuch2021simplicity, perron2019learned}.
JOB is known to have more complex queries than the standard TPC-H~\cite{TPCH}
and is preferable for stress-testing the optimizer.
There are a total of 113 queries in JOB with 91 of them returning non-empty results.
We use these 91 queries in our evaluation.
By default, we build a B+tree index for each primary key and foreign key appearing
in the schemas.
\item \textbf{Decision Support Benchmark (DSB)}
DSB extends the standard TPC-DS benchmark~\cite{TPCDS} with data skews~\cite{ding2021dsb}.
There are 52 queries (15 SPJ and 37 Non-SPJ) in DSB.
We set the scale factor to 5 in the evaluation.
\item \textbf{TPC-H}
TPC-H~\cite{TPCH} is the standard for benchmarking analytical queries by convention.
It follows a star-schema and contains 22 (Non-SPJ) queries.
We set the scale factor to 3 in the evaluation.
\end{itemize}

All experiments are performed on a server equipped with an
Intel\textregistered
Core\textregistered
i9-10900K CPU (3.70 GHz) and 128 GB of DRAM.
Similar to \ourname, all the baseline algorithms are also implemented in \postgres.
We use the same parameter configuration in \postgres across all solutions.
We set the \texttt{max\_parallel\_workers} to 0 to guarantee a serial execution
of the queries in the workload.
The \texttt{effective\_cache\_size} is set to 8 GB.
Other parameters follow the \postgres default.
The execution of each query times out after 1000 seconds.
We repeat each experiment three times and report the average measurements.

\subsection{Policies in QSA \& SSA}
\label{S52}

The default policy for subquery generation (i.e., QSA) in \ourname
is \textit{RCenter}.
To show its efficiency, we introduce two alternative strategies,
named \textit{ECenter} and \textit{MinSubquery}.
As the name suggests, ECenter is the dual of RCenter:
the directed join graph in ECenter has all edges reversed
(i.e., from a Relationship to an Entity), making the Entity
relation at the center of each generated subquery.
The other alternative MinSubquery refers to the strategy of dividing
the query into minimum-sized subqueries.
For each join predicate in the original query, we construct a subquery
containing only the involved relations with their corresponding filter
conditions, thus creating the smallest join units.

For the cost function $\Phi$ used to determine the subquery execution
order (i.e., SSA), besides the five candidates proposed in \cref{T2},
we introduce an additional baseline \textit{global\_deep} that orders the
subqueries according to the global physical plan.
At each \ourname iteration, we choose the deepest join operator in
the global plan tree and obtain the involved relation set $\textbf{R}$.
The algorithm then searches the subquery set and finds the one(s) whose
relation set is a superset of $\textbf{R}$.
If multiple subqueries satisfy the requirement, the algorithm randomly
picks a subquery to send for execution.
Notice that \ourname with the \textit{global\_deep} SSA policy is different from
executing the global plan directly because the subquery set is not
derived from the global plan.

\begin{table}[t!]
    \caption{JOB execution time for different combinations of QSA and SSA policies}
    \label{T3}
    \begin{tabular}{l|ccc}
        \toprule
        \diagbox{SSA}{Time(s)}{QSA} & RCenter & ECenter & MinSubquery\\
        \midrule
        $\Phi_1$: $\textbf{C}(q)$                            & 421 & 378 & 463 \\
        $\Phi_2$: $\textbf{C}(q) \cdot \log(\textbf{S}(q))$  & 327 & 349 & 428 \\         
        $\Phi_3$: $\textbf{C}(q) \cdot \sqrt{\textbf{S}(q)}$ & 328 & 339 & 418 \\
        $\Phi_4$: $\textbf{C}(q) \cdot \textbf{S}(q)$        & \textcolor{red}{\textbf{295}} & 350 & 427 \\
        $\Phi_5$: $\textbf{S}(q)$                            & 348 & 407 & 474 \\
        \texttt{global\_deep}                                & 356 & 413 & 401 \\
        \bottomrule
    \end{tabular}
\end{table}
    
\cref{T3} shows the execution time of the JOB workload using \ourname
with different combinations of QSA and SSA policies.
For QSA policies, RCenter consistently outperforms the others
(except for $\Phi_1$).
This result confirms that keeping more non-expanding operators in
subqueries is beneficial to re-optimization
(the USE paper presents similar findings~\cite{hertzschuch2021simplicity}).
The RCenter policy achieves the goal by preserving as much
primary-foreign-key joins as possible when splitting the
original query.
    
For SSA policies, $\Phi_3$ and $\Phi_4$ outperform the others in general.
The results indicate that subqueries with both short execution time and
small output cardinality should have execution priority in re-optimization,
and a more balanced weight assignment between these two metrics tends to
improve the query performance.
Overall, a combination of RCenter and $\Phi_4$ in \ourname leads to an
outstanding performance of the DBMS, as shown in \cref{T3}.

\begin{figure}[!t]
    \includegraphics[width=\linewidth]{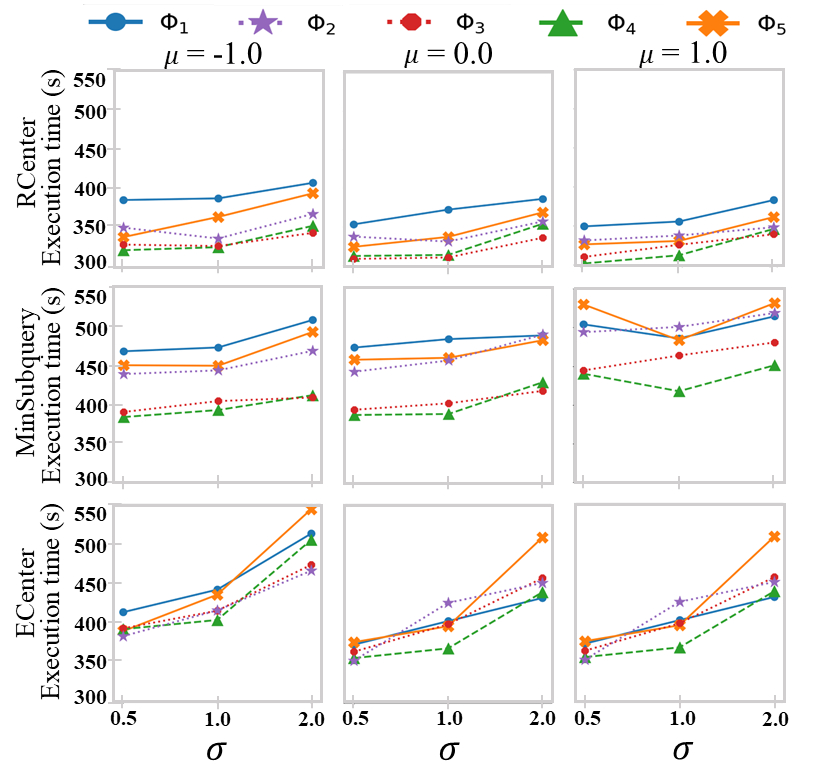}
    \centering
    \caption{Execution time of \ourname on JOB under erroneous cardinality estimations}
    \label{F9}
    \Description{}
\end{figure}
    
\textbf{Robustness test}
Because the cost functions $\Phi_1$ - $\Phi_5$ depend on the output
from the optimizer, we next test the robustness of these
cost functions against cardinality estimation (CE) errors.
The experiments are designed in the following way.
For each JOB query, we execute every valid subquery from it and
record its true output cardinality (\textit{true\_card}).
Next, we generate the erroneous cardinality (\textit{err\_card})
by adding a controlled noise to the true cardinality:
\[ err\_card = 2^{N(\mu, \sigma^2)} * true\_card \]

where $N(\mu, \sigma^2)$ represents a normal distribution with
$\mu$ as the mean and $\sigma$ as the standard deviation.
We then inject the erroneous cardinalities into \postgres's
optimizer via the method proposed by Cai et al.~\cite{cai2019pessimistic}
so that we can control the quality of the cardinality estimations (CE).

\cref{F9} shows the time of executing the JOB workload under the
aforementioned different \ourname policies with a varying mean
and standard deviation of the injected CE noise.

Regardless of the choice of cost functions, modest errors (i.e., $\sigma = 0.5, 1, 2$)
in cardinality estimation have a small negative impact on the query performance
when \ourname adopts the \RCenter or MinSubquery QSA policy.
The \ECenter policy is more sensitive to inaccurate cardinalities
because a \ECenter subquery is more likely to include multiple
large FK-relations, where a miscalculation of the join
cardinality could incur an unrecoverable penalty.
When the injected CE error becomes large (i.e., $\sigma = 4$, which means $Prob$(CE error > $1500\%$) > $30\%$),
\ourname is no longer robust no matter what QSA and SSA strategies it selects.
Because a policy combination of \RCenter and $\Phi_4$ outperforms
the other pairs consistently under modest CE errors, we set both policies as
the default in subsequent experiments.

\subsection{A Comparison to Baseline Solutions}
\label{S54}

In this section, we compare \ourname against the following baselines,
including four re-optimization algorithms and two approaches to improve
cardinality estimation.
\begin{itemize}[leftmargin = 15pt]
    \item \textbf{Default}: \postgres with the default optimizer.
    \item \textbf{Optimal}: \postgres with an ``ideal'' optimizer.
        We provide the optimizer with the accurate cardinality of every possible intermediate
        result so that it generates an optimal plan.
        This serves as the upper-bound for all the evaluated approaches. 
        Note that the ``optimality'' here is with respect to
        perfect cardinality estimates.
    \item \textbf{Reopt}: A re-optimization algorithm that triggers the optimizer
        at each pipeline breaker if it detects that the deviation between the true
        statistics and the estimation exceeds a threshold~\cite{kabra1998efficient}.
    \item \textbf{Pop}: A re-optimization algorithm extending \texttt{Reopt} where
        the optimizer is triggered aggressively in more situations including at
        nested-loop join operators~\cite{markl2004robust}.
    \item \textbf{IEF}: Incremental Execution Framework (IEF) by Neumann and Galindo-Legaria~\cite{neumann2013taking}
        is an adaptive query processing framework where query executions halt at pre-determined places
        in the global plan to remove uncertainty in cardinality estimation errors.
    \item \textbf{Perron19}: A most recent study on the effectiveness of re-optimization~\cite{perron2019learned}.
        In the original paper, the authors compare the true cardinality of each operator with
        the estimated value obtained from the \texttt{EXPLAIN} command. They then materialize
        the intermediate results of those operators with large CE errors and re-execute the
        query to study the performance trade-offs.
        Because it is impractical to get a global view of the true cardinalities by executing
        the query in advance, we modified the algorithm by setting a relative threshold
        (e.g., estimation error is $32\times$ off compared to collected statistics)
        as the run-time re-optimization trigger.
    \item \textbf{USE, Pessi.}: USE~\cite{hertzschuch2021simplicity} and
        Pessimistic Cardinality Estimation (Pessi.)~\cite{cai2019pessimistic}
        provide \textit{robust} cardinality estimation
        using sketches. Although USE forms subqueries during query optimization, its execution is non-adaptive.
    \item \textbf{FS}: A robust query processing technique that considers
        both cost and plan robustness (i.e., insensitive to cardinality estimation errors)
        during query optimization~\cite{wolf2018robustness}.
    \item \textbf{OptRange}: An algorithm that derives the range of
        estimated cardinalities where the current plan stays optimal~\cite{wolf2018calculation}.
        This can serve as a heuristic to reduce unnecessary re-optimizations.
    \item \textbf{NeuroCard, DeepDB, MSCN}: state-of-the-art \textit{learned}
        cardinality estimation algorithms, where NeuroCard~\cite{yang2020neurocard} and
        DeepDB~\cite{hilprecht2019deepdb} are trained directly against the data
        while MSCN~\cite{kipf2018learned} requires a set of training queries.
        These algorithms, however, have limited support to string columns.
    \end{itemize}

    For baselines that do not provide a \postgres integration,
    we implemented them according to the paper with the best effort.
    Implementation details can be found in \cref{A3}.
    For each algorithm, we evaluated two index states:
    (1) indexes are built for primary key (Pk) columns only, and
    (2) indexes are built for primary key (Pk) and foreign key (Fk) columns.

    \begin{figure}[!t]
        \includegraphics[width=\linewidth]{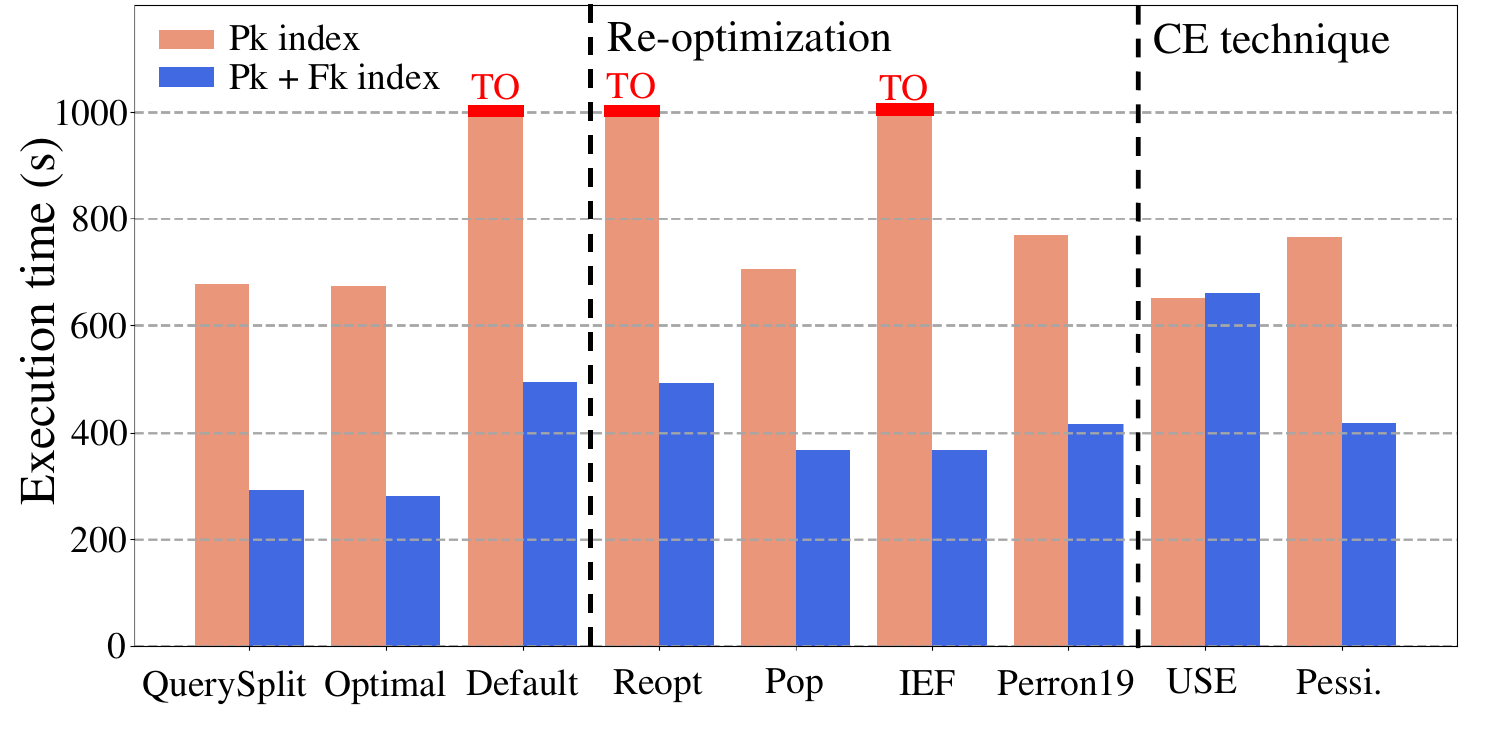}
        \centering
        \caption{Execution time of JOB for \ourname and different baselines (TO = timeout)}
        \label{F10}
        \Description{}
    \end{figure}
    
    \subsubsection{Join Order Benchmark}
    \cref{F10} reports the end-to-end execution time of the JOB workload for
    \ourname and the above baselines.
    In both \textit{Pk index only} and \textit{Pk + Fk index} cases,
    \ourname achieves the shortest execution time compared to the prior solutions.
    The performance improvements are more significant when both primary-key and
    foreign-key indexes are included (which is the default in JOB) because 
    the (sub)plan qualities between different algorithms diverge more with
    an enlarged optimization space.
    
    Notice that the performance difference between \ourname and \texttt{Optimal} is
    very small ($< 4\%$), indicating that \ourname is able to identify and quickly
    converge to an optimal plan.
    It also shows that the re-optimization overhead in \ourname is modest.
    The four re-optimization baselines achieve a certain amount of improvement
    over the \texttt{Default}, but they are still noticeably slower than the optimal.
    This is because they are all held back by the initial physical plan in the
    re-optimization process.
    We provide case studies in \cref{S57} to demonstrate \ourname's advantages
    in detail.
    
    Compared to the robust query processing baselines
    (i.e., USE \footnote{Note that \texttt{USE} has the same performance in both index configurations because it disables nest-loop join, and thus ignores indexes in its query planning.}, Pessi, FS, and OptRange),
    \ourname still exhibits significant performance advantages, especially when
    both primary key and foreign key columns have indexes built (i.e., the blue bars in~\cref{F10}).
    This is because robust query processing algorithms tend to settle down with
    plans that are insensitive to cardinality estimation errors and thus miss
    the opportunity to approach plans that are closer to the optimal.
    Meanwhile, learned cardinality estimation (i.e., NeuroCard, DeepDB, and MSCN)
    achieves limited performance improvement because JOB contains many string columns
    where the learned cardinality estimators fail to handle:
    they have to fall back to \postgres's default in those cases.
    These results confirm the finding in Perron et al.~\cite{perron2019learned} that
    re-optimization is likely to be more effective and efficient than refining CE
    in improving query performance.

    \begin{table}[!t]
        \small
        %\caption{Materialization condition of re-optimization techniques}
        \caption{Materialization frequency and memory usage of re-optimization algorithms}
        \label{T4}
        \begin{tabular}{c|c|c|c}
            \toprule
            \multirow{2}*{Algorithms} & Avg mem per & Avg mat. freq. & Total mem per \\
            ~ & subquery (MB) & per query & query (MB) \\
            \midrule
            \ourname & 5.79 & 2.66 & 15.40 \\
            \midrule
            Reopt & 43.31 & 0.21 & 9.09 \\
            Pop & 7.01 & 4.62 & 32.39 \\
            IEF & 14.59 & 3.11 & 45.37 \\
            Perron19 & 10.99 & 6.59 & 72.42 \\
            \bottomrule
        \end{tabular}\\
        \flushleft{\footnotesize{*``Total mem per query'' in the table refers to the total
        memory used to materialize the intermediate results. For reference,
        the peak memory usage for \texttt{Default} and \texttt{Optimal} is 572MB and 563MB, respectively.}}\\
    \end{table}
    
    \cref{T4} shows the materialization frequency and the associated memory usage
    of each of the re-optimization algorithms.
    Compared to the baselines, \ourname has the lowest memory consumption per subquery
    (i.e., per re-optimization iteration) because the RCenter-based QSA preserves as
    many non-expanding operators in the subqueries as possible.
    \ourname also has the second lowest re-optimization frequency per query.
    Reopt achieves the lowest because it only triggers re-optimization at pipeline breakers
    with large CE errors.
    Overall, except for Reopt that adopts an over-conservative strategy,
    the materialization memory cost of \ourname is significantly lower than that
    of the other competitors.
    
    \subsubsection{TPC-H}
\begin{figure}[!t]
            \setlength{\abovecaptionskip}{0pt}
            \includegraphics[width=\linewidth]{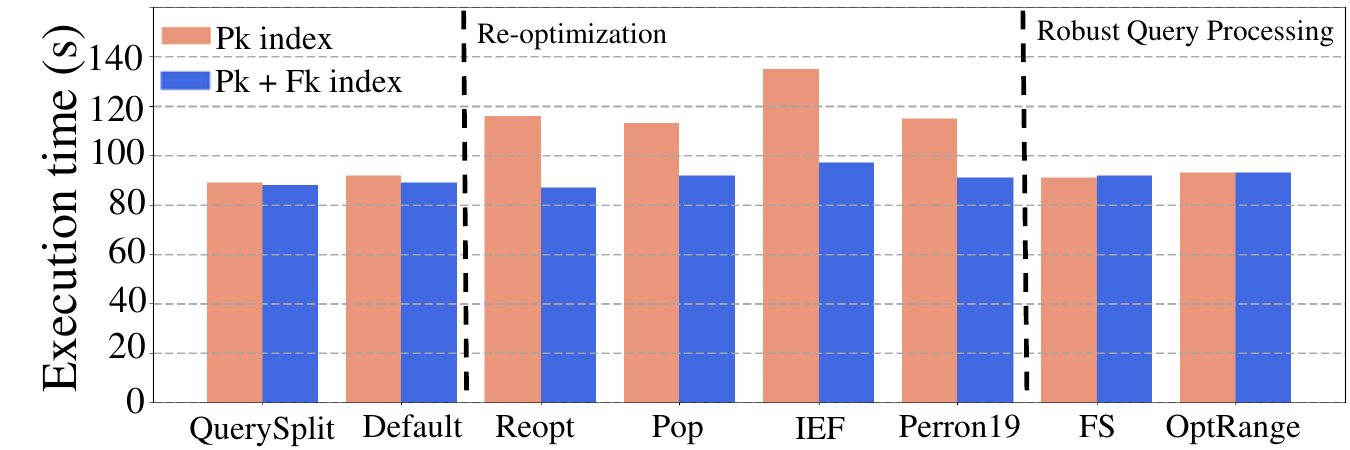}
            \centering
            \caption{TPC-H Execution Time}
            \label{F16}
            \Description{}
        \end{figure}

    Unlike JOB, TPC-H serves as a worst-case benchmark for \ourname
    because it follows the star-schema pattern and all of its queries are
    Non-SPJ queries. %(refer to \cref{S33} and \cref{S41}).
    \cref{F16} shows the results for TPC-H.
    Note that we only include the baseline approaches that support
    Non-SPJ queries in the figure.
    \ourname still produces the fastest plans among the baselines,
    although the performance improvements are much smaller than those in JOB.
    As discussed in~\cref{S41}, re-optimization is often unnecessary
    for star-schema queries because the optimizers are less likely to
    make serious mistakes.
    \ourname outperforms existing re-optimization algorithms, nonetheless,
    because \ourname triggers a (``useless'') re-optimization less frequently
    in TPC-H.

    \subsubsection{Decision Support Benchmark}
    
        \begin{figure}[!t]
            \setlength{\abovecaptionskip}{0pt}
            \includegraphics[width=\linewidth]{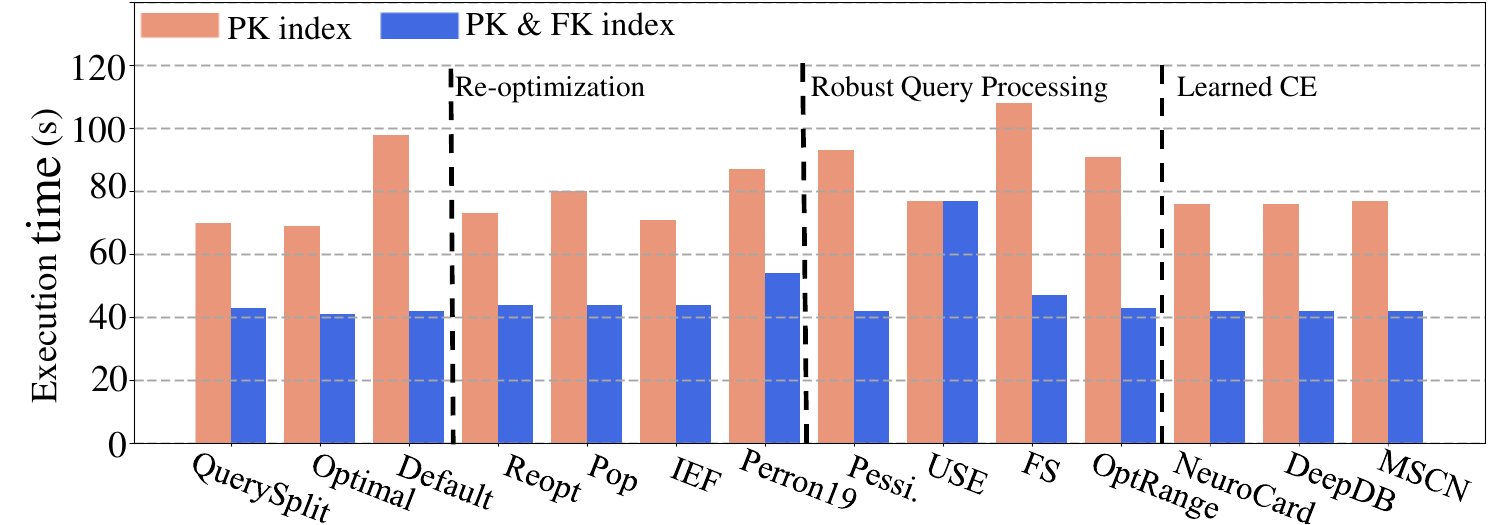}
            \centering
            \caption{DSB Execution Time -- SPJ Queries}
            \label{F13}
            \Description{}
        \end{figure}
        
        \begin{figure}[!t]
            \setlength{\abovecaptionskip}{0pt}
            \includegraphics[width=\linewidth]{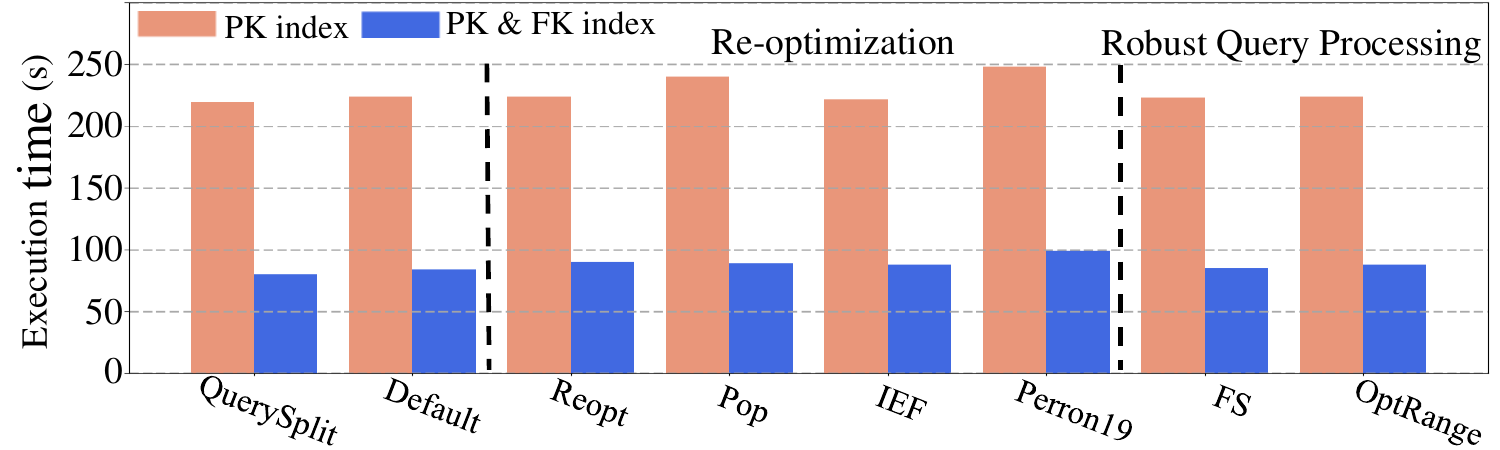}
            \centering
            \caption{DSB Execution Time -- Non-SPJ Queries}
            \label{F17}
            \Description{}
        \end{figure}
    
    Finally, we show the evaluation results for DSB, a mostly star-schema benchmark
    containing both SPJ and Non-SPJ queries.
    For SPJ queries, as shown in~\cref{F13}, \ourname is close to the ``optimal''
    and outperforms the baselines, especially in the Pk-index-only setting.
    Compared to the JOB results, 
    the overall benefit of re-optimization in DSB is less remarkable because of the star-schema pattern.
    Notice that the learned cardinality estimation algorithms become more effective
    because DSB has fewer string attributes involved in the queries.
    The results for Non-SPJ queries of DSB (\cref{F17}) are similar to those of TPC-H.
    Although \ourname targets at JOB-like workloads\footnote{With part of the queries following the inverse star-schema pattern.}
    that are proven to be challenging for optimizers~\cite{leis2015good},
    our experiments with TPC-H and DSB show that \ourname keeps a robust performance consistently,
    thanks to its low re-optimization overhead.

\subsection{Collecting Statistics Or Not?}
\label{S55}
    We continue with a follow-up study on whether collecting statistics on the
    materialized intermediate results is beneficial for each re-optimization algorithm.
    The statistics include the number of distinct values, most common values and their
    frequencies, equal-width/depth histograms, etc.
    Note that the basic row count is already obtained during the result materialization.
    
    Collecting the above statistics requires extra table scans in \postgres.
    Such an overhead may not exist if the DBMS is sophisticated enough to generate the
    statistics while materializing the intermediate results.
    Nevertheless, all the four re-optimization baselines choose to collect statistics
    at runtime by default.
    The reasoning is that collecting these statistics is going to help plan future
    subqueries because the optimizer has little knowledge of the newly materialized
    relation(s).
    
    We repeat the JOB experiments for the re-optimization algorithms in two different
    settings: (1) collecting the statistics for every materialized intermediate result,
    and (2) disabling the statistics collector and only passing along the row count
    to the optimizer.
    
    \begin{figure}[!t]
        \includegraphics[width=\linewidth]{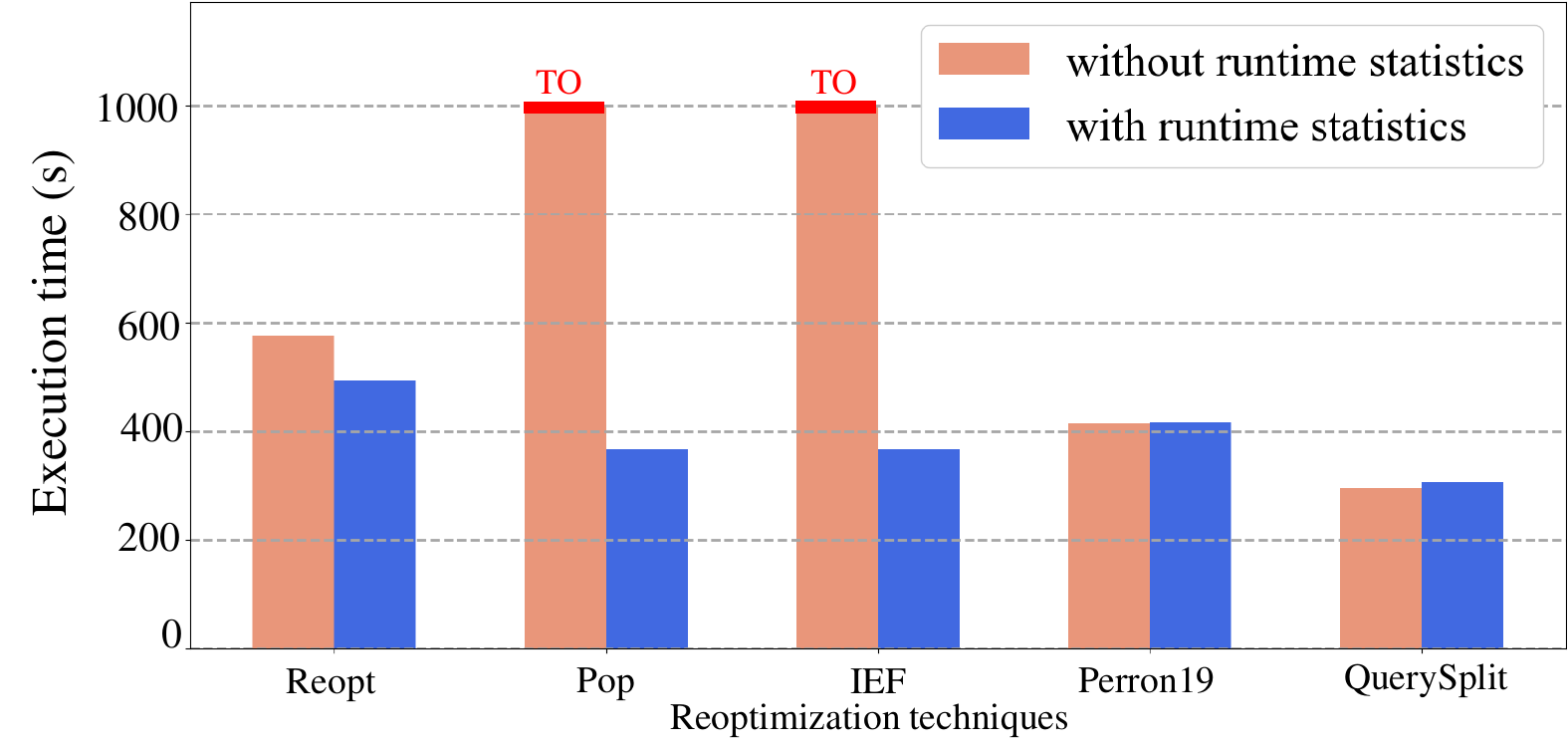}
        \centering
        \caption{JOB execution time with and without high-level statistics}
        \label{F12}
        \Description{}
    \end{figure}
    
    \cref{F12} shows the benchmark results.
    Surprisingly, collecting statistics at runtime has no effect
    (even a slightly negative effect) on the overall query performance
    in \texttt{Perron19} and \ourname, despite the performance
    of other algorithms depending heavily on those statistics.
    This is because each subquery in \texttt{Perron19} only involves at most
    two relations and is, therefore, less likely for the optimizer to make
    mistakes due to the lack of statistics.
    For \ourname, each subquery mostly contains primary-foreign-key joins.
    Because \postgres's optimizer does not use any additional statistics
    other than the row count of the primary-key table to estimate the
    cardinality of such a join, collecting the statistics provides little
    benefit for the optimizer to generate a better plan.
    
    The above experiments show that whether to collect statistics during
    re-optimization should not be a ``no-brainer''.
    The decision depends heavily on the re-optimization algorithm and the
    quality of the system's native optimizer.

\subsection{Existing Re-optimization Algorithms with New Cost Functions} 
\label{S56}

We investigate whether the cost functions proposed in~\cref{S42} can boost
the performance of existing re-optimization algorithms in this section.
We repeat the JOB experiments on Re-opt, Pop, IEF, and Perron19 and use
our cost functions (i.e., $\Phi_1 - \Phi_5$) to determine which subquery
to execute first. \cref{T7} presents the benchmark execution times.

\begin{table}[t!]
    \small
    \setlength{\abovecaptionskip}{1pt}
    \caption{JOB execution time for existing re-optimizations algorithms with cost functions from \ourname.}
    \label{T7}
    \begin{tabular}{l|cccc}
        \toprule
        \diagbox{SSA}{Time(s)}{QSA} & Reopt & Pop & IEF & Perron19 \\
        \midrule
        $\Phi_1$: $\textbf{C}(q)$                            & 488 & 406 & 877 & 412\\
        $\Phi_2$: $\textbf{C}(q) \cdot \log(\textbf{S}(q))$  & 482 & 479 & 835 & 420\\         
        $\Phi_3$: $\textbf{C}(q) \cdot \sqrt{\textbf{S}(q)}$ & 475 & 480 & 855 & 418\\
        $\Phi_4$: $\textbf{C}(q) \cdot \textbf{S}(q)$        & 483 & 485 & 857 & 417 \\
        $\Phi_5$: $\textbf{S}(q)$                            & 492 & 487 & 800 & 415\\
        \textbf{Original}                                    & \textbf{493} & \textbf{401} & \textbf{367} & \textbf{416}\\
        \bottomrule
    \end{tabular}
\end{table}

Overall, simply applying the proposed cost functions to existing re-optimization
algorithms brings little benefit.
IEF, in particular, experienced major performance degradation
because the new cost functions break IEF's optimization strategy
of prioritize subqueries with a high uncertainty in cardinality estimation.
These experiments demonstrate that a ``wise'' cost function alone is incapable of
compensating for a sub-optimal subquery division in a re-optimization algorithm.

\subsection{Insights into \ourname}
\label{S57}

In this section, we provide a deeper analysis on the reasons why
\ourname outperforms existing re-optimization algorithms.
An intuition is that \ourname prioritizes the execution of subqueries
that produce smaller intermediate results and, thus, postponing
potential large joins by as much as possible.

To verify this, we plot two sets of ``timelines'' for each JOB query,
where the X-axis is the count of completed re-optimization iterations.
For the first set of timelines, we monitor the size of the intermediate
result at each re-optimization iteration for each evaluated algorithm
(\texttt{Reopt} is omitted because of poor performance).
For the second set of timelines, we plot the execution time for the
corresponding subquery at each iteration.

When comparing \ourname against \texttt{Pop}, \texttt{IEF}, \texttt{Perron19},
and \texttt{Optimal} (for reference), we summarized four representative
categories out of the 91 JOB queries:

\begin{itemize}[leftmargin = 15pt]
    \item \textbf{Avoided Large Joins}: \ourname successfully avoided performing
    large join(s) that appears in other re-optimization algorithms.
    \item \textbf{Delayed Large Joins}: \ourname postponed the large join(s) to
    later iterations with (hopefully) a smaller input size and a smaller impact
    on the performance of other subqueries.
    \item \textbf{No Difference}: The timeline patterns and the performance
    are similar between \ourname and other re-optimization algorithms.
    \item \textbf{Worse}: A rare case where \ourname unexpectedly produced large
    intermediate results and performed worse than the others.
\end{itemize}

We next provide detailed case studies for each category.

    \begin{figure}[t!]
        \subfigure[Result size at each iteration]
        {
            \begin{minipage}[t]{0.47\linewidth}
                \includegraphics[width=\linewidth]{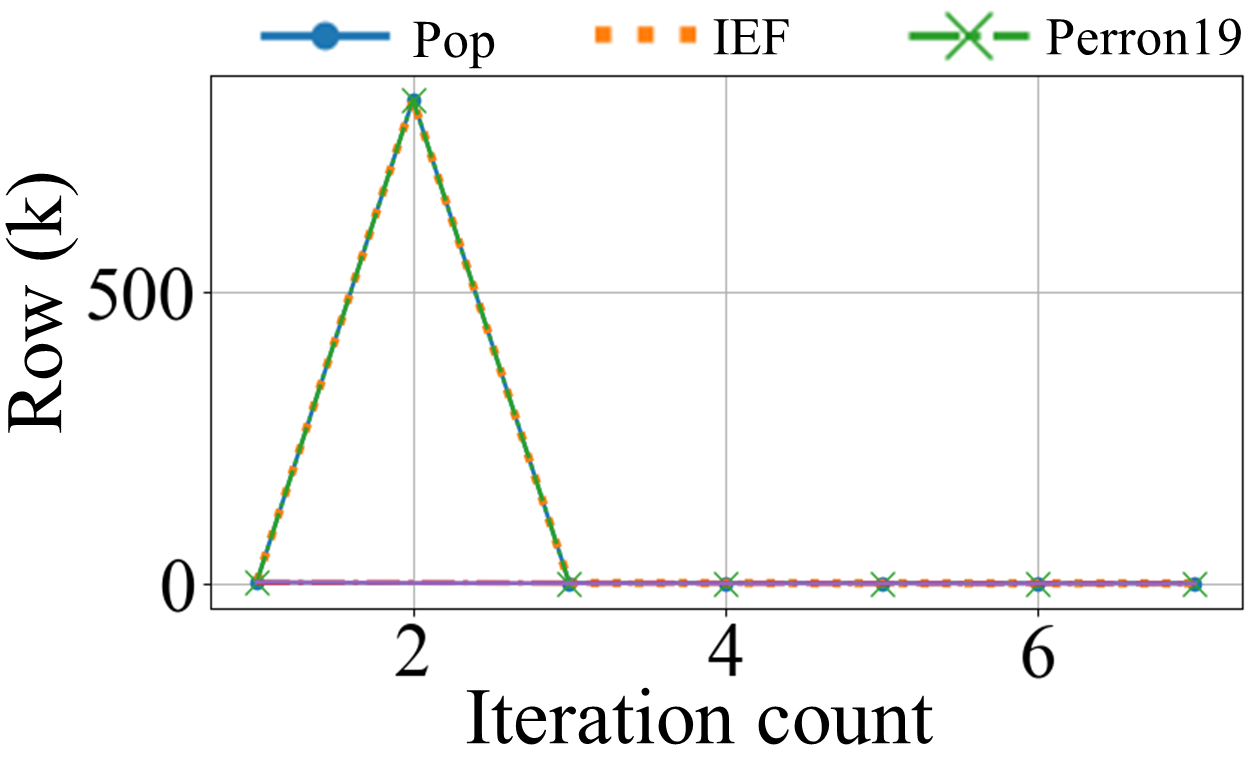}
                \label{F23a}
            \end{minipage}
        }
        \subfigure[Execution time at each iteration]
        {
            \begin{minipage}[t]{0.47\linewidth}
                \includegraphics[width=\linewidth]{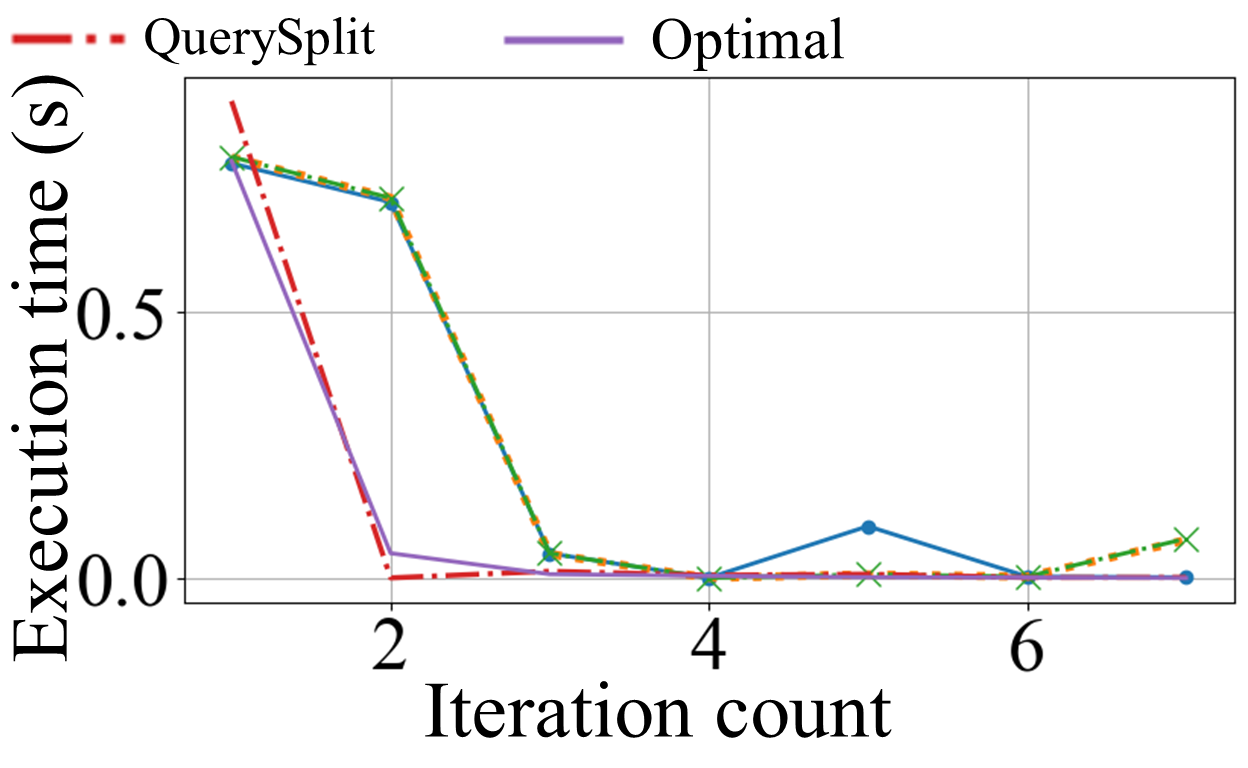}
                \label{F23b}
            \end{minipage}
        }
        \centering
        \caption{Re-optimization Timelines for the ``Avoided Large Join'' Category}
        \label{F23}
        \Description{}
    \end{figure}
    
\subsubsection{Avoided Large Join}
\cref{F23} shows the intermediate result size and execution time at each
re-optimization iteration for a representative query in this category.
We can see that \texttt{Pop}, \texttt{IEF} and \texttt{Perron19} execute a subquery
that generates a large intermediate result close to 1M rows in an early
(2nd) iteration.
The execution time of this subquery is also large, as shown in \cref{F23b}.
The reason is that both algorithms rely on a bad initial plan (generated by
the \postgres's default optimizer) that decides to execute this large join
early because of cardinality estimation errors.
On the other hand, \ourname successfully avoided the large join by first
executing simple subqueries that imposed highly-selective filters on
the large input relations.
As expected, these wise choices perfectly overlap with the optimal plan.

     \begin{figure}[t!]
        \subfigure[Result size at each iteration]
        {
            \begin{minipage}[t]{0.47\linewidth}
                \includegraphics[width=\linewidth]{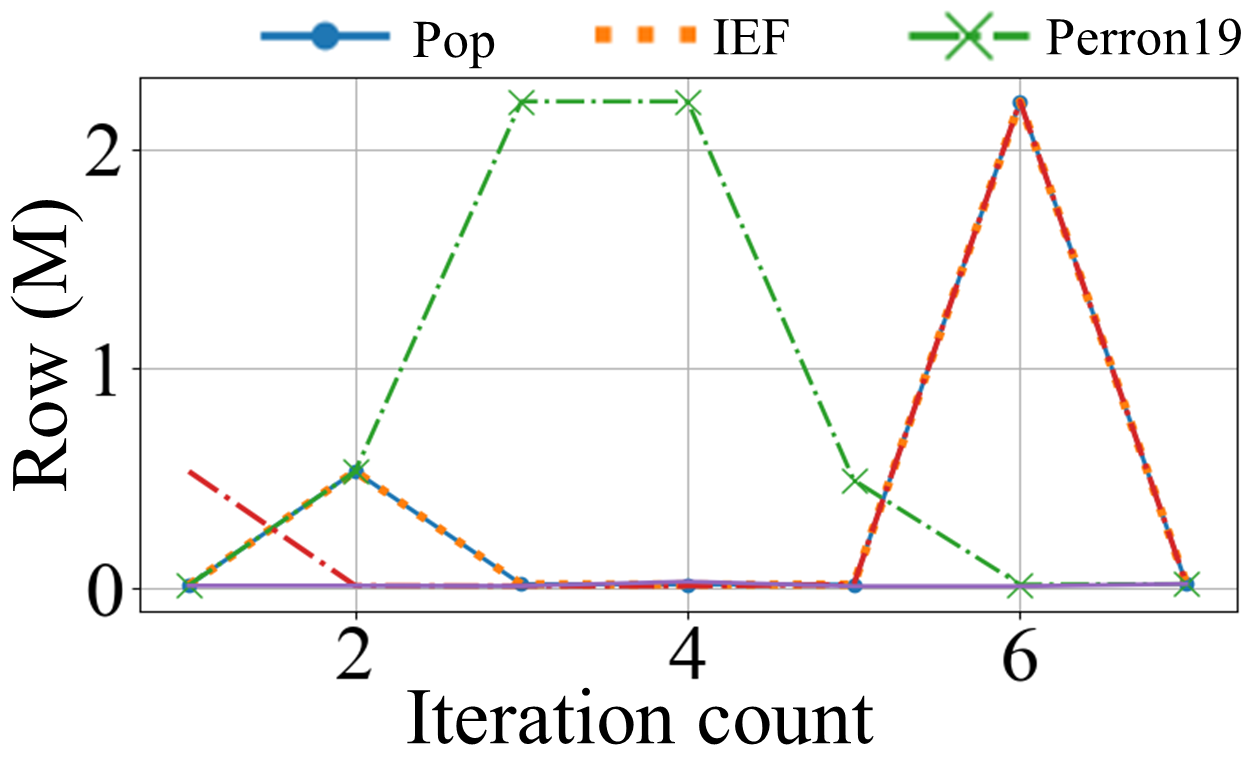}
                \label{F24a}
            \end{minipage}
        }
        \subfigure[Execution time at each iteration]
        {
            \begin{minipage}[t]{0.47\linewidth}
                \includegraphics[width=\linewidth]{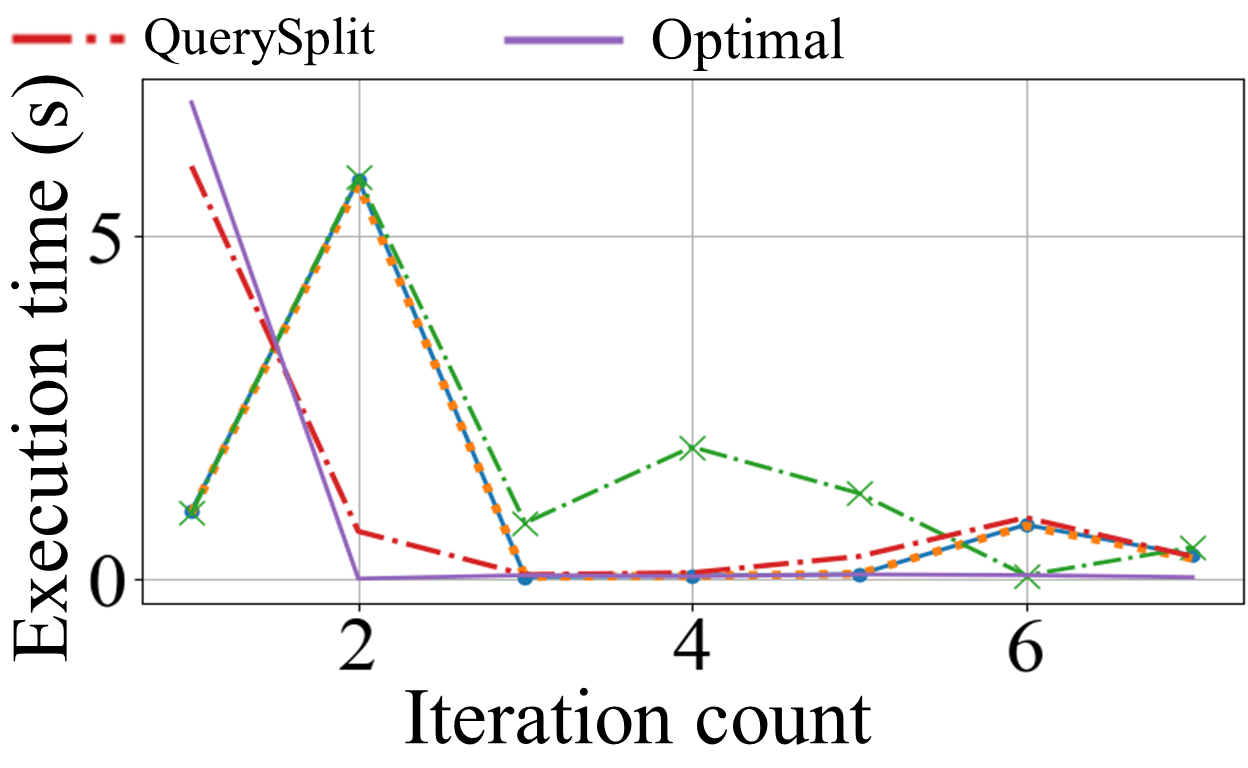}
                \label{F24b}
            \end{minipage}
        }
        \centering
        \caption{Re-optimization Timelines for the ``Delayed Large Join'' Category}
        \label{F24}
        \Description{}
    \end{figure}
    
    \begin{figure}[!t]
        \subfigure[Result size at each iteration]
        {
            \begin{minipage}[t]{0.47\linewidth}
                \includegraphics[width=\linewidth]{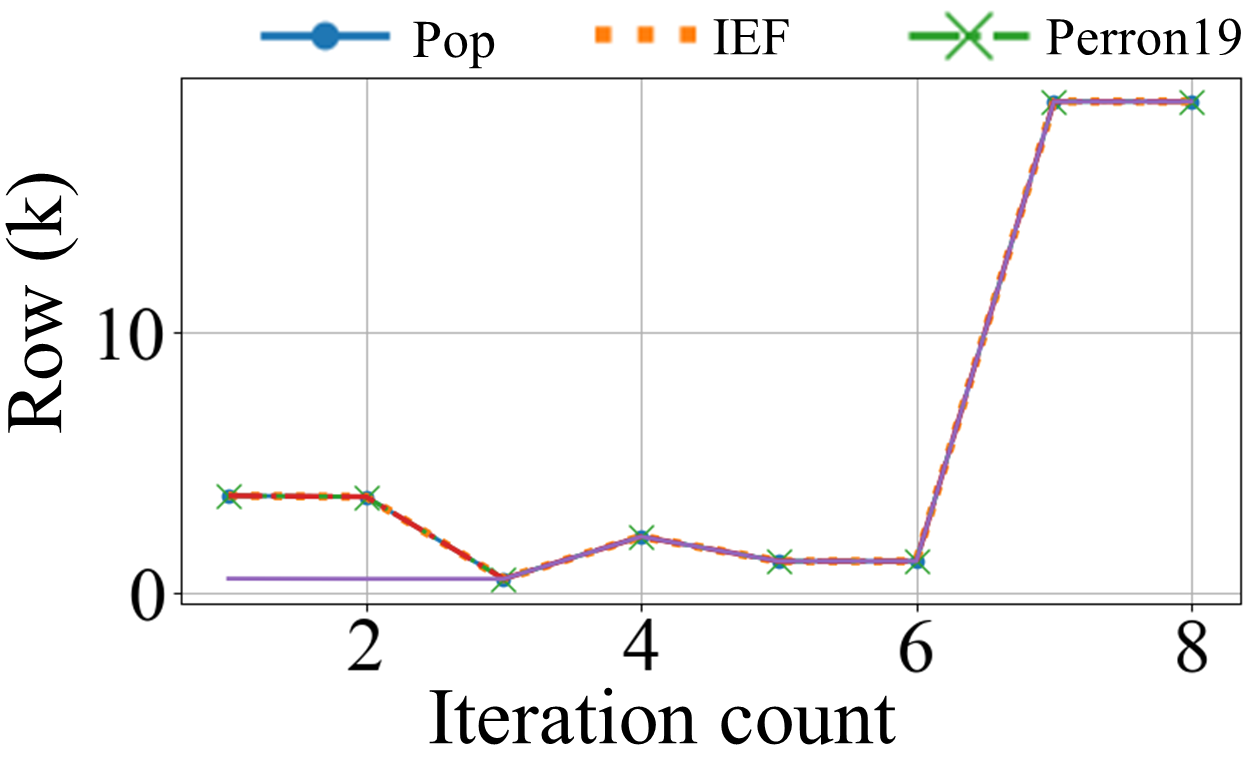}
            \end{minipage}
        }
        \subfigure[Execution time at each iteration]
        {
            \begin{minipage}[t]{0.47\linewidth}
                \includegraphics[width=\linewidth]{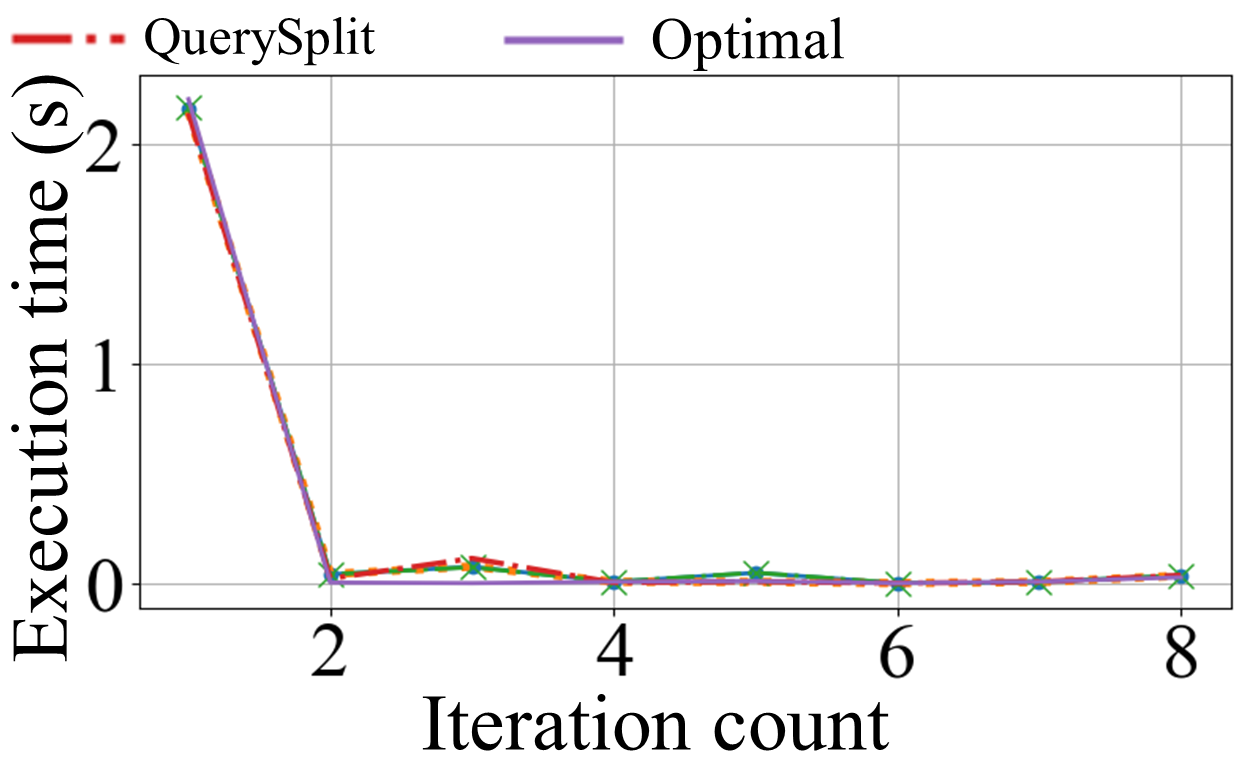}
            \end{minipage}
        }
        \centering
        \caption{Re-optimization Timelines for the ``No Difference'' Category}
        \label{F28}
        \Description{}
    \end{figure}
    
    \begin{figure}[!t]
        \subfigure[Result size at each iteration]
        {
            \begin{minipage}[t]{0.47\linewidth}
                \includegraphics[width=\linewidth]{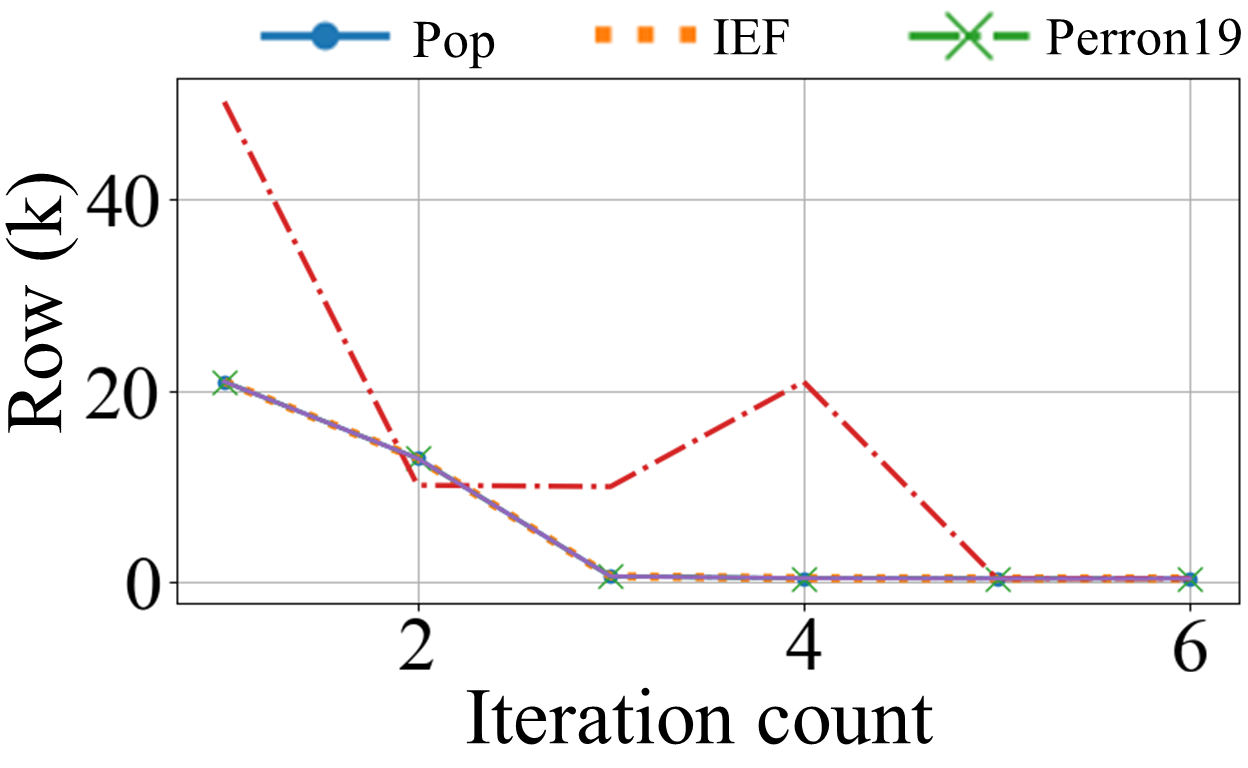}
                \label{F25a}
            \end{minipage}
        }
        \subfigure[Execution time at each iteration]
        {
            \begin{minipage}[t]{0.47\linewidth}
                \includegraphics[width=\linewidth]{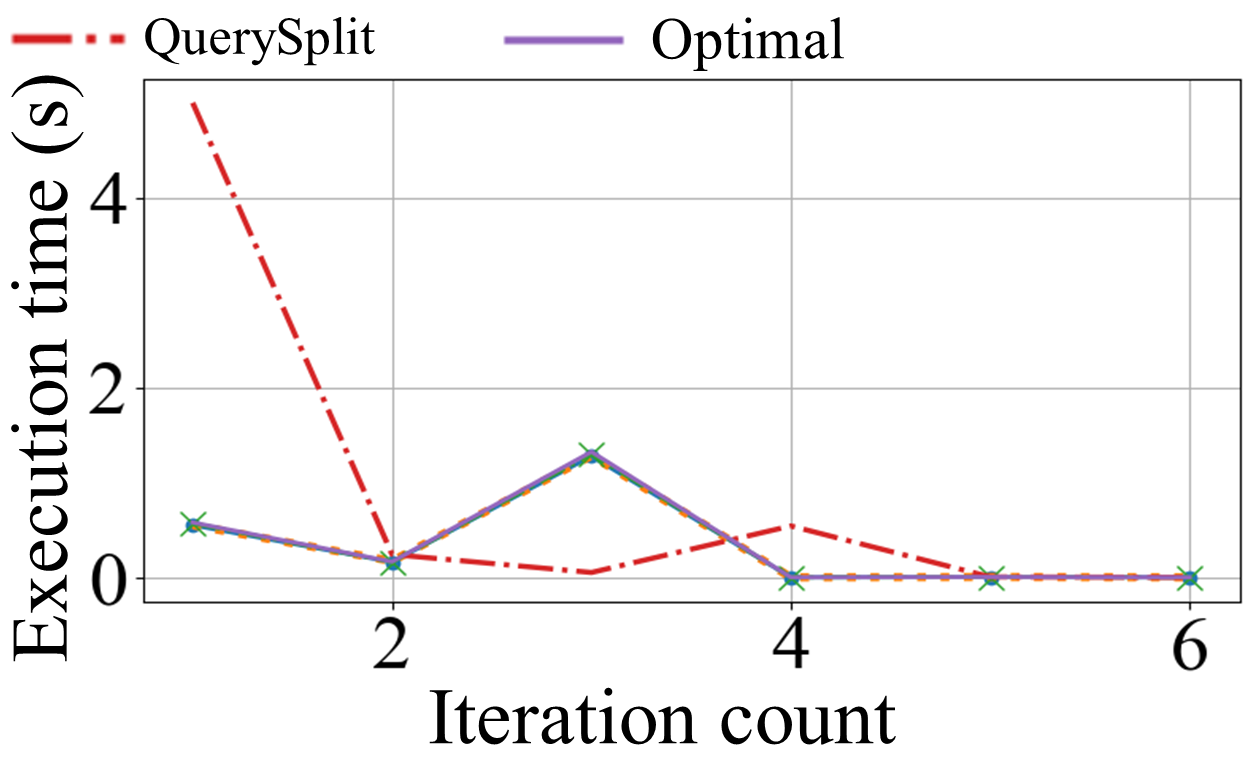}
                \label{F25b}
            \end{minipage}
        }
        \centering
        \caption{Re-optimization Timelines for the ``Worse'' Category}
        \label{F25}
        \Description{}
    \end{figure}
    
\subsubsection{Delayed Large Join}
A representative set of re-optimization timelines for this category
is presented in \cref{F24}.
The observation is that all of the re-optimization algorithms execute
at least one subquery that generates a large intermediate result.
\ourname, however, delays such an execution by as much as possible
because the cost function $\Phi_4$ used in the SSA module
has a strong preference for subqueries that are ``easy'' and produce
small outputs.
Delaying the execution of large joins reduces the probability of
letting them slow down additional subqueries.
As shown in \cref{F24a}, because \texttt{Perron19} executes a large join
as early as in the third iteration, its succeeding iteration suffers
from large input and output sizes again due to the ripple effect.
    
    \begin{figure*}[!t]
        \subfigure[Join Graph of JOB query $\#$9c]
        {
            \begin{minipage}[t]{0.23\linewidth}
                \includegraphics[width=\linewidth]{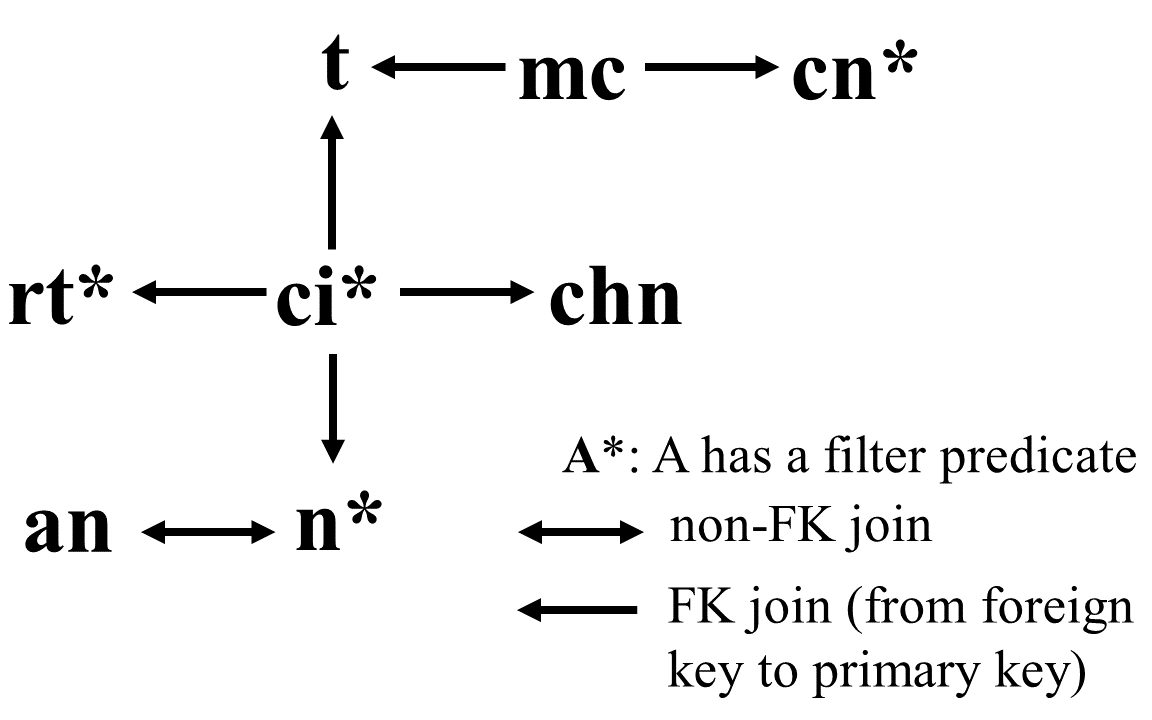}
                \label{F27a}
            \end{minipage}
        }
        \subfigure[Perron19]
        {
            \begin{minipage}[t]{0.23\linewidth}
                \includegraphics[width=\linewidth]{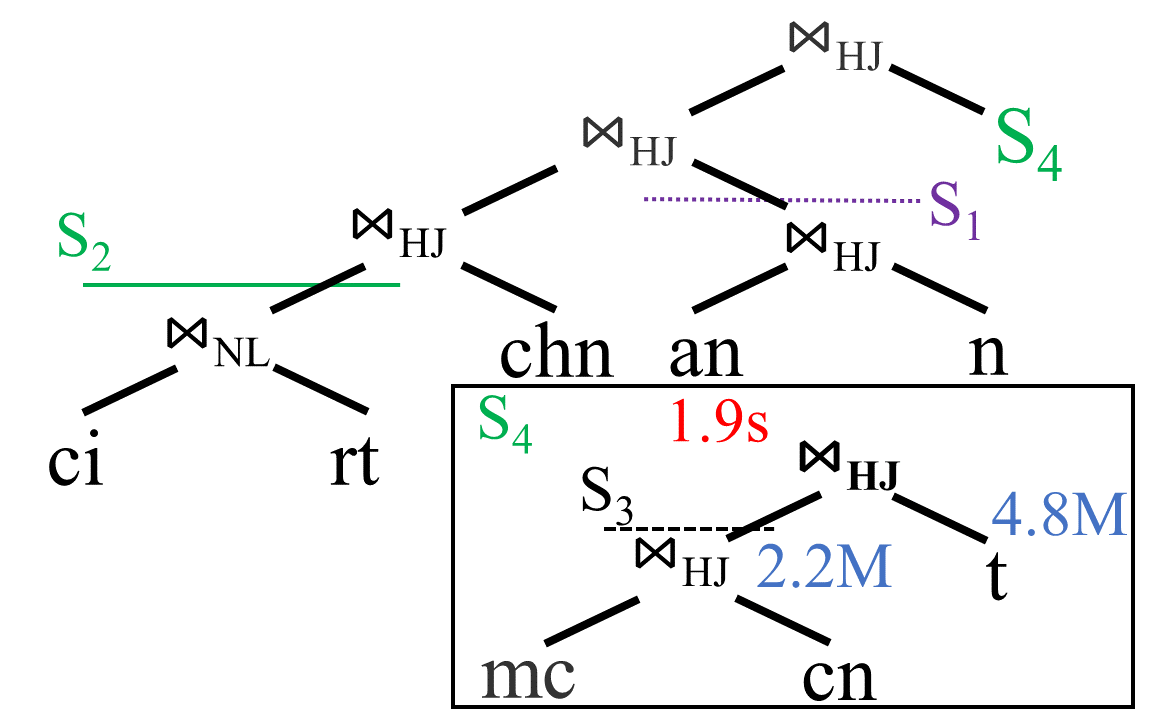}
                \label{F27b}
            \end{minipage}
        }
        \subfigure[\ourname]
        {
            \begin{minipage}[t]{0.23\linewidth}
                \includegraphics[width=\linewidth]{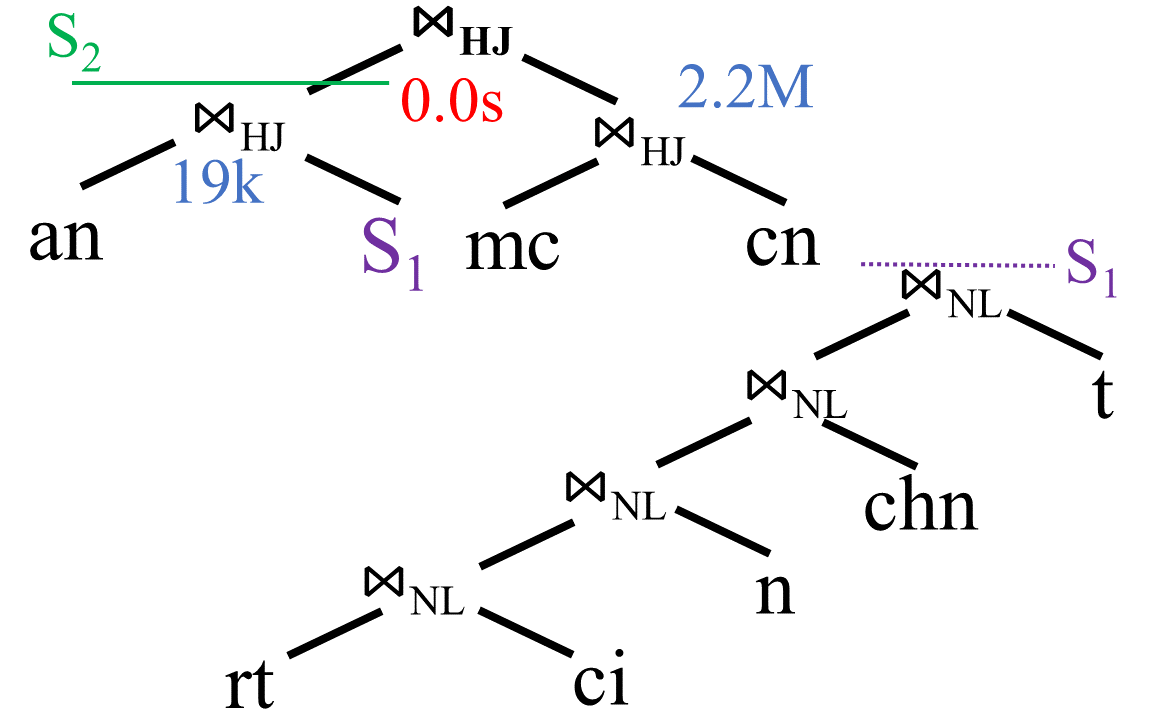}
                \label{F27c}
            \end{minipage}
        }
        \subfigure[Optimal]
        {
            \begin{minipage}[t]{0.23\linewidth}
                \includegraphics[width=\linewidth]{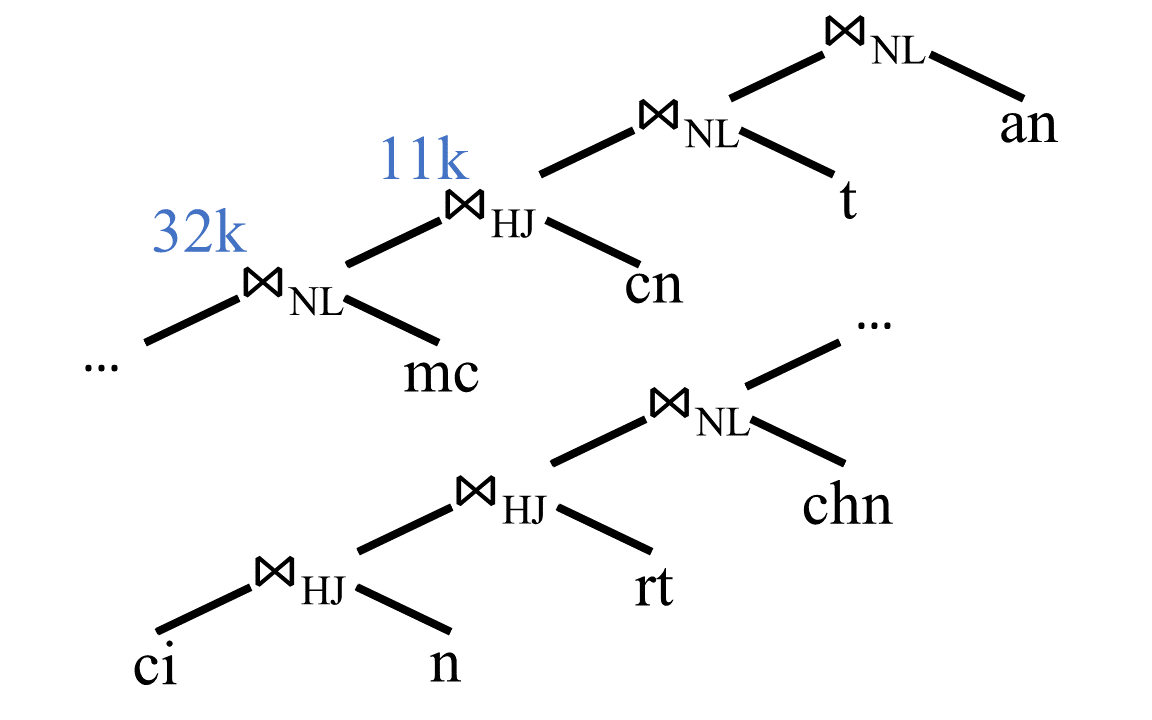}
                \label{F27d}
            \end{minipage}
        }
        \centering
        \caption{(a)The join graph of JOB query \#9c and (b)(c) execution processes of the example query in Delayed Large Joins (The blue and red text represents the actual cardinality and the execution time respectively, and the cutting lines represent where the re-optimization is triggered)}
        \label{F27}
        \Description{}
    \end{figure*}
    
We show the concrete query in \cref{F27} for a case study.
\cref{F27a} shows the join graph of the JOB query.
The (effective) execution plan for \texttt{Perron19}, \ourname,
and \texttt{Optimal} are illustrated in \cref{F27b,F27c,F27d}, respectively.
Both \texttt{Perron19} and \ourname encounter the large join
\texttt{mc} $\Join$ \texttt{cn} that produces an output relation
\textbf{x} with a size of $2.2$M)
in one of their subqueries.
However, because \texttt{Perron19} performs this join too early,
its subsequent join (i.e., \textbf{x} $\Join$ \texttt{t}) becomes
even larger ($2.2$M $\times$ $4.8$M) and takes $1.9$s to complete.
On the contrary, \ourname delays the execution of \texttt{mc} $\Join$ \texttt{cn}
towards the end and gets rewarded by having a much smaller-scale
subsequent join ($2.2$M $\times$ $19$K) that can be finished in
less than $0.1$s.

An interesting observation is that the large \texttt{mc} $\Join$ \texttt{cn}
is completely avoided in the optimal plan.
By comparing the plans between \ourname and \texttt{Optimal} carefully,
we found that the decisive mistake made by \ourname is choosing to
execute \texttt{an} $\Join$ \texttt{$S_1$} first instead of \texttt{mc} $\Join$ \texttt{$S_1$}.
This is because \texttt{an} $\Join$ \texttt{$S_1$} has a much smaller
estimated cost from the optimizer (i.e., $46$K vs. $132$K for $\textbf{C}(q)$)
and a similar estimated output size (i.e., $19$K vs. $18$K for $\textbf{S}(q)$)
compared to \texttt{cn} $\Join$ \texttt{mc} $\Join$ \texttt{$S_1$}.
A more sophisticated cost function $\Phi$ might be able to further
reduce these undesirable decisions.
Nevertheless, we notice that the differences in query execution time
between \ourname and \texttt{Optimal} are small despite our
algorithm generating a larger intermediate result.

\subsubsection{No Difference}
As shown in \cref{F28}, in this category, all the re-optimization algorithms
converge to the same (effective) execution plan.
This is because the cardinality estimation of these queries is relatively
accurate to prevent the optimizer and the re-optimizing process from making
mistakes.

\subsubsection{Worse}
This is a relatively infrequent category where \ourname is slower
than the other competitors because of a bad decision leading to
a large intermediate result.
The re-optimization timelines of a representative query are shown
in \cref{F25}.
\ourname consumes more time and produces larger outputs than the
others in the first and fourth iterations.

\begin{figure*}[!t]
        \subfigure[Join graph of JOB query $\#$15c]
        {
            \begin{minipage}[t]{0.32\linewidth}
                \includegraphics[width=\linewidth]{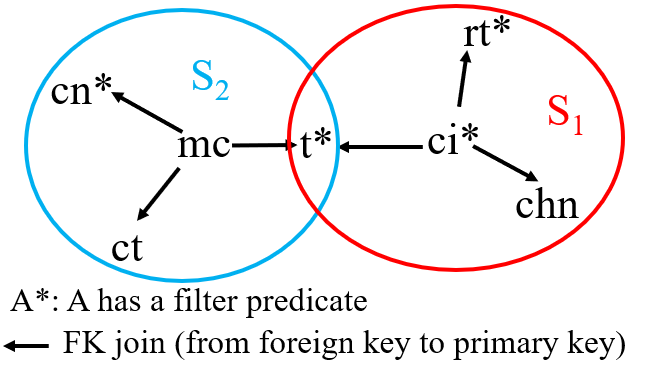}
                \label{F26a}
            \end{minipage}
        }
        \subfigure[Execution process of \ourname]
        {
            \begin{minipage}[t]{0.32\linewidth}
                \includegraphics[width=\linewidth]{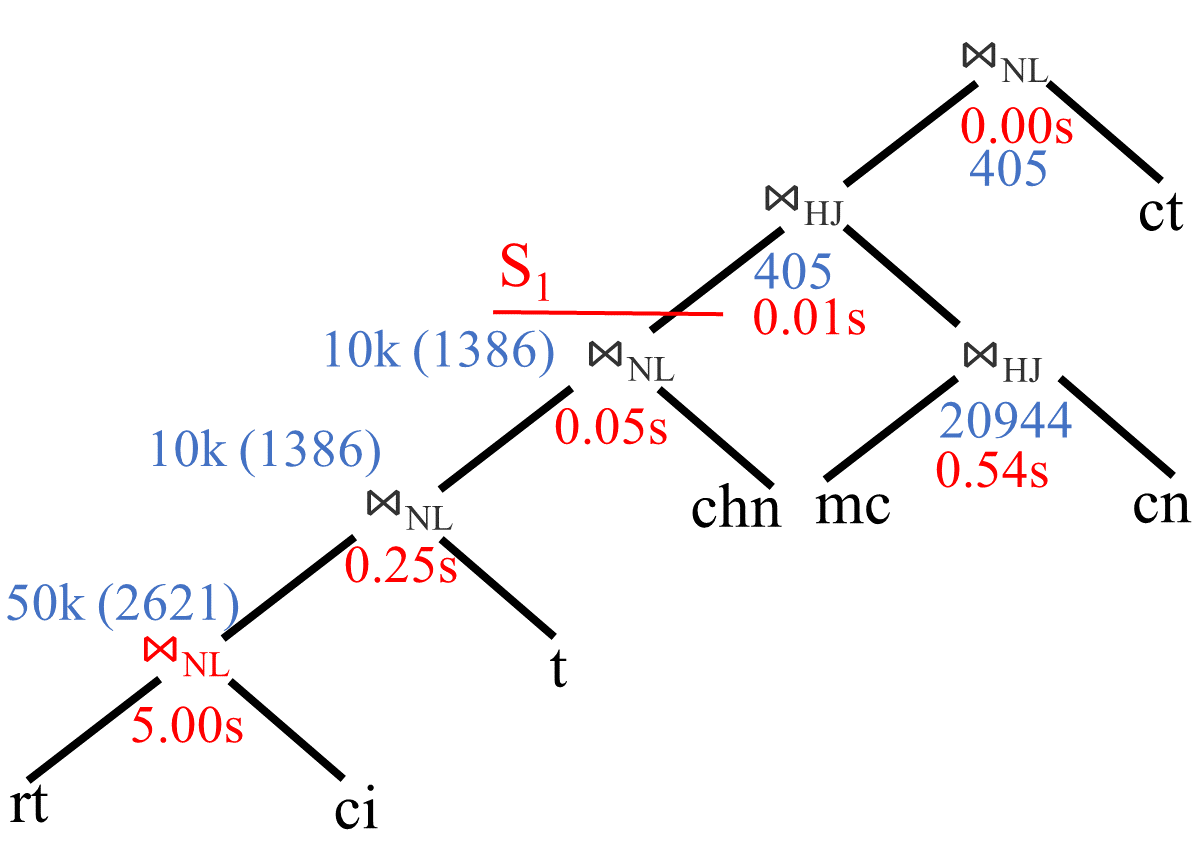}
                \label{F26b}
            \end{minipage}
        }
        \subfigure[Execution process of IEF]
        {
            \begin{minipage}[t]{0.32\linewidth}
                \includegraphics[width=\linewidth]{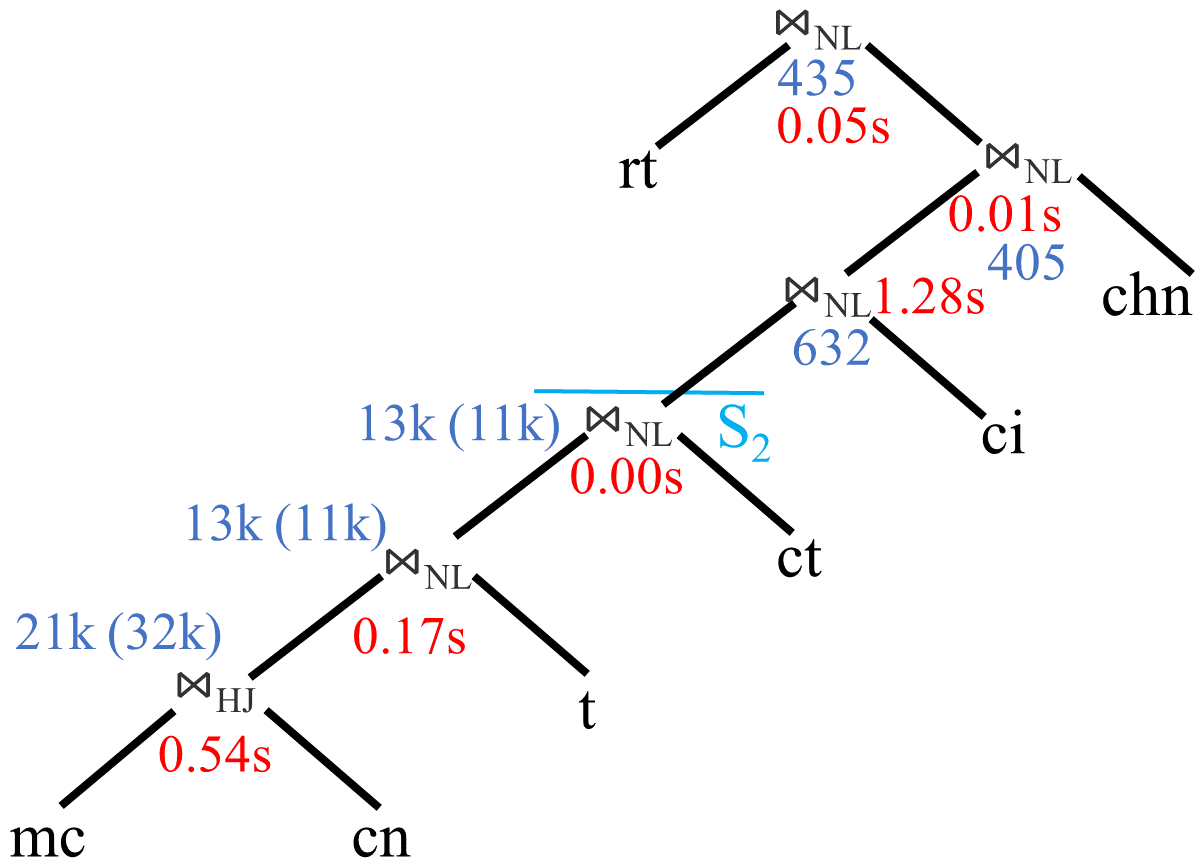}
                \label{F26c}
            \end{minipage}
        }
        \centering
        \caption{(a) The join graph of JOB query \#15c and (b)(c) the execution processes of the example queries in Worse (The blue text outside the bracket and in the bracket represents the actual cardinality and the estimated cardinality respectively, the red text represents the execution time, and the cutting lines represent where the re-optimization is triggered)}
        \label{F26}
        \Description{}
    \end{figure*}

We found that almost all the bad cases are small queries, each
getting split into only two subqueries in \ourname.
\cref{F26} shows an example.
Again, \cref{F26a} shows the join graph of the JOB query.
We depict the (effective) execution plan for \ourname and \texttt{IEF}
in \cref{F27b,F27c}, respectively.
Other alternative algorithms have the same plan as \texttt{IEF}.

We observe that both \ourname and \texttt{IEF} generate the identical
subqueries: \texttt{S$_1$} $=$ \texttt{ci} $\Join$ \texttt{rt} $\Join$ \texttt{t} $\Join$ \texttt{chn}
and \texttt{S$_2$} $=$ \texttt{mc} $\Join$ \texttt{cn} $\Join$ \texttt{t} $\Join$ \texttt{ct}.
The difference is that \ourname chose to execute \texttt{S$_1$} first,
while \texttt{IEF} chose to prioritize \texttt{S$_2$}.
This is because \postgres's optimizer makes a huge mistake in estimating
the cardinality of \texttt{S$_1$}.
Such a mistake ``tricks'' \ourname into believing that \texttt{S$_1$} has a
much smaller output size than \texttt{S$_2$} ($1386$ vs. $11$K),
an advantage outweighing the difference between their execution cost
($5.3$s vs. $0.71$s).
Considering the true cardinality for \texttt{S$_1$} and \texttt{S$_2$}
are similar ($10$K vs. $13$K),
\ourname made a bad decision in executing the heavier \texttt{S$_1$} first.

The lesson learned (for future improvements) is that fine-grained subqueries
are preferred in re-optimization because they are less likely to cause
devastating cardinality estimation errors even with a mediocre optimizer.

\subsubsection{Summary}

     \begin{table}[!t]
        \small
        \caption{Frequencies and the average performance effect of the four
        categories of JOB queries}
        \label{T6}
        \begin{tabular}{c|c|c}
            \toprule
            Category & Frequency & Average Perf. Effect \\
            \midrule
            Avoided Large Join & 40 / 91 & 40.5\% \\
            Delayed Large Join & 23 / 91 & 21.7\% \\
            No difference & 18 / 91 & 3.8\%  \\
            Worse & 10 / 91 & -39.5\%  \\
            \bottomrule
        \end{tabular}
    \end{table}
    
\cref{T6} presents the query count of each of the above categories out of
the 91 JOB queries.
The average performance effect refers to \ourname's relative performance
improvement over the best alternative algorithm.
The query counts show that $\approx 70\%$ of the queries belong to the
first two categories where \ourname outperforms alternative re-optimization
algorithms by a sizeable gap.
Although queries get slowed down significantly in the \texttt{Worse} category,
it has a small effect on the overall benchmark performance because
such a query is infrequent and the query itself is small.

\section{Related Work}
\label{S:rel}
There are two research directions related to our work:
(a) adaptive query processing and (b) cardinality estimation techniques.
We review existing work in these directions
in the following two subsections.
\subsection{Adaptive Query Processing} \label{S81}
    Adaptive query processing is a research direction with a long history. Babu and Bizarro~\cite{babu2005adaptive} have conducted a comprehensive review of existing works in this direction. According to their investigation, adaptive query processing techniques can be broadly categorized into three families: plan-based system (re-optimization), routing-based system and continuous-query-based system.
    
    \subsubsection{Re-optimization}
    A re-optimization system monitors the execution of the current plan and re-optimizes the plan whenever the actual condition differs significantly from the estimations made by the optimizer.
    
    As far as we know, Reopt is the first research that proposed the idea of re-optimization. Reopt~\cite{kabra1998efficient} adds statistics collecting operators after pipeline breakers (e.g., hash or sort) in the physical plan. When a deviation is detected, the database calculates the benefit of re-planning the remaining part of the query and compares it with the cost of re-optimization. Pop~\cite{markl2004robust} is very similar to Reopt, except that Pop can trigger re-optimization in more join nodes, like the outer side of nest-loop join. Instead of deciding when to materialize based on the node type, incremental execution framework (IEF)~\cite{neumann2013taking} chose the node in the global physical plan which has the maximal estimation error on cardinality to materialize. The estimation error on cardinality is estimated based on the statistics and assumptions used. Recently, Perron et al.~\cite{perron2019learned} conducts a simulation study to investigate the effectiveness of re-optimization. They use the \texttt{EXPLAIN} command to evaluate cardinality estimation error, and materialize the intermediate results that deviate too much from estimation as a temporary table. Their result shows that re-optimization can sharply improve the execution time in PostgreSQL. Moreover, Databricks and Spark extended re-optimization to the Mapreduce background~\cite{AQPSpark}. They re-optimized at shuffle or broadcast exchange, and optimized not only join order and physical operator selection, but also shuffle partitions.
    
    Compared to the \ourname framework, the above methods choose the subtree of the global physical plan to execute. Such a strategy can go wrong when the referencing global plan deviates largely from an optimal one. And if a bad subplan is chosen, the damage often influences later execution.

    \subsubsection{Routing-based system} Routing-based systems behave differently compared to traditional RDBMS. They process queries by routing tuples through a pool of operators. The idea of the routing-based system can be traced back to INGRES~\cite{wong1976decomposition}. The most representative work is Eddies~\cite{avnur2000eddies}, which adds a new operator called ripple join and can change the join order in ripple joins.
    Compared to re-optimization, routing-based systems totally abandon the optimizer, making routing algorithms highly dependent on the greedy algorithm and therefore unsuitable for complex queries~\cite{ioannidis1997parametric, trummer2021skinnerdb}.
    
    \subsubsection{Continuous-query-based System} Continuous-Query-based, or CQ-based, systems are used for queries that will run many times or a very long time, which is prevalent in data stream systems.
    Compared to other adaptive query processing, CQ-base systems pay attention to the runtime change of stream characteristics and system conditions, rather than cardinality estimation errors of a given query.
    
\subsection{Cardinality Estimation} \label{S82}
    Cardinality estimation techniques are relevant but orthogonal to our work. \ourname benefits from the improvement of cardinality estimation on small joins. Making an optimal plan in each subquery can significantly enhance overall performance.
    
    Cardinality estimation techniques can be categorized into traditional methods and learned methods~\cite{sun2021learned}, depending on whether machine learning techniques are used. 
    
    \subsubsection{Traditional Methods}
    Traditional methods include sketch~\cite{rusu2008sketches, cai2019pessimistic, hertzschuch2021simplicity}, histogram~\cite{gunopulos2005selectivity} and sampling~\cite{leis2017cardinality, wu2016sampling}. One particularly related work in this part is USE~\cite{hertzschuch2021simplicity}, in which the idea of using non-expanding operators (e.g., filter and Pk-Fk join) is independently proposed. In USE, these operators are prioritized to form subqueries (which is similar to subqueries formed in RCenter). Then, sketch-based cardinality estimation techniques are used to decide the join order between subqueries. However, USE is not an adaptive query processing method, and it uses standard query optimization after conducting the above query transformation.

    \subsubsection{Learned Methods}
    Learned methods can be further divided into two categories: data-driven cardinality estimator~\cite{yang13deep, tzoumas2011lightweight, gunopulos2005selectivity, hilprecht2019deepdb, sun2019end, yang2020neurocard, kipf2019estimating, hasan2020deep} and query-driven cardinality estimator~\cite{sun2019end, stillger2001leo, heimel2015self, kipf2018learned, park2020quicksel, marcus12neo, dutt2020efficiently, ortiz2018learning}. The data-driven cardinality estimator approximates the data distribution of a table by mapping each tuple to its probability of occurrence in the table. The query-driven cardinality estimator uses some models to learn the mapping between queries and cardinalities.
    Although learned methods are indeed more accurate than traditional methods, they often suffer from high training and inference costs~\cite{sun2021learned}.

\section{Conclusion}
\label{S:con}
In this paper, we propose \ourname, a re-optimization framework which ignores the potentially misleading global plans and instead extracts subqueries directly from the original logical plan. We proposed a cost function that prioritizes the execution of simple subqueries with small output sizes. Experimental results on Join Order Benchmark showed that \ourname outperforms other re-optimization methods and state-of-the-art sketch-based cardinality estimation techniques, and reaches near-optimal execution time.
    
\bibliographystyle{ACM-Reference-Format}
\bibliography{ref}

\appendix
\section{Proof of Theorem 1}
    This appendix shows the proof of Theorem 1.
    \begin{Theorem}
        Let $q(\textbf{\textit{R}}, \textbf{\textit{P}})$ be an SPJ query, $\textbf{\textit{Q}}$ be a set of subqueries of $q$.
        \ourname produces the same output as $q$ if $\textbf{\textit{Q}} \rightharpoonup_c q$.
    \end{Theorem}
    
    \begin{Proof}
        \ \newline 
        \indent To begin with, let us introduce several notations:
        \begin{enumerate}[leftmargin = 15pt]
            \item For a set of subqueries $\textbf{\textit{Q}}=\{q_1(\textbf{\textit{R}}_1,\textbf{\textit{P}}_1),...,q_n(\textbf{\textit{R}}_n,\textbf{\textit{P}}_n)\}$, we denote $R(\textbf{\textit{Q}})=\cup_{i=1}^n \textbf{\textit{R}}_i$, $P(\textbf{\textit{Q}})=\cup_{i=1}^n \textbf{\textit{P}}_i$.
            \item For a set of relations $\textbf{\textit{R}}=\{r_1,...,r_n\}$, we denote $\mathsf{X}_{r \in \textbf{\textit{R}}}=r_1 \times ... \times r_n$.
            \item For a SPJ query $q(\textbf{\textit{R}},\textbf{\textit{P}})$ and a set of subqueries $\textbf{\textit{Q}}$ of $q$, we denote the result of $q$ as $E(q)=\sigma_{\textbf{\textit{S}}}(\mathsf{X}_{r \in \textbf{\textit{R}}})$, and the output of the \ourname algorithm as $E(\textbf{\textit{Q}})$.
        \end{enumerate}\par
        \indent Under these notations, we can rewrite the theorem as: Given a SPJ query $q(\textbf{\textit{R}},\textbf{\textit{P}})$, and a set of subqueries $\textbf{\textit{Q}}$ of $q$, such that $\textbf{\textit{Q}} \rightharpoonup_c q$. Then we have $E(q)=E(\textbf{\textit{Q}})$.\newline
        \indent Without loss of generality, we assume that the names of all attributes in $\textbf{\textit{R}}$ are unique. Under such assumption, we do not need to consider the rename operation when modifying subqueries, and for simplicity we assume that the rename step is skipped.\newline
        \indent Now, we start to prove the rewritten theorem by induction on $|\textbf{\textit{Q}}|$.\newline
        \indent First, We prove the statement holds when $|\textbf{\textit{Q}}|=1$, in which case $\textbf{\textit{Q}}=\{q_1(\textbf{\textit{R}}_1,\textbf{\textit{P}}_1)\}$. Apparently the only way that $\textbf{\textit{Q}} \rightharpoonup_c q$ is $q_1=q$. So the statement clearly holds for $|\textbf{\textit{Q}}|=1$.\newline
        \indent Now, assume that the statement holds when $|\textbf{\textit{Q}}|=n-1$. We consider the case of $|\textbf{\textit{Q}}|=n$, $\textbf{\textit{Q}}=\{q_1(\textbf{\textit{R}}_1,\textbf{\textit{P}}_1),...,q_n(\textbf{\textit{R}}_n,\textbf{\textit{P}}_n)\}$.\newline
        \indent Without loss of generality, we denote the first executed subquery as $q_1(\textbf{\textit{R}}_1,\textbf{\textit{P}}_1)$ and discuss two cases: (1) $\forall i > 1, \textbf{\textit{R}}_1 \cap \textbf{\textit{R}}_i=\emptyset$ and (2) $\exists i > 1, s.t. \textbf{\textit{R}}_1 \cap \textbf{\textit{R}}_i \neq \emptyset$.\newline
        \textbf{Case 1}: We first execute $q_1$ and materialize its result as relation $m_1=E(q_1)$. Then, because $\forall i > 1, \textbf{\textit{R}}_1 \cap \textbf{\textit{R}}_i=\emptyset$, according to the algorithm, we have to add $m_1$ to the subquery result set $\textbf{\textit{L}}$. After that, we remove $q_1$ from $\textbf{\textit{Q}}$ and have a new subquery set $\textbf{\textit{Q}}'=\{q_2(\textbf{\textit{R}}_2,\textbf{\textit{P}}_2),...,q_n(\textbf{\textit{R}}_n,\textbf{\textit{P}}_n)\}$ for the next iteration.\newline
        \indent We construct a SPJ query $q'(\textbf{\textit{R}}',\textbf{\textit{P}}')$, where $\textbf{\textit{R}}'= \cup_{i=2}^{n} \textbf{\textit{R}}_i$ and $\textbf{\textit{P}}'=\cup_{i=2}^{n} \textbf{\textit{P}}_i$. Apparently, $\textbf{\textit{Q}}' \rightharpoonup_c q'$ and as $|\textbf{\textit{Q}}'|=n-1$, by induction hypothesis, $E(q')=E(\textbf{\textit{Q}}')$.\newline
        \indent  The final result is the Cartesian product on the elements in $\textbf{\textit{L}}$, so we have:
        $$E(\textbf{\textit{Q}})=\mathsf{X}_{r \in \textbf{\textit{L}}}=m_1 \times \mathsf{X}_{r \in (\textbf{\textit{L}} \setminus \{m_1\})}=E(q_1) \times E(\textbf{\textit{Q}}')=E(q_1) \times E(q')$$
        $$=\sigma_{\textbf{\textit{P}}_1}(\mathsf{X}_{r \in \textbf{\textit{R}}_1}) \times \sigma_{\textbf{\textit{P}}'}(\mathsf{X}_{r \in \textbf{\textit{R}}'})=\sigma_{\textbf{\textit{P}}_1 \cup \textbf{\textit{P}}'}(\mathsf{X}_{r \in \textbf{\textit{R}}})=\sigma_{\textbf{\textit{P}}}(\mathsf{X}_{r \in \textbf{\textit{R}}})=E(q)$$
        \textbf{Case 2}: We denote the set of subqueries that need to be modified after executing $q_1$ as $\textbf{\textit{W}}=\{q_k(\textbf{\textit{R}}_k,\textbf{\textit{P}}_k) \in \textbf{\textit{Q}}:k > 1,\textbf{\textit{R}}_1 \cap \textbf{\textit{R}}_k \neq \emptyset\}$.\newline
        \indent After we execute $q_1$ and materialize its result as relation $m_1$, we modify each $q_i \in \textbf{\textit{W}}$ and keep $\textbf{\textit{L}}=\emptyset$. We denote these new-formed subqueries as $q'_i(\textbf{\textit{R}}'_i, \textbf{\textit{P}}'_i)$ and $\textbf{\textit{R}}'_i=\textbf{\textit{R}}_i \setminus \textbf{\textit{R}}_1 \cup \{m_1\}$, $\textbf{\textit{P}}'_i=\textbf{\textit{P}}_i$. These new-formed subqueries form a new set $\textbf{\textit{W}}'$.\newline
        \indent Now, $\textbf{\textit{Q}}$ becomes a new subquery set $\textbf{\textit{Q}}'=\textbf{\textit{Q}} \cup \textbf{\textit{W}}' \setminus \{q_1\} \setminus \textbf{\textit{W}}$. Because $\textbf{\textit{L}} = \emptyset$ at this point, when reconstruction finishes, we have $E(\textbf{\textit{Q}})=E(\textbf{\textit{Q}}')$.\newline
        \indent We construct a new query $q'(\textbf{\textit{R}}',\textbf{\textit{P}}')$, where $\textbf{\textit{R}}'=\textbf{\textit{R}} \setminus \textbf{\textit{R}}_1 \cup \{m_1\}$ and $\textbf{\textit{P}}'=\cup_{i=2}^{n} \textbf{\textit{P}}_i$. We will prove that $E(q')=E(\textbf{\textit{Q}}')$ and $E(q')=E(q)$, hence finishes the proof. To prove $E(q')=E(\textbf{\textit{Q}}')$, by induction hypothesis, we only need to show that $\textbf{\textit{Q}}' \rightharpoonup_c q'$ and $|\textbf{\textit{Q}}'|=n-1$.
        \begin{itemize}[leftmargin = 15pt]
            \item Apparently, $R(\textbf{\textit{Q}}')=R(\textbf{\textit{Q}}) \setminus \textbf{\textit{R}}_1 \cup \{m_1\}=\textbf{\textit{R}} \setminus \textbf{\textit{R}}_1 \cup \{m_1\}=\textbf{\textit{R}}'$. And because $P(\textbf{\textit{Q}}')=\cup_{i=2}^{n} \textbf{\textit{P}}_i=\textbf{\textit{P}}'$, $P(\textbf{\textit{Q}}')$ logical implies $\textbf{\textit{P}}'$, so $\textbf{\textit{Q}}' \rightharpoonup_c q'$.
            \item Notice that $\textbf{\textit{W}} \subseteq \textbf{\textit{Q}}$, $\{q_1\} \subseteq \textbf{\textit{Q}}$, $\textbf{\textit{W}}' \cap \textbf{\textit{Q}}=\emptyset$ and $|\textbf{\textit{W}}|=|\textbf{\textit{W}}'|$, so $|\textbf{\textit{Q}}'|=n-1$.
        \end{itemize}\par
        \indent By using the induction hypothesis, we get $E(q')=E(\textbf{\textit{Q}}')$.\newline
        \indent At last, we need to prove $E(q')=E(q)$:
        $$E(q')=\sigma_{\textbf{\textit{P}}'}(\mathsf{X}_{r \in \textbf{\textit{R}}'})=\sigma_{\textbf{\textit{P}}'}(\mathsf{X}_{r \in (\textbf{\textit{R}}' \setminus \{m_1\})} \times m_1)=\sigma_{\textbf{\textit{P}}'}(\mathsf{X}_{r \in (\textbf{\textit{R}} \setminus \textbf{\textit{R}}_1)} \times m_1)$$
        \indent Since $m_1$ is the execution result of $q_1$, $m_1=\sigma_{\textbf{\textit{S}}_1}(\mathsf{X}_{r \in \textbf{\textit{R}}_1})$, so we have:
        $$E(q')=\sigma_{\textbf{\textit{P}}'}(\mathsf{X}_{r \in (\textbf{\textit{R}} \setminus \textbf{\textit{R}}_1)} \times \sigma_{\textbf{\textit{P}}_1}(\mathsf{X}_{r \in \textbf{\textit{R}}_1}))=\sigma_{\textbf{\textit{P}}}(\mathsf{X}_{r \in \textbf{\textit{R}}})=E(q)$$
        \indent Thus, $E(q)=E(\textbf{\textit{Q}})$ and the statement holds for $|\textbf{\textit{Q}}|=n$.
    \end{Proof} \label{A2}
\section{The Implementation of Perron19'}
    This appendix describes our implementation of Perron19'. Although Perron et al.~\cite{perron2019learned} compare the performance between PostgreSQL, optimal query execution strategy, and a re-optimization strategy, they only do it in a simulation way. In their simulation, they examine the ``EXPLAIN ANALYZE" output of the query and compare the true cardinalities to the PostgreSQL cardinality estimation. For the first join operator in the query plan with a q-error over the threshold, they rewrite this subquery to create a temporary table instead. For the remainder of the query, they replace all the tables in the above join with the temporary table and re-plan. They repeat this procedure until no join operators in the query plan have a q-error over the threshold.

    However, it is impossible to calculate the q-error of each join node in the pipeline during execution in practice. This is because each join operator does not produce all its results until the final result is available. Hence, when we know the exact tuple that a join operator produces, the query has been executed, and it is too late to re-optimize.

    Hence, to implement the simulation of Perron et al. in practice, we have to break the tuple pipeline in the executor thoroughly. To do so, we materialize the result of each intermediate join operator. When the q-error between the cardinality of materialized result and estimated one is over the threshold, which is set to 32 according to the previous result~\cite{perron2019learned}, we first use the routine of the ``analyze" command to collect the statistics of the temporary table. And then we use the temporary table to replace the tables that have been used and re-optimize the modified query. \label{A3}

\end{document}